\documentclass[
aps,%
12pt,%
final,%
notitlepage,%
oneside,%
onecolumn,%
nobibnotes,%
nofootinbib,% 
superscriptaddress,%
noshowpacs,%
floatfix,%
centertags]%
{revtex4}

\usepackage{longtable}
\usepackage{array}
\usepackage{rotate}
\usepackage{rotating}
\usepackage{graphicx}
\usepackage{pdflscape}
\usepackage{epstopdf}
\usepackage{aas_macros}

\newcommand{\kms}{km\,s$^{-1}$}

\input maik.sty 
\usepackage{xcolor}
\usepackage[utf8]{inputenc}
\def\refitem#1{\relax}

\begin{document}
\selectlanguage{english}
%\UseRawInputEncoding

\begin{flushleft}
{\bf УДК 524.527 }
\end{flushleft}

\title{A SURVEY OF HIGH MASS STAR FORMING REGIONS \\ IN THE LINE OF SINGLY DEUTERATED AMMONIA NH$_2$D}% Разбиение на строки осуществляется командой \\

\author{\firstname{E.}~\surname{Trofimova}}
% Здесь разбиение на строки осуществляется автоматически или командой \\
\email{tea@ipfran.ru}
\affiliation{%
Federal Research Center A.V. Gaponov-Grekhov Institute of Applied Physics of the Russian Academy of Sciences (IAP RAS), Nizhny Novgorod, Russia
}%
\author{\firstname{I.}~\surname{Zinchenko}}
\email{zin@ipfran.ru}
\affiliation{%
Federal Research Center A.V. Gaponov-Grekhov Institute of Applied Physics of the Russian Academy of Sciences (IAP RAS), Nizhny Novgorod, Russia
}%

\author{\firstname{P.}~\surname{Zemlyanukha}}
\email{petez@ipfran.ru}
%\noaffiliation % если у автора место работы не указывается
\affiliation{%
Federal Research Center A.V. Gaponov-Grekhov Institute of Applied Physics of the Russian Academy of Sciences (IAP RAS), Nizhny Novgorod, Russia
}%

\author{\firstname{M.}~\surname{Thomasson}}
\email{magnus.thomasson@chalmers.se }
\affiliation{%
Chalmers University of Technology, Onsala Space Observatory, Onsala, Sweden
}%

\date{\today}
%\today печатает cегодняшнее число

\begin{abstract}
The present survey represents a continuation of our study of high mass star forming regions in the lines of deuterated molecules, the first results of which were published in \cite{Trofimova20}. This paper presents the results of observations of 50 objects in the line of ortho modification of singly deuterated ammonia NH$_2$D $1_{11}^s - 1_{01}^a$ at frequency 85.9 GHz, carried out with the 20-m radio telescope of the Onsala Space Observatory (Sweden). This line is detected in 29 sources. The analysis of obtained data, as well as the fact that gas density in the investigated sources, according to independent estimates, is significantly lower than the critical density for this NH$_2$D transition, indicate non-LTE excitation of NH$_2$D. Based on non-LTE modeling, estimates of the relative content of the NH$_2$D molecule and the degree of deuterium enrichment were obtained, and the dependencies of these parameters on temperature and velocity dispersion were analyzed with and without taking into account detection limits assuming the same gas density in all sources. An anti-correlation between the NH$_2$D relative abundances and the kinetic temperature is revealed in the temperature range 15--50~K. At the same time, significant decrease in the ratio of the NH$_2$D/NH$_3$ abundances with increasing temperature, predicted by the available chemical models, is not observed under the adopted assumptions. An anti-correlation was also revealed between the relative content of the main isotopologue of ammonia NH$_3$ and the velocity dispersion, while no statistically significant correlation with the kinetic temperature of sources in the same temperature range was found.

\end{abstract}

\maketitle

\section{Introduction}

The effect of deuterium fractionation in interstellar clouds is due to the exothermicity of proton-deutron replacement reactions in molecules (first of all H$_3^+$), which underlie the chains of chemical reactions leading to the formation of most other molecules (e.g., \cite{Roueff07}). Besides, an important role  in this effect is played by the freezing of molecules such as СО on dust grains at low temperatures which destroy $\mathrm{H_2D^+}$, as well as a decrease of the gas ionization degree, which reduces recombination rate of $\mathrm{H_2D^+}$. This effect is most noticeable and has been actively studied primarily in dark cold clouds, but recently a number of works on its study in high mass star forming regions have appeared (e.g., \cite{Pillai07, Pillai11, Fontani11,  Miettinen11, Gerner15}). Also recently a survey of massive clumps from the ATLASGAL in deuterated ammonia lines was released \cite{Wienen21}.

In 2017--2018 years our survey of several dozens of high mass star forming regions in lines of a number of deuterated molecules in the wavelength range 3--4 mm with the 20-m radio telescope of the Onsala Space Observatory has been completed. The observation results of DCN, DNC, DCO$^+$ и N$_2$D$^+$ were published in the paper \cite{Trofimova20}. In the present work we report the results of the survey in the line of ortho modification of singly deuterated ammonia NH$_2$D. The results are compared with available chemical models. The most detailed model of ammonia deuteration is presented in the work \cite{Roueff05}. Section \ref{sec:obs} gives a selection of observed sources and describes observation and data processing procedures. Section \ref{sec:results} presents the results of observations and estimates of molecular abundances. Section \ref{sec:disc} provides an analysis of the results obtained.

\section{Observations and data processing} \label{sec:obs}
The observations analyzed in this work were described in the paper \cite{Trofimova20}. The main parameters of these observations are summarized below. A total of 50 objects were studied. These sources were selected due to the presence of signs of high mass stars formation and were observed previously by us in other lines \cite{Trofimova20}. A list of objects with their coordinates and alternative names is presented in Table~\ref{tab:sources}. Observations of these sources were carried out in 2017--2018 with the 20-m radio telescope of the Onsala Space Observatory (Sweden). The $1_{11}^s - 1_{01}^a$ ortho-NH$_2$D line (hereafter the designation $1_{11} - 1_{01}$ is used) at a frequency of 85.926278 GHz falls within one of the observed frequency bands. Table \ref{tab:sources} shows only those objects that were observed in this band. For some of them maps in NH$_2$D line were obtained, which are not discussed here. At the observed positions in the sources Per4, G202.99+2.11 and NGC2264 the NH$_2$D line or other lines that could be used to estimate the detection limits of NH$_2$D, were not detected. Therefore, these sources are not used in the further analysis and are not included in Tables \ref{tab:param_NH2D} and \ref{tab:N}.

The NH$_2$D is an asymmetric top molecule. The transitions of this molecule undergo hyperfine splitting due to the quadrupole moment of both the nitrogen nucleus N and the deuterium nucleus D. The latter, however, is very small and is not resolved in astronomical observations usually \cite{Daniel16}. The hyperfine structure of the observed transition, caused by nitrogen nucleus, contains 6 components, but 2 central components merge in our observations, because the interval between them is only 0.16~\kms\ \cite{Tine00, Coudert06}.

The obtained NH$_2$D spectra were processed by approximating them using the HFS method in the CLASS program, which is a part of the GILDAS package developed at the Institute of Millimeter Radio Astronomy\footnote{http://www.iram.fr/IRAMFR/GILDAS}. The estimates of column densities of molecules were carried out using the offline version of the RADEX\footnote{http://var.sron.nl/radex/radex.php} \cite{vdTak07} code. The estimates of correlation coefficients and linear regression taking into account detection limits were performed using the ASURV Rev. 1.2 code \cite{ASURV90, ASURV92}, while without limits using standard routines from Numerical Recipes \cite{Press2007}. For the frequencies of transitions and hyperfine components the databases SPLATOLOGUE\footnote{http://www.cv.nrao.edu/php/splat/} and NIST\footnote{https://physics.nist.gov/cgi-bin/micro/table5/start.pl} were used.

{
\section{Results} \label{sec:results}

\subsection{Observation results} \label{sec:res1}
}
NH$_2$D lines were detected in 29 sources. Figure \ref{ris:NH2D_spec_sm} shows spectra of all sources, where this line was registered. Figure \ref{ris:NH2D_pr_sm} shows spectra of those sources where this line was not registered. As a result of data processing the parameters of the $1_{11} - 1_{01}$ {NH$_2$D} line (velocity, line width and intensity) were determined, which are given in Table \ref{tab:param_NH2D}. Line intensities are presented in the main beam brightness temperature scale. Also in Table \ref{tab:param_NH2D} values of upper limits of intensities of the NH$_2$D lines are given for sources, where they were not detected. The upper limit of NH$_2$D line intensity was determined as 3$\sigma$/$\sqrt{N}$, where $\sigma$ is root mean square (rms) noise value in channels, and $N$ is the number of spectral channels,    which is defined as the ratio of the adopted line width for the limit to the width of the spectral channel. The average widths of narrow lines in source of such molecules as H$_2$CO, HC$_3$N and H$^{13}$CN were taken as line width for the limits. A comparison of the narrow line widths ($\langle \Delta V \rangle$) in source and NH$_2$D line width shows on average the ratio $\langle \Delta V \rangle /\Delta V(\mathrm{NH_2D}) \approx  1.52$ (Fig. \ref{ris:dV-dV}). This means that estimates of upper limits of NH$_2$D line intensities may be underestimated by $\sim$25\%. Then the estimates of upper limits of integral line intensity, which are used for calculations of the column densities, will be overestimated by the same factor, that is not very significant.

The last column of Table \ref{tab:param_NH2D} shows the estimates of optical depth in the line obtained in the CLASS program from the intensity ratios of hyperfine structure line components by HFS method. This value is the sum of optical depths for all components. The optical depth of the central component is 1/2 from this value. For cases where the value of optical depth is too small and can't be defined, Table \ref{tab:param_NH2D} shows the upper limits which are three times values of the uncertainties of the optical depth. It should be noted that in the GILDAS documentation one of the assumptions for the HFS method is the absence of overlap of hyperfine splitting components. For our sources this condition  is not fulfilled in most cases as can be seen in Figure~\ref{ris:NH2D_spec_sm}. However, an analysis of the GILDAS documentation shows that this assumption is probably not mandatory. For checking, we estimated the optical depths for some sources using the approach described in \cite{Pazukhin2023}. The results are similar. 

\bigskip
{
\subsection{The estimates of NH$_2$D column densities}
}
To estimate the column densities of NH$_2$D molecules and determine the degree of deuterium enrichment in the sources under research it is necessary to know the gas kinetic temperature in these regions. The temperatures of several sources were taken from publications \cite{Malafeev05, Pirogov16, Schreyer96, Pirogov93, Harju93, Jijina99, Pillai06, Zin09, Zin97}, the main instrument for estimation of which is the main isotopologue of ammonia NH$_3$. The temperatures of a part of the observed objects were determined by the rotational diagrams of a symmetric-top molecule CH$_3$CCH, the lines of which are visible in these objects. This molecule is a good indicator of temperature of a sufficiently dense gas \cite{Malafeev05}. The temperatures obtained by this method are taken from our previous publication \cite{Trofimova20}. Recently the paper was published with comparison of the methods of temperature estimation from the lines of the main ammonia isotopologue NH$_3$ and from the lines of the methyl acetylene CH$_3$CCH \cite{Pazukhin22}. The results of this research showed a good agreement of estimation results from both methods. For sources, whose temperatures are unknown and no molecular lines have been detected from which the temperature of objects can be determined, the gas kinetic temperatures were taken equal to 20 K. There are 9 such sources in total -- 4 sources with detected NH$_2$D lines and 5 sources where the NH$_2$D line was not detected. It corresponds to the typical dust temperature for several sources of our sample \cite{Pazukhin22}. The gas kinetic temperature of the observed objects used for calculations are given in Table \ref{tab:N}.        
  
The column densities of ortho-NH2D molecules were estimated in the non-LTE model using the offline version of the RADEX code using the line parameters obtained as a result of spectra processing which are given in Table \ref{tab:param_NH2D}. We considered different gas densities from 1$\times$10$^{4}$~см$^{-3}$ to 1$\times$10$^{6}$~см$^{-3}$. This range of densities corresponds to estimates, which were obtained for some sources from this sample from the observations of two transitions of DCN, DNC and DCO$^+$ molecules \cite{Pazukhin2023}. The necessity to use the non-LTE model is due to the following factors. For matching the estimates of the optical depths in the lines with the measured values of antenna temperature there are several possibilities. Firstly, it can be assumed, that the gas kinetic temperature in the regions of NH$_2$D emission is very low, just a few K. This assumption seems completely unrealistic, because it is unclear how such regions could appear in these sources. Secondly, it can be assumed, that the beam filling factor is very low. But this assumption does not agree with the data about the source sizes (\cite{Zinchenko22x,Pazukhin2023}). In principle, the low filling factor may be related to possible inhomogeneous structure of sources. However, available data of small-scale inhomogeneity in such sources (e.g., \cite{Pirogov2008, Pirogov2018}) do not agree with the required very low filling factor. It remains to be assumed, that the excitation temperature of this transition is much lower than the gas kinetic temperature, which makes it necessary to use non-LTE modeling. It should also be noted, that the critical gas density (at which the rates of radiative and collisional transitions are compared \cite{Shirley15}) for this NH$_2$D transition is very high, $\sim 4\times 10^6$~см$^{-3}$ at temperatures 5--20~K according to \cite{Feng19}. For warmer sources with the gas kinetic temperature T$_k$ $\sim40$~K the critical gas density decreases and is close to $\sim$1$\times$10$^{6}$~см$^{-3}$. These values are significantly higher than the above estimates of the gas densities in such sources, that also indicates a necessity to use non-LTE modeling. The estimates of the NH$_2$D column densities, obtained for the critical density, are accepted as lower limits of these values and are given in Table \ref{tab:N}. As shown in \cite{Trofimova20}, exactly at the gas densities close to the critical density the minimum estimates of the column densities are obtained. 

However, the conclusion about non-LTE excitation of NH$_2$D may cast doubt on the estimates of the optical depths in this line, because they are based on the assumption about equality of the excitation temperatures of the hyperfine splitting components. In our opinion, there are no reasons for these doubts. For these excitation temperatures to be equal the LTE isn't required in general. Actually, there are good known cases of ``anomalies'' of the hyperfine structure, where the excitation temperatures of the components vary greatly, for example, in the transitions $J=1-0$ of HCN and (1,1) of NH$_3$. They are explained by some peculiarities of the energy level structure of these molecules and the overlapping of the components \cite{Guilloteau1981, Zinchenko1987, Goicoechea2022, Stutzki1985}. In case of NH$_3$ these anomalies do not hinder for estimations of the optical depth. For NH$_2$D such anomalies are unknown, so the estimates of the optical depth seem reliable.

It is worth noting, that when modeling the NH$_2$D emission by RADEX, the hyperfine splitting of this transition is not taken into account. It can lead to mistakes of the estimates of the column densities. As the input parameters we give the brightness temperature of the central component and the line width. Whereas the satellites, probably, are not taken into account, the obtained value of the column density may be underestimated by a factor of 2. For checking this assumption, we compared the LTE estimates of the column densities, obtained with well-known formulas (e.g., \cite{Busquet10}), with the estimates by RADEX for the very high gas densities ($n = 10^9$~см$^{-3}$), at which the LTE conditions must be met. It turned out, that the RADEX estimates are about 2 times lower, indeed. Therefore, all estimates of the column densities, obtained with RADEX modeling, were multiplied by 2.    

For further estimates of the total column densities of NH$_2$D the calculated column densities of ortho-NH$_2$D were multiplied by the coefficient 1.33, which corresponds to the ratio of the abundances of the ortho- and para- NH$_2$D equal to 3 from spin statistics \cite{Sipila15}. This value agrees well with observational data. The most strict analysis of the observations with taking into account the optical depth in the lines gives the value 3.7$\pm$1.2 for the ratio of the column densities \cite{Wienen21} (for massive clumps from the ATLASGAL survey). The simpler estimates for the ratio of the integral line intensities of ortho- and para-NH$_2$D gave the values 2.6$\pm$0.6 \cite{Fontani2015}, 2.6$\pm$0.8 \cite{Wienen21}, 2.6$\pm$0.7 \cite{Pazukhin2023}. The lower values as compared with the ratio of the column densities are, probably, explained by a higher optical depth in the lines of ortho-NH$_2$D in comparison with the lines of para-NH$_2$D. The values of the total column densities for the molecule of singly deuterated ammonia NH$_2$D taking into account the correction for para-NH$_2$D are given in Table \ref{tab:N}. For several observed objects the information about column densities of ammonia NH$_3$ wasn't found, therefore these objects are not taken into account in this research.    

In the source W75(OH) 2 NH$_2$D lines with different velocities were detected. Each of these lines was processed separately and data obtained as a result of processing are presented for each line in a separate string in Tables~\ref{tab:param_NH2D} and \ref{tab:N}. The line with the velocity --3.84~km/s as the more intense and corresponding to the main component of the source was chosen for further analysis (e.g., \cite{Pazukhin2023}).

The upper limits of the NH$_2$D column densities were calculated in the same way as the column densities of the detected sources, but using the upper limits of the NH$_2$D line intensities and the average widths of the narrow lines of molecules, detected in a specific source as described above. Because these average widths are somewhat higher than the NH$_2$D line widths (Figure~\ref{ris:dV-dV}), the upper limits of the column densities may be overestimated by $\sim$25\%, which is not important. The narrow lines were not detected in the source G81.50+0.14, and the average value of the line widths of the (1,1) and (2,2) NH$_3$ transitions, taken from work \cite{Pillai06}, were used as the line width. The kinetic temperatures of the sources, which were used for calculations of the detection limits of singly deuterated ammonia, are given in Table~\ref{tab:N}.      

When modeling the measured spectra using RADEX, we obtained model values of optical depth in lines, too. At the assumed gas density from 1$\times$10$^{4}$~см$^{-3}$ to 1$\times$10$^{6}$~см$^{-3}$ these model values are significantly lower than the estimates of optical depth in Table~\ref{tab:param_NH2D}. In order to match the model optical depths with the measured ones, there are several possibilities. Firstly, lower gas densities can be assumed. The required value for this is $n\sim3\times10^{3}$~см$^{-3}$. It is at least 3 times lower than the estimates of the gas densities mentioned above in such sources. The estimates from ammonia observations also give higher values usually. However the regions, which make the main contribution to NH$_2$D and NH$_3$ emission, could well be different. It is worth noting, that the uncertainties of the optical depth estimates from observational data in most cases are very large, which doesn't allow us to judge about possible dependence of gas densities, at which it's possible to match the model and measured values of the optical depth, on the gas kinetic temperature given in Table~\ref{tab:N}. Other versions can be a lower gas kinetic temperature in regions of NH$_2$D emission, or a low beam filling factor. These versions were considered and rejected above. Thus, the estimates of the optical depth, most likely, indicate a quite low gas density in those regions where this optical depth is significant. However, we need to consider the fact, that modeling in RADEX doesn't take into account the hyperfine splitting of the transition as noted above. It leads to uncertainties in estimates. For further estimates we adopt $n = 10^4$~см$^{-3}$ as the most likely value of gas density, which corresponds to the estimates mentioned above from the analysis of the molecule excitation and is close enough to that, which is required for matching the measured and model values of optical depth in the line. Nevertheless, this value is at the lower limit of the range of the density values, obtained from the analysis of the molecule excitation. Thus, the values of the column densities obtained by us under the assumption of $n = 10^4$~см$^{-3}$ constitute in fact the upper limits of the true values.

\bigskip
{%\color{red}
\subsection{Estimates of the relative abundances of the NH$_2$D molecules and the ratio of the NH$_2$D/NH$_3$ abundances }
}
The next step of the research was the calculations of the relative abundances of NH$_2$D, i.e. the ratio of the column densities of NH$_2$D to the column densities of H$_2$. The data about the C$^{18}$O molecules, taken from work \cite{Zin00}, were used for calculations of the column densities of H$_2$. The relative abundance of C$^{18}$O is assumed equal to $1.7\times 10^{-7}$ \cite{Frer82}. Thus, by deriving from this ratio the column density of the molecular hydrogen H$_2$ we can calculate the relative abundances of deuterated ammonia NH$_2$D. It's known that the relative abundance of C$^{18}$O varies along the galactocentric radius due to the presence of the gradient in the abundance ratio $^{16}$O/$^{18}$O \cite{Wilson94,Liu13}. We don't take into account this variation, because for the sources of our sample with the sufficiently reliable distance estimates it doesn't exceed $\sim$30\% (the galactocentric distances for them lie in the range from $\sim$8 to $\sim$11~kpc -- Table~\ref{tab:sources}). It's significantly lower than the variations discussed further. For about 20\% of sources the distances are unknown. In some cases we calculated kinematic distances (noted in Table~\ref{tab:sources}). However in directions of the center and anticenter of the Galaxy, and also in the tangential directions the kinematic distances have very large errors and their usage is not possible.   

For determination of the deuterium enrichment degree for the molecule of singly deuterated ammonia NH$_2$D it's necessary to know the column densities of ammonia NH$_3$. They were taken from publications \cite{Schreyer96, Harju93, Jijina99, Pillai06, Zin97, Li16, Kohno22}. For several sources the data of the observations of ammonia NH$_3$ lines were taken from the work \cite{Torrelles86} and the column densities for them were calculated using formulas presented in \cite{Kohno22}.  

\bigskip
{%\color{red}
\subsection{An analysis of the dependencies of the NH$_2$D and NH$_3$ relative abundances and the NH$_2$D/NH$_3$ abundance ratio on the kinetic temperature and on the velocity dispersion} }
Further we analyzed the dependencies of the obtained estimates of the NH$_2$D relative abundances and the NH$_2$D/NH$_3$ abundance ratio both with and without taking into account the upper limits of these values on the kinetic temperature and on the line width in the source.

The presence of correlation taking into account the limits was determined by Kendall's method \cite{Cor86}, and of the linear regression coefficients (they are the slope coefficient) were determined using Buckley-James method \cite{Cor86}. All calculations of these parameters were performed using the ASURV code. Without taking into account the limits, besides the Kendall rank correlation the Pearson correlation coefficient and the significance indicator (``$p$-value'') were calculated, too. The criteria of presence of correlation was taken to be the significance indicator $p < \alpha$, where $\alpha$ is a threshold of a significance level, set equal to 0.05, which means the acceptable probability of the first type error (false positive conclusion) no more than 5\%. The obtained values of the significance levels for all dependencies, which were considered in this work both with and without taking into account the limits, are given in Table \ref{tab:stat}. 
Table \ref{tab:stat} also shows the values of linear regression coefficients for all considered cases. When determining the significance levels and the linear regression coefficients, the sources for which the kinetic temperatures are unknown, were excluded.   
 
The plots of the relative abundance NH$_2$D in dependence on the kinetic temperature of the gas and on the line width in the source are presented in Figs.~\ref{ris:N/H-Tk}, \ref{ris:N/H-dV}. The average widths of other narrow lines in the source (HC$_3$N, H$_2$CO and H$^{13}$CN) were used for the upper limits of the relative abundances as noted above. For the dependence of relative abundance of singly deuterated ammonia on the kinetic temperature of the source, presented in Fig.~\ref{ris:N/H-Tk}, the statistically significant correlation was found. This figure shows the estimates of the relative abundances of NH$_2$D depending on the kinetic temperature of the source for the gas density equal to 1$\times$10$^{4}$~см$^{-3}$. In Figure \ref{ris:N/H-Tk} red circles show the obtained estimates, and blue triangles show the values calculated using the detection limits of NH$_2$D. Figure \ref{ris:N/H-Tk} shows that the relative abundance of NH$_2$D decreases with increasing the kinetic temperature of the source. The significance indicator for this dependence, taking into account the limits, is $p$=0.073, which slightly exceeds the accepted significance level threshold. The solid line in Figure \ref{ris:N/H-Tk} is a line of the linear regression taking into account the limits. Also an analysis of the dependence of the relative abundance of NH$_2$D on the kinetic temperature of the source without taking into account the detection limits of this molecule was performed. The Pearson significance level in this case is $p$=0.029, which is lower than the accepted significance level threshold. The dashed line in Figure \ref{ris:N/H-Tk} is a linear regression line without taking into account the limits. 

Figure ~\ref{ris:N/H-dV} shows the dependence of the relative abundance of singly deuterated ammonia NH$_2$D on the line width in the source for the case when the gas density is adopted equal to 10$^{4}$~см$^{-3}$. The line widths of NH$_2$D in all observed sources except one are in the range from 0.5 km/s to 4 km/s. The significance indicator for this ratio is equal to 0.014, which is below the accepted significance level threshold. Here the solid line is a linear regression line taking into account the limits. The Pearson significance level for this dependence without taking into account the limits is $p$=0.149, which significantly exceeds the accepted significance level threshold.
 
Figure \ref{ris:N/N-Tk} shows the dependence of the ratio of the column densities of NH$_2$D and NH$_3$ on the kinetic temperature of the source for the adopted gas density of 10$^4$~см$^{-3}$. There are no the statistically significant change of this ratio in the considered kinetic temperature range of 15--50~K. The spread of this ratio in the studied objects is from $\sim$0.08 to $\sim$1.6 for the assumed gas density of 10$^{4}$~см$^{-3}$ and from $\sim$0.009 to $\sim$0.25 for the assumed gas density of 10$^{5}$~см$^{-3}$. The average value of the ratio of the column densities of NH$_2$D and NH$_3$ is equal to $\sim$0.5 in the temperature range of 10--30 K for the assumed gas density of 10$^{4}$~см$^{-3}$. The ammonia deuteration model \cite{Roueff05,Roueff07} predicts the ratio of the column densities of NH$_2$D and NH$_3$ $\sim$0.1 at the gas kinetic temperatures of 10--20 K, which is lower than the values obtained by us (which are essentially the upper limits as noted above).  

In a similar way, the dependence of the ratio of the column densities of singly deuterated ammonia NH$_2$D and the main isotopologue NH$_3$ on the velocity dispersion was considered. This dependence for the assumed gas density 10$^4$~см$^{-3}$ is presented in Figure \ref{ris:N/N-dV}. No statistically significant correlation between these values was found (Table~\ref{tab:stat}). In figures \ref{ris:N/N-Tk} and \ref{ris:N/N-dV} there are no points corresponding to sources for which the kinetic temperatures are unknown and accepted to be equal to 20~K.
 
We also considered the dependence of the relative abundances of ammonia NH$_3$ on the source kinetic temperature and on the velocity dispersion. These dependencies are presented in Figures \ref{ris:NH3-Tk} and \ref{ris:NH3-dV}, respectively. The relative abundance of ammonia NH$_3$ was determined similarly to the relative abundance of singly deuterated ammonia NH$_2$D described above. In Figure \ref{ris:NH3-Tk} there is a trend to decreasing the relative NH$_3$ abundance with increasing the gas kinetic temperature in the sources, but it's not statistically significant. At the same time there is the statistically significant decrease of the ammonia relative abundance with increasing the velocity dispersion (Fig. \ref{ris:NH3-dV}).

\section{Discussion} \label{sec:disc}
Our results show the absence of the statistically significant change of the NH$_2$D/NH$_3$ abundance ratio in the studied temperature range of 15--50~K, although the scatter of the estimates is quite large (it exceeds an order of magnitude) and some tendency to decreasing this ratio at the temperatures $T>30$~К in Fig. ~\ref{ris:N/N-Tk} can be seen. At the same time existing theoretical models predict a strong temperature dependence in this range. The deuteration model of ammonia presented in the works \cite{Roueff05,Roueff07} predicts the ratio NH$_2$D/NH$_3$ $\sim$0.1 at 10--20~K and its decreasing to $\sim 2\times 10^{-3}$ at 50~K. A strong dependence of the deuteration degree on temperature is predicted also in other models (e.g., \cite{Sipila15}). However, a significant drawback of the correlation analysis in the present work is the assumption of the same gas density in all sources. This assumption is most likely false. According to some estimates there may be a positive correlation between the temperature and the gas density in some of these sources (A. Pazukhin, private communication). Because with increasing gas density estimates of the NH$_2$D column density decrease, such a correlation will translate into the anticorrelation between these estimates and temperature, which may strengthen the trends discussed here and provide agreement with the theoretical model.     

Recently a survey of 50 high mass star forming regions in the para-NH$_2$D $1_{11}^a - 1_{01}^s$ line at a frequency of 110~GHz using the 30-m radio telescope of the Institute of millimeter radio astronomy \cite{Li22} was done. The line was detected in 36 sources, which is 72\% of the observed objects. Data analysis was performed in the LTE approximation. The estimates of deuteration degree in this work were on average approximately the same as ours. At the same time in the work \cite{Li22} it was found (at the level of 3.5$\sigma$), that the NH$_2$D/NH$_3$ abundance ratio decreases with increasing velocity dispersion in the source, which is not confirmed by our data taking into account the detection limits of NH$_2$D. Also in the work \cite{Fontani2015} such anti-correlation wasn't found. In this work the line widths do not exceed $\sim$3.8~km/s.  
      
In \cite{Wienen21} it was found that the NH$_2$D detection rate in ATLASGAL clumps decreases at temperatures $>$20~K. In principle it agrees with decreasing of the NH$_2$D relative abundance with increasing temperature found by us. The dependencies of the NH$_2$D/NH$_3$ ratio on temperature (in the range of 10--30~K) in the work \cite{Wienen21} was not found as in ours. 

In the work \cite{Pazukhin2023}, based on the analysis of the maps of several sources obtained with the 30-m IRAM radio telescope, a decrease of the NH$_2$D/NH$_3$ ratio with increasing temperature higher than $\sim$30~K was found. At the same time the values of this ratio are several times higher than predicted by the model \cite{Roueff07} as in the present work. It should be noted that the amount of analyzed data in the work \cite{Pazukhin2023} is significantly higher than here. 

Concerning the found trend for decreasing the relative abundance of ammonia with increasing gas kinetic temperature in the sources, as well as its statistically significant decrease with increasing velocity dispersion, it's worth noting that in the work \cite{Zin97} a decrease of relative abundance of ammonia in sources with the highest luminosity was found. A gas temperature around sources with a high luminosity as well as turbulence naturally increase. At the same time the values of  NH$_3$ relative abundance at the temperatures $\lesssim$30~К are close to typical values \cite{Harju93}.

\section{Conclusions}

As a result of the survey of the high mass star forming regions the emission of singly deuterated ammonia NH$_2$D was detected in 29 of 50 observed objects. An analysis of obtained data and also the fact that the gas density in the studied sources according to independent estimates is significantly lower than the critical density for this NH$_2$D transition, indicate non-LTE excitation of NH$_2$D. Estimates of the NH$_2$D column densities are obtained based on non-LTE modeling using available data about the gas kinetic temperature for several values of the assumed gas density. Statistically significant anti-correlations were found between the relative abundance of NH$_2$D and the gas kinetic temperature in the temperature range of 15--50~K, as well as between the relative abundance of NH$_2$D and the velocity dispersion in the source. 
  
At the same time a significant decrease of the NH$_2$D/NH$_3$ abundance ratio with increasing temperature, predicted by existing chemical models, is not observed (assuming the same gas density in all sources). It's $\sim$0.5 at the assumed gas density of $n\sim 10^{4}$~см$^{-3}$ in this temperature range, although the scatter of the estimates is quite large and these estimates represent upper limits in fact. In the work \cite{Roueff07} the NH$_2$D/NH$_3$ ratio is predicted to be $\sim$0.1 at 10--20~K and it decreases to $\sim 2\times 10^{-3}$ at 50~K. In the recently published work \cite{Pazukhin2023} in the temperature range of 10--20~K this ratio is $\sim$0.2, however there is a decrease of NH$_2$D/NH$_3$ ratio at temperatures $>$30~K in agreement with the model, although it remains higher than the model values. It should be noted, however, that the assumption of the same gas density in all sources is most likely false. If there is a positive correlation between the gas density and the temperature, the estimates of NH$_2$D/NH$_3$ ratio will decrease with increasing temperature. Thus, taking into account the uncertainties of our estimates of the column densities, the results of this work don't contradict the chemical model in the works \cite{Roueff05, Roueff07}. Herewith the dependence of the NH$_2$D/NH$_3$ abundance ratio on the velocity dispersion is not found. Also an analysis of available data revealed a statistically significant anti-correlation between the relative abundance of ammonia NH$_3$ and the velocity dispersion, at the same time the obvious anti-correlation between the relative abundance of ammonia and the gas kinetic temperature in the source in the range from 15~K to 50~K was not found. However there is a tendency to decreasing the relative NH$_3$ abundance with increasing gas temperature.

\section*{FUNDING}
The work was supported by the Russian Science Foundation grant 22-22-00809.

\section*{Acknowledgments}
The authors acknowledge support from Onsala Space Observatory for the provisioning of its facilities/observational support. The Onsala Space Observatory national research infrastrcuture is funded through Swedish Research Council grant No 2017-00648.
\textcolor{red}{We are grateful to the referees for their helpful comments.}

\bibliographystyle{maik}
\selectlanguage{english}
%\section*{maik.bst}
%\nocite{*}
\bibliography{ammon}

\newpage

\selectlanguage{english}

\newpage

\begin{center}
%\footnotesize
\begin{longtable}[h]{|l|l|l|r|c|c|l|}
\caption{Sources list} \label{tab:sources}
\\

\hline \hline \multicolumn{1}{|c|}{{Source}} & \multicolumn{1}{|c|}{{RA(J2000)}} & \multicolumn{1}{|c|}{{Dec(J2000)}} & \multicolumn{1}{|c|}{{$V_{\mathrm{LSR}}$ }} & \multicolumn{1}{|c|}{{D}} & \multicolumn{1}{|c|}{R$_g$} & \multicolumn{1}{|c|}{{Notes}} \\
\multicolumn{1}{|c|}{name} & (h)(m)(s)& ($^\circ$)(')(") & \multicolumn{1}{|c|}{(km/s)} & kpc & kpc & \\ 
\hline
\endfirsthead

\multicolumn{6}{c} {{\bfseries \tablename\ \thetable{}} -- continuation} \\ \hline

\hline \hline \multicolumn{1}{|c|}{{Source}} & \multicolumn{1}{|c|}{{RA(J2000)}} & \multicolumn{1}{|c|}{{Dec(J2000)}} & \multicolumn{1}{|c|}{{$V_{\mathrm{LSR}}$ }} & \multicolumn{1}{|c|}{{D}} & \multicolumn{1}{|c|}{R$_g$} & \multicolumn{1}{|c|}{{Notes}} \\
\multicolumn{1}{|c|}{name} &(h)(m)(s)& ()(')(")&\multicolumn{1}{|c|}{(km/s)} & kpc & kpc &  \\ 
\hline
\endhead

%($^\circ$)(')(")

\hline 
\endfoot

\hline
\endlastfoot
G121.30+0.66 & 00:36:47.50 & +63:29:02.1 & -17.7 & 0.85 & 9.0 & IRAS00338+6312 \\
S184 & 00:52:25.15 & +56:33:53.3 & -30.4 & 2.1 & 9.8 & G123.07-6.31, IRAS00494+5617 \\
S187(N$_2$H$^+$) & 01:23:15.38 & +61:49:43.1 & -14.0 & 3 & 10.6 & G126.68-0.81, IRAS01194+6136 \\
G133.69+1.22 & 02:25:28.23 & +62:06:57.7 & -43.1 & 2.3 & 10.2 &  \\
G133.71+1.22 & 02:25:40.56 & +62:05:53.2 & -38.9 & 2.3 & 10.2 & IRAS02219+6152, AFGL326 \\
G133.75+1.20 & 02:25:53.50 & +62:04:10.7 & -38.9 & 2.3 & 10.2 &  \\
G133.95+1.07 & 02:27:04.68 & +61:52:25.5 & -47.7 & 2.3 & 10.2 & IRAS02232+6138, AFGL3314 \\
S199 & 02:57:35.60 & +60:17:22.0 & -38.0 & 2.1 & 10.1 & IRAS02575+6017, AFGL4029 \\
S201 & 03:03:17.90 & +60:27:52.0 & -37.0 & 2.1 & 10.2 & G138.5+1.6, IRAS02593+6016 \\
AFGL490 & 03:27:31.51 & +58:44:28.8 & -12.0 & 0.35 & 8.8 & IRAS03236+5836 \\
G142.00+1.83 & 03:27:38.77 & +58:47:00.1 & -13.9{\cite{Wouterloot89}} & 0.35 & 8.8 &  \\
Per4 & 03:29:18.00 & +31:27:31.0 & 7.6 & 0.35 & 8.8 & G158.27-20.37 \\
G170.66-0.27 & 05:20:16.14 & +36:37:21.1 & -18.8 &  &  & IRAS05168+3634 \\
%14.4 
G174.20-0.08 & 05:30:45.62 & +33:47:51.6 & -3.5 & 1.8 & 10.3 & AFGL5142, IRAS05274+3 \\
G173.17+2.35 & 05:37:57.85 & +35:58:40.5 & -19.5 & 2.3 & 10.8 & IRAS05345+3556 \\
S231 & 05:39:12.90 & +35:45:54.0 & -16.6 & 2.3 & 10.8 & G173.48+2.45, IRAS05358+3543 \\
G173.58+2.44 & 05:39:27.94 & +35:40:41.4 & -16.0 & 2.3 & 10.8 & IRAS05361+3539 \\
S235 & 05:40:53.32 & +35:41:48.7 & -17.0 & 2.3 & 10.8 & G173.72+2.70, IRAS05375+3540 \\
G205.11-14.11 & 05:47:05.45 & +00:21:50.0 & 9.8 & 0.5 & 9.0 & AFGL818, NGC2071, IRAS05445+0016 \\
G189.78+0.35 & 06:08:35.41 & +20:39:02.9 & 9.1 & 1.5 & 10.0 &  \\
AFGL6366 & 06:08:41.00 & +21:31:01.0 & 3.0 & 1.5 & 10.0 & G189.03+0.78, IRAS06056+2131 \\
S247 & 06:08:53.94 & +21:38:36.6 & 3.3 & 2 & 10.5 & G188.95+0.89, IRAS06058+2138 \\
S255N & 06:12:53.64 & +18:00:26.8 & 7.1 & 2.5 & 11.0 & G192.58-0.04 \\
S255IR & 06:12:54.00 & +17:59:23.1 & 7.1 & 2.5 & 11.0 & G192.60-0.05 \\
G202.99+2.11 & 06:40:44.59 & +09:48:12.6 & 18.0 & 0.8 & 9.2 &  \\
NGC2264 & 06:40:58.00 & +09:53:42.0 & 18.0 & 0.8 & 9.2 & G202.94+2.19 \\
W217 & 06:41:10.96 & +09:29:31.8 & 18.0 & 0.8 & 9.2 & IRAS06384+0932, AFGL989 \\
W40 & 18:31:15.75 & $-$02:06:49.3 & 5.0 & 0.5 & 8.1 &  \\
G58.47+0.43 & 19:38:58.12 & +22:46:32.2 & 37.3 &  &  & IRAS19368+2239 \\ %7.1
S100 & 20:01:45.59 & +33:32:41.1 & -23.8 &  &  & G70.29+1.60, IRAS19598+3324 \\ %9.7
G65.78-2.61 & 20:07:06.74 & +27:28:52.9 & 8.0 &  &  & IRAS20050+2720 \\ %8.2
G69.54-0.98 & 20:10:09.13 & +31:31:37.3 & 11.8 & 1.4 & 8.1 & IRAS20081+3122 \\
G77.46+1.76 & 20:20:38.54 & +39:38:18.9 & 3.1 & 4 & 8.6 & IRAS20188+3928, JC20188+3928 \\
G75.78-0.34 & 20:21:43.89 & +37:26:38.6 & -0.4 & 0.5 & 8.4 &  \\
G79.27+0.39 & 20:31:57.50 & +40:18:30.0 & 1.2 & 1 & 8.4 &  \\
G79.34+0.33 & 20:32:21.80 & +40:20:08.0 & 0.0 & 1 & 8.4 &  \\
W75N & 20:38:36.93 & +42:37:37.0 & 10.7 & 1.7 & 8.4 & G81.87+0.78 \\
W75(OH) & 20:39:00.60 & +42:22:48.8 & -3.8 & 3 & 8.6 & G81.72+0.57 \\
W75S3 & 20:39:03.43 & +42:25:53.0 & 2.1 & 3 & 8.6 & G81.77+0.60 \\
G81.50+0.14 & 20:40:08.30 & +41:56:26.0 & -4.5 & 1.3 & 8.4 &  \\
G92.67+3.07 & 21:09:21.74 & +52:22:37.6 & -15.2 &  &  &  \\ %9.2
G99.98+4.17 & 21:40:42.36 & +58:16:09.7 & 1.8 & 0.75 & 8.7 & IRAS21391+5802, L1121 \\
S140 & 22:19:18.20 & +63:18:51.2 & -7.0 & 0.91 & 8.8 & G106.80+5.31 \\
G109.87+2.11 & 22:56:18.10 & +62:01:49.4 & -7.0 &  &  &  \\ %8.8
G108.76-0.95 & 22:58:42.71 & +58:47:09.2 & -50.4 & 3.5 & 10.2 &  JC22566+5830 \\
S153 & 22:58:47.66 & +58:45:00.7 & -51.0 & 4.3 & 10.7 & G108.76-0.99, IRAS22566+5828 \\
S152(OH) & 22:58:49.60 & +58:45:15.3 & -52.7 & 3.5 & 10.2 & G108.77-0.98 \\
S156 & 23:05:09.90 & +60:14:31.0 & -50.6 & 3.5 & 10.3 & G110.11+0.04 \\
G111.54+0.78 & 23:13:44.72 & +61:28:09.7 & -57.6 & 3.5 & 10.3 & IRAS23116+6111, NGC7538 \\
S158 & 23:13:44.84 & +61:26:50.7 & -55.5 & 3.5 & 10.3 & G111.53+0.76 \\
\hline \hline
\end{longtable}
\normalsize
\end{center}

\newpage

\begin{center}
\begin{longtable}[h]{|l|c|c|c|c|}
\caption{The parameters of detected lines of singly deuterated ammonia NH$_2$D. The rms errors for the last digits are given in the brackets. The average widths of other narrow lines, found in a specific source, were used for the sources not detected in the NH$_2$D line (see p.3 Results).}
\label{tab:param_NH2D} 
\\

\hline \hline \multicolumn{1}{|c|}{\textmd{Source}} & \multicolumn{1}{|c|}{\textmd{V,}} & \multicolumn{1}{|c|}{\textmd{T$_m$$_b$,}} & \multicolumn{1}{|c|}{\textmd{$\Delta V$,}} & \multicolumn{1}{|c|}{\textmd{$\tau$}} \\

\multicolumn{1}{|c|}{\textmd{name}} & \multicolumn{1}{|c|}{\textmd{km/s}} & \multicolumn{1}{|c|}{\textmd{mK}} & \multicolumn{1}{|c|}{\textmd{km/s}} &  \\ \hline
\endfirsthead

\multicolumn{5}{c} {{\bfseries \tablename\ \thetable{}} -- continuation} \\ \hline
%--\multicolumn{6}{c} {{\bfseries \tablename\ \thetable{}} -- продолжение} \\ \hline

\hline \multicolumn{1}{|c|}{\textmd{Source}} & \multicolumn{1}{|c|}{\textmd{V,}} & \multicolumn{1}{|c|}{\textmd{T$_m$$_b$,}} & \multicolumn{1}{|c|}{\textmd{$\Delta V$,}} & \multicolumn{1}{|c|}{\textmd{$\tau$}} \\ 

\multicolumn{1}{|c|}{\textmd{name}} & \multicolumn{1}{|c|}{\textmd{km/s}} & \multicolumn{1}{|c|}{\textmd{mK}} & \multicolumn{1}{|c|}{\textmd{km/s}} &  \\ \hline

\endhead

\hline 
\endfoot

\hline
\endlastfoot

G121.30+0.66 & -17.50(05) & 308(18) & 1.99(12) & 1.20(54) \\
S184 & -29.80(18) & 168(43) & 2.15(54) & 1.72(1.79) \\
S187(N$_2$H$^+$) & -13.30(04) & 225(21) & 0.88(08) & 1.82(98) \\
G133.69+1.22 & & <81 & 4.98(28) &  \\
G133.71+1.22 & & <85 & 3.29(20) &  \\
G133.75+1.20 & & <119 & 2.82(1.31) &  \\
G133.95+1.07 & & <82 & 4.40(72) &  \\
S199 & -37.90(13) & 158(27) & 1.26(21) & 0.83(2.17) \\
S201 & & <38 & 2.96(27) &  \\
AFGL490 & -12.20(03) & 335(24) & 0.84(06) & 2.35(83) \\
G142.00+1.83 & & <89 & 2.02(50) &  \\
%Per4 & & <265 & &  \\
G170.66-0.27 & & <78 & 1.52(14) &  \\
G174.20-0.08 & -3.77(09) & 344(69) & 1.16(23) & 0.67(2.26) \\
G173.17+2.35 & & <55 & 2.12(81) &  \\
S231 & -16.40(12) & 149(28) & 2.17(41) & 1.00(1.29) \\
G173.58+2.44 & -16.40(03) & 254(13) & 1.1(06) & 2.39(68) \\
S235 & & <74 & 2.23(31) &  \\
G205.11-14.11 & 9.49(06) & 254(29) & 1.00(12) & 4.57(1.71)  \\
G189.78+0.35 & 8.75(30) & 117(19) & 2.55(41) & 0.47(2.75) \\
%G189.78+0.35 & 8.73(21) & 143(53) & 2.63(44) & 2.23(1.61) \\
AFGL6366 & 3.83(10) & 149(47) & 0.75(24) & <14.00 \\
S247 & & <46 & 3.05(39) &  \\
S255N & & <58 & 3.07(18) &  \\
S255IR & & <148 & 2.25(37) &  \\
%G202.99+2.11 & & <240 & &  \\
%NGC2264 & & <191 & &  \\
W217 & 7.49(13) & 427(77) & 2.21(40) & 2.14(1.31) \\
W40 & 4.76(01) & 1231(38) & 0.64(02) & 3.80(41) \\
%W40 & 4.73(01) & 1204(131) & 0.68(02) & 3.81(41) \\
G58.47+0.43 & 36.50(26) & 144(32) & 1.83(40) & <9.00 \\
%G58.47+0.43 & & <67 & 2.6 &  \\
S100 & & <88 & 5.99(30) &  \\
G65.78-2.61 & 6.41(30) & 195(64) & 2.80(92) & 1.75(2.48) \\
%G65.78-2.61 & & <65 & 3.07 &  \\
G69.54-0.98 & 10.90(23) & 222(48) & 2.52(54) & 2.78(1.85) \\
G77.46+1.76 & 1.72(05) & 667(33) & 1.99(10) & 2.38(50) \\
G75.78-0.34 & & <28 & 3.88(53) &  \\
G79.27+0.39 & 1.86(09) & 443(49) & 1.38(15) & 4.37(1.56) \\
G79.34+0.33 & 0.26(08) & 436(58) & 1.26(17) & 1.24(1.42) \\
W75N & 8.30(02) & 134(36) & 2.88(78) & 0.82(1.71) \\
%W75N & 8.26(08) & 257(73) & 1.45(14) & 3.75(1.25) \\
W75(OH) line 1 & -3.84(27) & 302(192) & 1.39(89) & 1.12(10) \\
W75(OH) line 2 & -2.58(27) & 140(25) & 5.02(89) & 0.52(10) \\
%?W75(OH) line2 & -2.86(27) & 176(10) & 4.55(89) & 0.26(10) \\
W75S3 & -4.32(01) & 964(12) & 1.59(02) & 2.84(14) \\
%?W75S3 line2 & -2.33(02) & 235(44) & 0.80(05) & 3.91(72) \\
G81.50+0.14 & & <68 & 1.18(73) &  \\
G92.67+3.07 & & <53 & 2.32(37) &  \\
G99.98+4.17 & 0.43(32) & 97(26) & 2.05(56) & 5.02(3.25) \\
S140 & -7.51(25) & 59(10) & 2.32(38) & <2.50 \\
G109.87+2.11 & & <106 & 3.81(65) &  \\
G108.76-0.95 & -50.80(09) & 241(19) & 1.88(15) & <2.50 \\
S153 & -51.90(21) & 338(92) & 1.76(48) & 5.06(2.18) \\
%S153 line1 & -50.30(17) & 160(32) & 1.31(22) & 0.60(36) \\
%S153 line2 & -52.00(07) & 286(100) & 1.40(14) & 5.14(1.47) \\
S152(OH) & -51.30(05) & 584(35) & 1.56(09) & 2.96(66) \\
S156 & & <193 & 5.21(40) &  \\
G111.54+0.78 & -56.30(09) & 28(8) & 3.76(1.07) & 1.63(0.28) \\
S158 & -55.30(36) & 125(36) & 4.19(1.19) & 0.36(3.21) \\
%&&&&&&&&&&&&\\
 
\hline \hline
\end{longtable}
\end{center}

\newpage

\begin{landscape}
\begin{center}
\footnotesize
\begin{longtable}{|l|c|c|c|c|c|c|c|c|}
\caption{Column densities} 
\label{tab:N} \\
\hline \hline \multicolumn{1}{|c|}{\textmd{Source}} & \multicolumn{1}{|c|}{\textmd{NH$_2$D}} & \multicolumn{1}{|c|}{\textmd{NH$_2$D}} & \multicolumn{1}{|c|}{\textmd{NH$_2$D}} & \multicolumn{1}{|c|}{\textmd{NH$_3$}} & \multicolumn{1}{|c|}{\textmd{NH$_2$D/NH$_3$}} & \multicolumn{1}{|c|}{\textmd{NH$_2$D/NH$_3$}} & \multicolumn{1}{|c|}{\textmd{NH$_2$D/NH$_3$}}& \multicolumn{1}{|c|}{\textmd{T$_{kin}$,}} \\ 

\multicolumn{1}{|c|}{\textmd{}} & \multicolumn{1}{|c|}{\textmd{$\times$ 10$^{13}$, cm$^{-2}$}} & \multicolumn{1}{|c|}{\textmd{$\times$ 10$^{13}$, cm$^{-2}$}} & \multicolumn{1}{|c|}{\textmd{$\times$ 10$^{13}$, cm$^{-2}$}}& \multicolumn{1}{|c|}{\textmd{$\times$ 10$^{14}$,}} & \multicolumn{1}{|c|}{\textmd{$\times$ 10$^{-2}$}} & \multicolumn{1}{|c|}{\textmd{$\times$ 10$^{-2}$}} & \multicolumn{1}{|c|}{\textmd{$\times$ 10$^{-2}$}} & \multicolumn{1}{|c|}{\textmd{}} \\ 

\multicolumn{1}{|c|}{\textmd{name}} & \multicolumn{1}{|c|}{\textmd{n=1$\times$10$^4$ cm$^{-3}$}} & \multicolumn{1}{|c|}{\textmd{n=1$\times$10$^5$ cm$^{-3}$}} & \multicolumn{1}{|c|}{\textmd{n=1$\times$10$^6$ cm$^{-3}$}} & \multicolumn{1}{|c|}{\textmd{cm$^{-2}$}} & \multicolumn{1}{|c|}{\textmd{n=1$\times$10$^4$ cm$^{-3}$}} & \multicolumn{1}{|c|}{\textmd{n=1$\times$10$^5$ cm$^{-3}$}} & \multicolumn{1}{|c|}{\textmd{n=1$\times$10$^6$ cm$^{-3}$}} & \multicolumn{1}{|c|}{\textmd{K}} \\ \hline

\endfirsthead

\multicolumn{9}{c} {{\bfseries \tablename\ \thetable{}} -- continuation} \\ \hline

\hline \hline \multicolumn{1}{|c|}{\textmd{Source}} & \multicolumn{1}{|c|}{\textmd{NH$_2$D}} & \multicolumn{1}{|c|}{\textmd{NH$_2$D}} & \multicolumn{1}{|c|}{\textmd{NH$_2$D}} & \multicolumn{1}{|c|}{\textmd{NH$_3$}} & \multicolumn{1}{|c|}{\textmd{NH$_2$D/NH$_3$}} & \multicolumn{1}{|c|}{\textmd{NH$_2$D/NH$_3$}} & \multicolumn{1}{|c|}{\textmd{NH$_2$D/NH$_3$}} & \multicolumn{1}{|c|}{\textmd{T$_{kin}$,}} \\ 

\multicolumn{1}{|c|}{\textmd{}} & \multicolumn{1}{|c|}{\textmd{$\times$ 10$^{13}$, cm$^{-2}$}} & \multicolumn{1}{|c|}{\textmd{$\times$ 10$^{13}$, cm$^{-2}$}} & \multicolumn{1}{|c|}{\textmd{$\times$ 10$^{13}$, cm$^{-2}$}} & \multicolumn{1}{|c|}{\textmd{$\times$ 10$^{14}$,}} & \multicolumn{1}{|c|}{\textmd{$\times$ 10$^{-2}$}} & \multicolumn{1}{|c|}{\textmd{$\times$ 10$^{-2}$}} & \multicolumn{1}{|c|}{\textmd{$\times$ 10$^{-2}$}} & \multicolumn{1}{|c|}{\textmd{}} \\ 

\multicolumn{1}{|c|}{\textmd{name}} & \multicolumn{1}{|c|}{\textmd{n=1$\times$10$^4$ cm$^{-3}$}} & \multicolumn{1}{|c|}{\textmd{n=1$\times$10$^5$ cm$^{-3}$}} & \multicolumn{1}{|c|}{\textmd{n=1$\times$10$^6$ cm$^{-3}$}} & \multicolumn{1}{|c|}{\textmd{cm$^{-2}$}} & \multicolumn{1}{|c|}{\textmd{n=1$\times$10$^4$ cm$^{-3}$}} & \multicolumn{1}{|c|}{\textmd{n=1$\times$10$^5$ cm$^{-3}$}} & \multicolumn{1}{|c|}{\textmd{n=1$\times$10$^6$ cm$^{-3}$}} & \multicolumn{1}{|c|}{\textmd{K}} \\ \hline

\hline

\endhead

\hline 
\endfoot

\hline
\endlastfoot

G121.30+0.66 & 50.00(2.97) & 4.79(28) & 1.97(12) & 10.00{\cite{Jijina99}} & 50.00(2.97) & 4.79(28) & 1.97(12) & 21.1{\cite{Jijina99}}; 34.4$^a$ \\
S184 & 22.60(5.72) & 2.21(56) & 1.04(26) & 7.94{\cite{Jijina99}} & 28.50(7.21) & 2.78(70) & 1.31(33) & 29{\cite{Jijina99}}; 30$^a$ \\
S187(N$_2$H$^+$) & 23.60(2.25) & 2.15(20) & 0.77(07) & 6.48{\cite{Torrelles86}} & 36.50(3.47) & 3.31(32) & 1.18(11) & 15{\cite{Jijina99}} \\
G133.69+1.22 & <8.57 & <0.83 & <1.14 & 1.58{\cite{Jijina99}} & <54.00 & <5.25 & <7.20 & 30.7{\cite{Malafeev05}} \\
G133.71+1.22 & <20.30 & <1.89 & <0.84 &  &  &  &  & 25.1{\cite{Schreyer96}} \\
G133.75+1.20 & <9.98 & <1.41 & <1.49 & 1.58{\cite{Jijina99}} & <62.90 & <8.90 & <9.40 & 55.2$^a$ \\
G133.95+1.07 & <7.42 & <0.66 & <1.22 & 5.01{\cite{Jijina99}} & <14.80 & <1.32 & <2.44 & 18.6$^a$ \\
S199 & 13.60(2.29) & 1.30(22) & 0.59(10) & 2.47{\cite{Jijina99}} & 55.10(9.28) & 5.27(89) & 2.38(40) & 26.4{\cite{Schreyer96}} \\

S201 & <7.05 & <0.67 & <0.32 &  &  &  &  & 29.7{\cite{Schreyer96}} \\

AFGL490	& 24.00(1.74) & 2.29(17) & 0.92(07) & 20.0{\cite{Torrelles86}} & 12.00(87) & 1.15(08) & 0.46(03) & 20{\cite{Jijina99}} \\
G142.00+1.83 & <16.20 & <1.46 & <0.59 &  &  &  &  & 20$^b$ \\

%Per4 & < & < &  &  &  & 20$^b$ \\

G170.66-0.27 & <10.70 & <0.96 & <0.39 &  &  &  &  & 20$^b$ \\

G174.20-0.08 & <26.00 & <2.59 & <1.18 & 3.98{\cite{Jijina99}} & <65.20 & <6.49 & <2.95 & 27{\cite{Schreyer96}} \\

G173.17+2.35 & <10.60 & <0.94 & <0.38 &  &  &  &  & 20$^b$ \\

S231 & 22.10(4.17) & 2.11(40) & 0.95(18) & 7.94{\cite{Jijina99}} & 27.80(5.25) & 2.65(50) & 1.20(23) & 26.5{\cite{Jijina99}}; 40.2$^a$ \\
G173.58+2.44 & 26.10(1.36) & 2.43(13) & 0.95(05) & 8.50{\cite{Harju93}} & 30.70(1.59) & 2.86(15) & 1.12(06) & 18.6{\cite{Jijina99}} \\
S235 & <2.93 & <0.33 & <0.44 & 2.77{\cite{Schreyer96}} & <10.60 & <1.17 & <1.60 & 40.4$^a$ \\
G205.11-14.11 & 10.60(1.22) & 1.21(14) & 0.69(08) & 2.00{\cite{Jijina99}} & 53.10(6.12) & 6.05(70) & 3.43(40) & 40.5$^a$ \\
G189.78+0.35 & 17.80(2.89) & 1.75(28) & 0.85(14) &  &  &  &  & 30.6{\cite{Schreyer96}} \\
AFGL6366 & 5.32(1.67) & 0.57(18) & 0.31(10) & 2.00{\cite{Jijina99}} & 26.70(8.37) & 2.85(90) & 1.53(48) & 37.1$^a$ \\
S247 & <1.52 & <0.14 & <0.40 & 3.16{\cite{Jijina99}} & <4.80 & <0.45 & <1.28 & 28.5{\cite{Schreyer96}} \\
S255N & <4.81 & <0.49 & <0.49 & 4.79{\cite{Zin97}} & <10.10 & <1.03 & <1.02 & 34.8$^a$ \\
S255IR & <3.40 & <0.35 & <0.92 & 2.45{\cite{Zin97}} & <13.90 & <1.43 & <3.75 & 34.5{\cite{Zin09}} \\

%G202.99+2.11 & < & < & & & & 20$^b$ \\
%NGC2264 & < & < & & & & 20$^b$ \\

W217 & 64.90(11.60) & 6.49(1.16) & 2.85(51) & 7.94{\cite{Jijina99}} & 81.70(14.60) & 8.17(1.46) & 3.58(64) & 25{\cite{Jijina99}}; 37.1$^a$ \\
W40 & 62.50(1.94) & 6.65(21) & 2.69(08) &  &  &  &  & 20$^b$ \\

G58.47+0.43 & 23.40(5.17) & 2.14(47) & 0.86(19) &  &  &  &  & 20$^b$ \\
S100 & <47.60 & <4.28 & <1.72 &  &  &  &  & 20$^b$ \\
G65.78-2.61 & 47.90(15.70) & 4.44(1.46) & 1.78(59) &  &  &  &  & 20$^b$ \\

G69.54-0.98 & 47.10(10.10) & 4.42(95) & 1.80(39) & 15.90{\cite{Jijina99}} & 29.60(6.36) & 2.78(60) & 1.13(24) & 20.8{\cite{Jijina99}}; 32.5$^a$ \\
G77.46+1.76 & 78.20(3.88) & 8.27(41) & 3.88(19) & 5.01{\cite{Jijina99}} & 156.00(7.74) & 16.50(82) & 7.75(38) & 29$^a$ \\

G75.78-0.34 & <5.29 & <0.55 & <0.30 &  &  &  &  & 36.8$^a$ \\

G79.27+0.39 & 51.60(5.65) & 5.00(55) & 2.01(22) & 20.50{\cite{Pillai06}} & 25.00(2.76) & 2.44(27) & 0.98(11) & 20$^b$ \\
G79.34+0.33 & 64.60(8.67) & 6.17(83) & 2.18(29) & 11.90{\cite{Pillai06}} & 54.30(7.29) & 5.19(70) & 1.82(25) & 14.6{\cite{Pillai06}} \\
W75N & 16.00(4.31) & 1.81(49) & 1.04(28) &  &  &  &  & 41.2$^a$ \\
W75(OH) line 1 & 25.30(16.10) & 2.54(1.62) & 1.16(74) & 25.10{\cite{Jijina99}} & 10.10(6.42) & 1.01(65) & 0.46(30) & 29.6$^a$ \\

W75(OH) line 2 & 35.40(6.27) & 3.67(0.65) & 1.94(35) & 25.10{\cite{Jijina99}} & 14.10(2.50) & 1.46(26) & 0.77(14) & 29.6$^a$ \\

%W75(OH) line 2 & 24.6(1.4) & 2.4(0.1) & 25.10{\cite{Jijina99}} & 9.8 & 1.0 & 29.6$^a$ \\
W75S3 & 59.60(76) & 7.37(09) & 4.23(05) & 3.71{\cite{Jijina99}} & 7.50(10) & 0.93(01) & 0.53(01) & 41.6$^a$ \\
%W75S3 line 2 & 3.8(0.7) & 0.4(0.1) & 3.71{\cite{Jijina99}} & 0.5 & 0.1 & 41.6$^a$ \\
G81.50+0.14 & <11.10 & <0.96 & <0.30 & 7.60{\cite{Pillai06}} & <14.60 & <1.26 & <0.40 & 15.4{\cite{Pillai06}} \\

G92.67+3.07 & <7.39 & <0.72 & <0.35 &  &  &  &  & 30.7$^a$ \\

G99.98+4.17 & 11.00(2.98) & 1.10(30) & 0.55(15) & 2.51{\cite{Jijina99}} & 43.60(11.90) & 4.37(1.19) & 2.20(60) & 33$^a$ \\
S140 & 7.69(1.26) & 0.76(13) & 0.38(06) & 7.94{\cite{Jijina99}} & 9.68(1.59) & 0.96(16) & 0.48(08) & 32.6$^a$ \\
G109.87+2.11 & <11.10 & <1.01 & <1.32 & 18.80{\cite{Li16}} & <5.90 & <0.53 & <0.70 & 20$^b$ \\
G108.76-0.95 & 46.60(3.62) & 4.31(34) & 1.62(13) & 12.60{\cite{Jijina99}} & 36.90(2.87) & 3.42(27) & 1.28(10) & 17{\cite{Jijina99}} \\
S153 & 62.50(17.10) & 5.88(1.61) & 2.18(60) &  &  &  &  & 16.4{\cite{Harju93}} \\
%?S153 line 2 & 21.3(7.4) & 2.0(0.7) & 15.90{\cite{Jijina99}} & 13.4 & 1.2 & 16.4{\cite{Harju93}} \\
S152(OH) & 91.50(5.48) & 9.02(54) & 3.35(20) & 15.90{\cite{Jijina99}} & 57.50(3.45) & 5.67(34) & 2.11(13) & 16.4{\cite{Jijina99}}; 31.4$^a$ \\
S156 & <8.01 & <0.73 & <0.34 & 3.98{\cite{Jijina99}} & <20.10 & <1.84 & <0.85 & 18.8{\cite{Harju93}} \\
G111.54+0.78 & 3.62(1.02) & 0.45(13) & 0.28(08) & 4.32{\cite{Li16}} & 8.37(2.37) & 1.03(29) & 0.65(18) & 47.7$^a$ \\
S158 & 23.10(6.57) & 2.54(72) & 1.42(40) & 1.00{\cite{Jijina99}} & 231.00(65.7) & 25.40(7.22) & 14.2(4.03) & 39.3$^a$ \\

\hline \hline
\end{longtable}
\end{center}
a --- The temperatures obtained from the rotational diagrams of CH$_3$CCH molecules. \\
b --- Adopted value. \\

\end{landscape}

\newpage
\footnotesize
\begin{longtable}[h]{|l|c|c|c|c|c|}
%{|m{6cm}|m{2cm}|m{2cm}|m{2cm}|m{2.5cm}|m{2.5cm}|}
\caption{Significance indicators of correlations and slope coefficients of linear regression.} 
\label{tab:stat}
\\

\hline \hline \multicolumn{1}{|c|}{{}} & \multicolumn{3}{|m{6cm}|}{{Significance indicator (p)}} & \multicolumn{2}{|m{5cm}|}{\raggedright{Slope coefficient \mbox{linear} regression}} \\
\cline{2-6}
\multicolumn{1}{|c|}{{Dependency}} & \multicolumn{1}{|m{2cm}|}{\raggedright{with \mbox{limits}}} & \multicolumn{2}{|c|}{\raggedright{without \mbox{limits}}} & \multicolumn{1}{|m{2.5cm}|}{\raggedright{with \mbox{limits}}} & \multicolumn{1}{|m{2.5cm}|}{\raggedright{without \mbox{limits}}} \\
\cline{3-4}
%\multicolumn{1}{|c|}{{}} & \multicolumn{1}{|c|}{{пределов}} & \multicolumn{1}{|c|}{{пределов}} & \multicolumn{1}{|c|}{{пределов}} & \multicolumn{1}{|c|}{{с учетом}} & \multicolumn{1}{|c|}{{без учета}} \\
\multicolumn{1}{|c|}{{}} & \multicolumn{1}{|c|}{{}} & \multicolumn{1}{|c|}{{(Kendall)}} & \multicolumn{1}{|c|}{{(Pearson)}} & \multicolumn{1}{|c|}{{}} & \multicolumn{1}{|c|}{{}} \\ \hline
\endhead

\hline 
\endfoot

N(NH$_2$D)/N(H$_2$) T$_k$ n=1e4 cm$^3$ & 0.073 & 0.061 & 0.029 & --0.036$\pm$0.012 & --0.029$\pm$0.012 \\
N(NH$_2$D)/N(H$_2$) T$_k$ n=1e5 cm$^3$ & 0.097 & 0.099 & 0.053 & --0.032$\pm$0.012 & --0.025$\pm$0.012 \\
N(NH$_2$D)/N(H$_2$) $\Delta V$ n=1e4 cm$^3$ & 0.014 & 0.509 & 0.149 & --0.397$\pm$0.159 & --0.235$\pm$0.152 \\
N(NH$_2$D)/N(H$_2$) $\Delta V$ n=1e5 cm$^3$ & 0.010 & 0.378 & 0.182 & 
--0.399$\pm$0.159 & --0.212$\pm$0.149 \\
N(NH$_2$D)/N(NH$_3$) T$_k$ n=1e4 cm$^3$ & 0.204 & 0.194 & 0.254 & --0.010$\pm$0.009 & --0.010$\pm$0.009 \\
N(NH$_2$D)/N(NH$_3$) T$_k$ n=1e5 cm$^3$ & 0.417 & 0.427 & 0.444 & --0.007$\pm$0.009 & --0.007$\pm$0.009 \\
N(NH$_2$D)/N(NH$_3$) $\Delta V$ n=1e4 cm$^3$ & 0.076 & 0.932 & 0.517 & --0.122$\pm$0.107 & 0.065$\pm$0.098 \\
N(NH$_2$D)/N(NH$_3$) $\Delta V$ n=1e5 cm$^3$ & 0.062 & 0.842 & 0.418 & --0.118$\pm$0.106 & 0.080$\pm$0.096 \\
N(NH$_3$)/N(H$_2$) T$_k$ & -- & 0.283 & 0.249 & -- & --0.014$\pm$0.012 \\
N(NH$_3$)/N(H$_2$) $\Delta V$ & -- & 0.029 & 0.008 & -- & --0.263$\pm$0.089 \\
 
\hline \hline
\end{longtable}

\newpage

\begin{figure}[h]
\begin{minipage}[h]{0.3\linewidth}
\center{\includegraphics[width=1\linewidth]{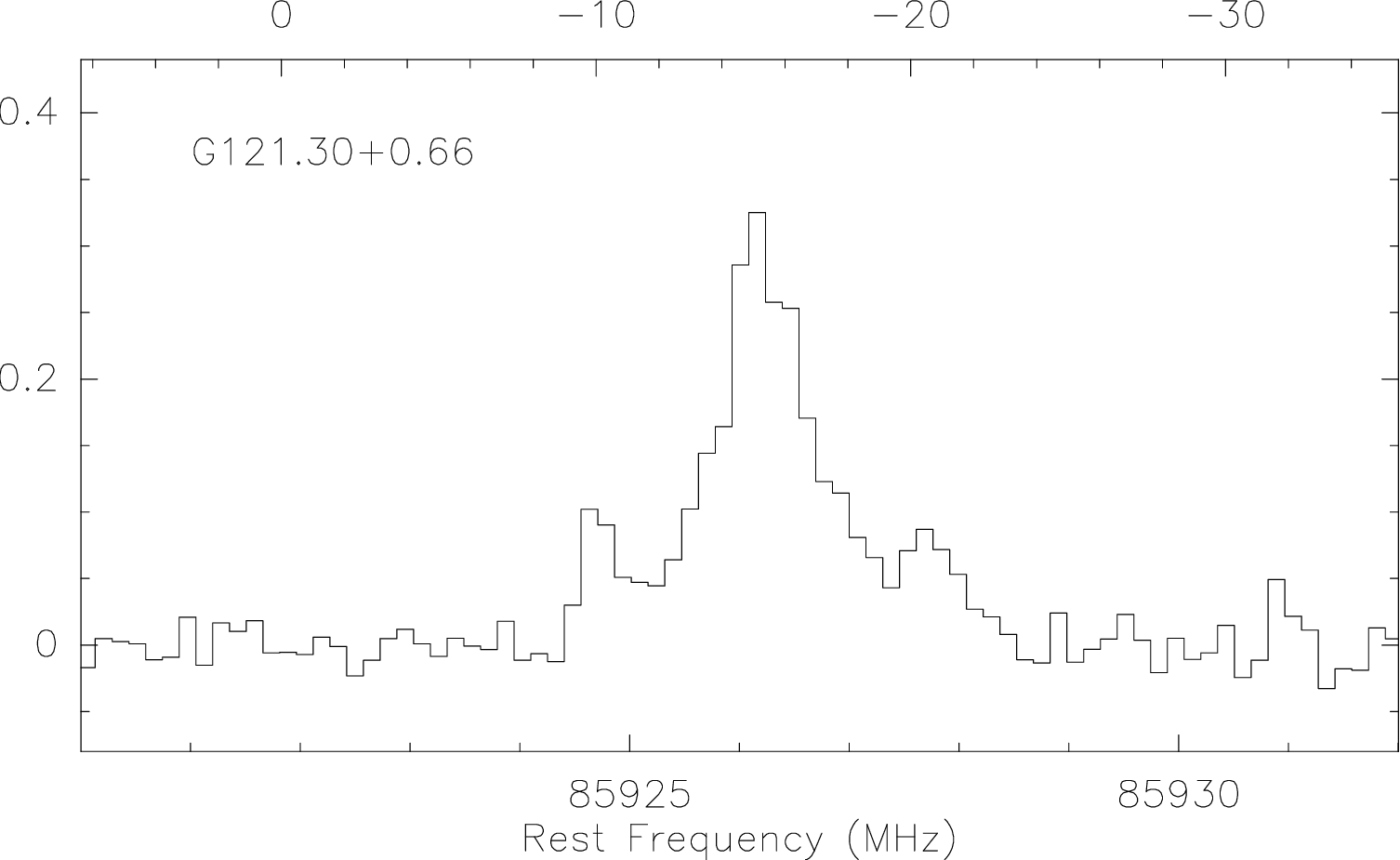}}  \\ \vspace{1cm}
\end{minipage}
\hfill
\begin{minipage}[h]{0.3\linewidth}
\center{\includegraphics[width=1\linewidth]{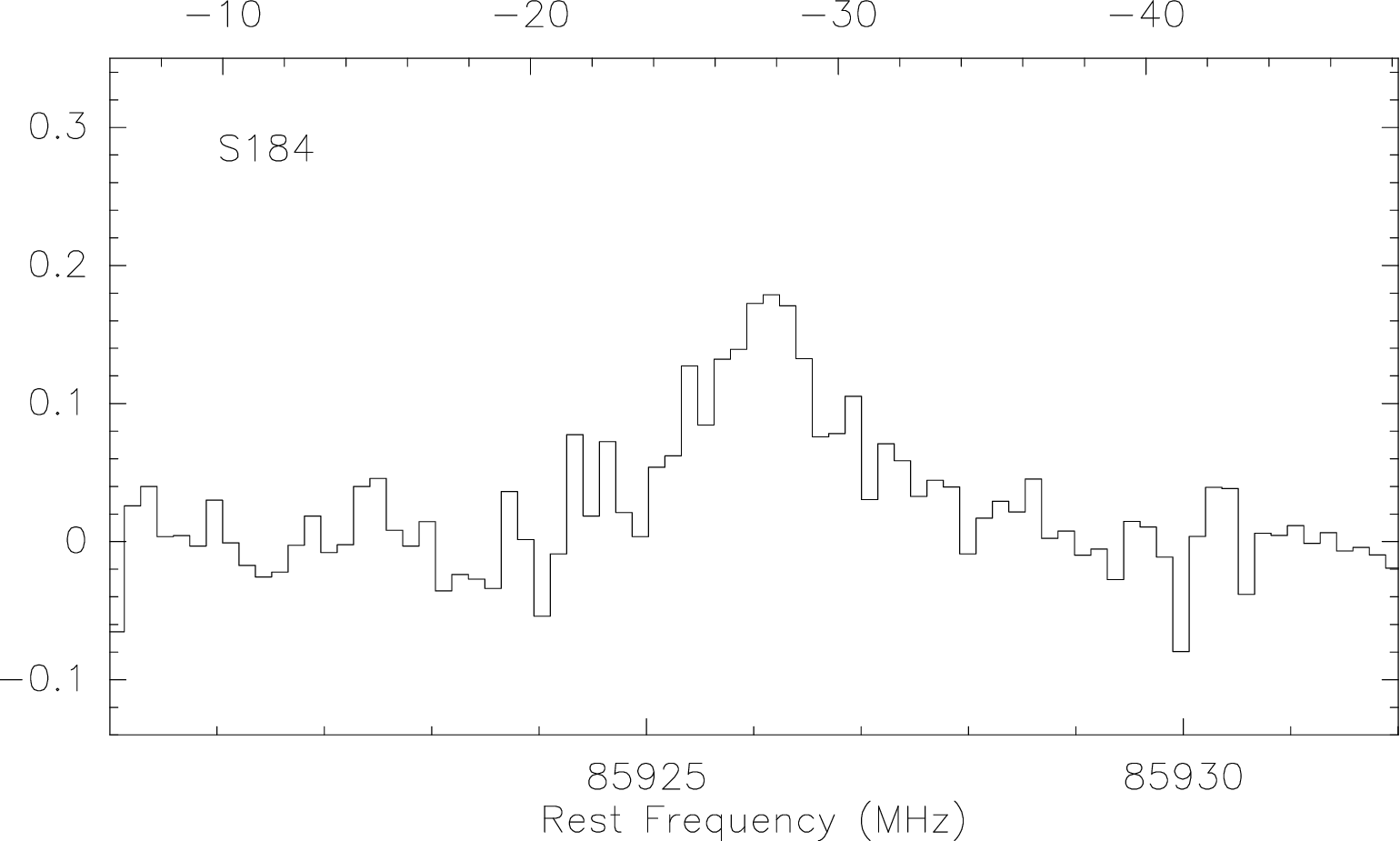}} 
\\ \vspace{1cm}
\end{minipage}
\hfill
\begin{minipage}[h]{0.3\linewidth}
\center{\includegraphics[width=1\linewidth]{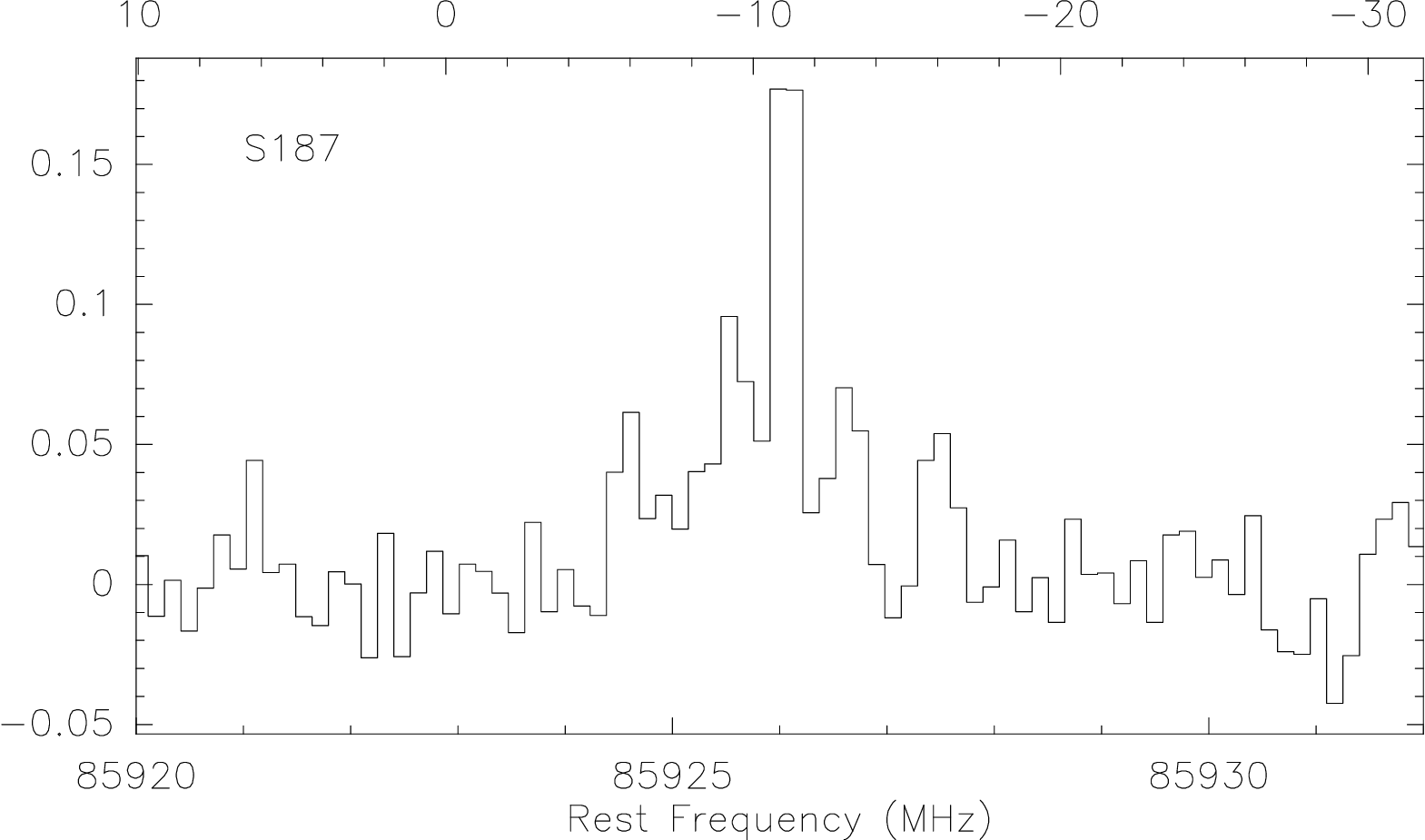}} 
\\ \vspace{1cm}
\end{minipage}

\vfill
\begin{minipage}[h]{0.3\linewidth}
\center{\includegraphics[width=1\linewidth]{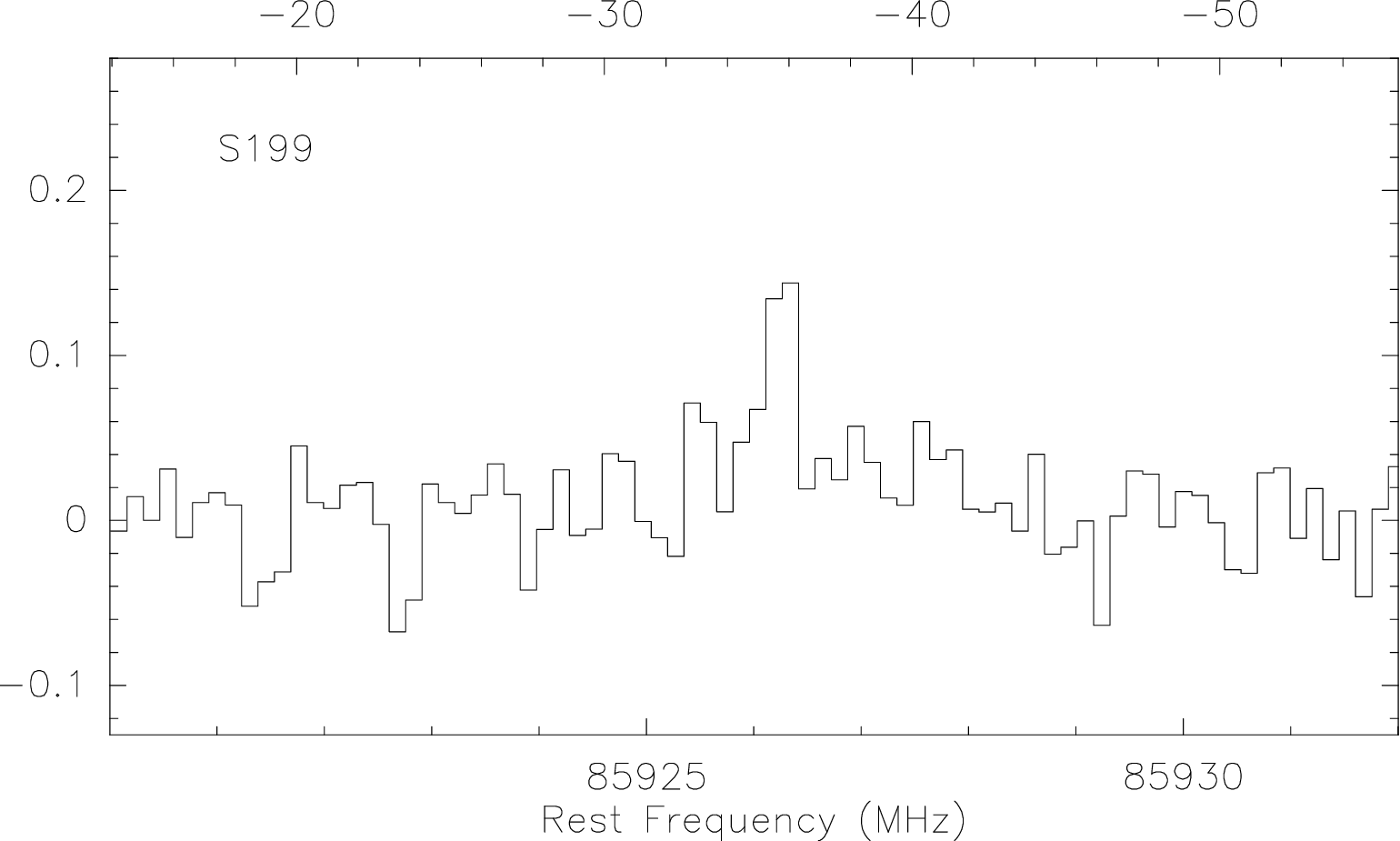}}  \\ \vspace{1cm}
\end{minipage}
\hfill
\begin{minipage}[h]{0.3\linewidth}
\center{\includegraphics[width=1\linewidth]{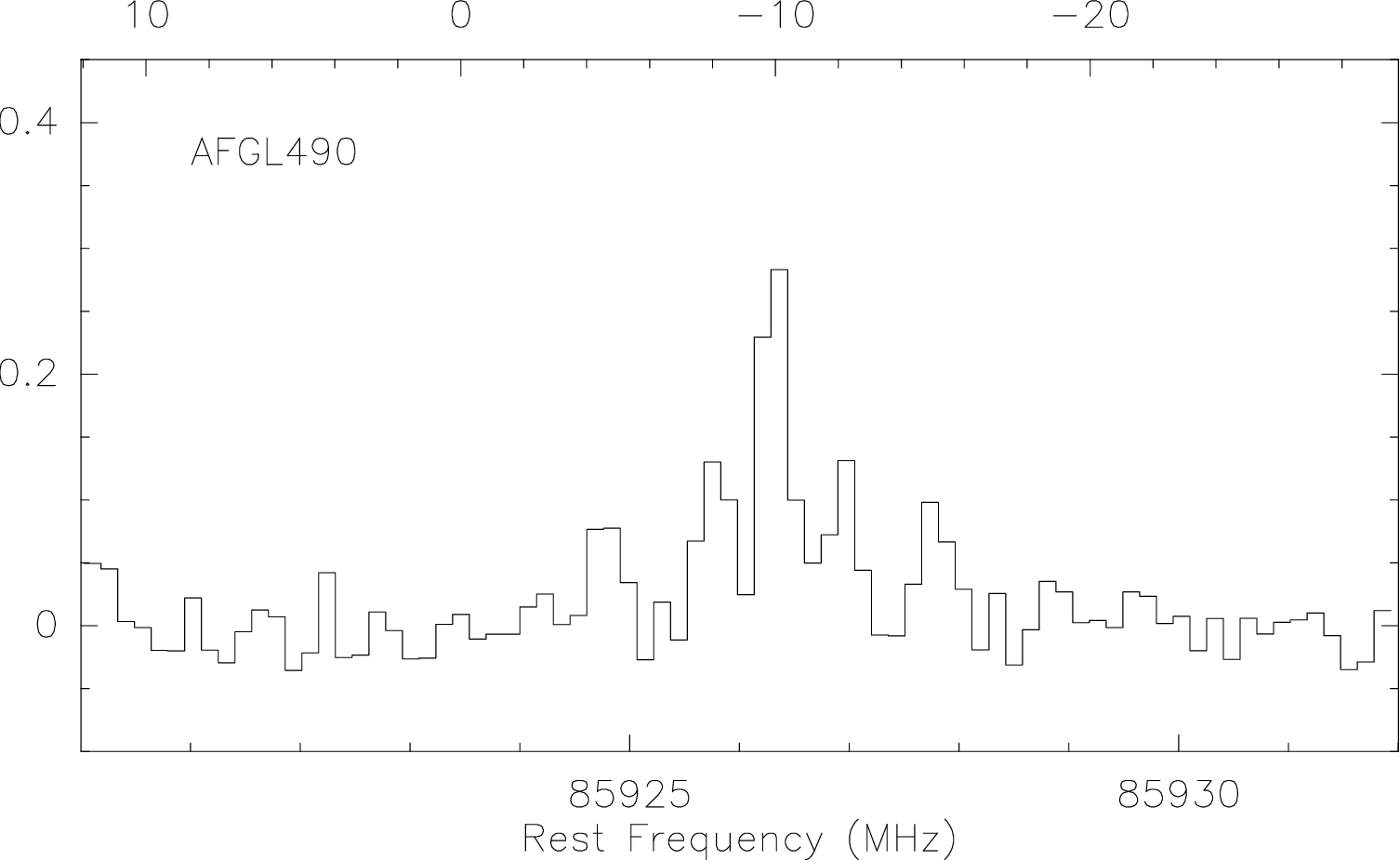}} 
\\ \vspace{1cm}
\end{minipage}
\hfill
\begin{minipage}[h]{0.3\linewidth}
\center{\includegraphics[width=1\linewidth]{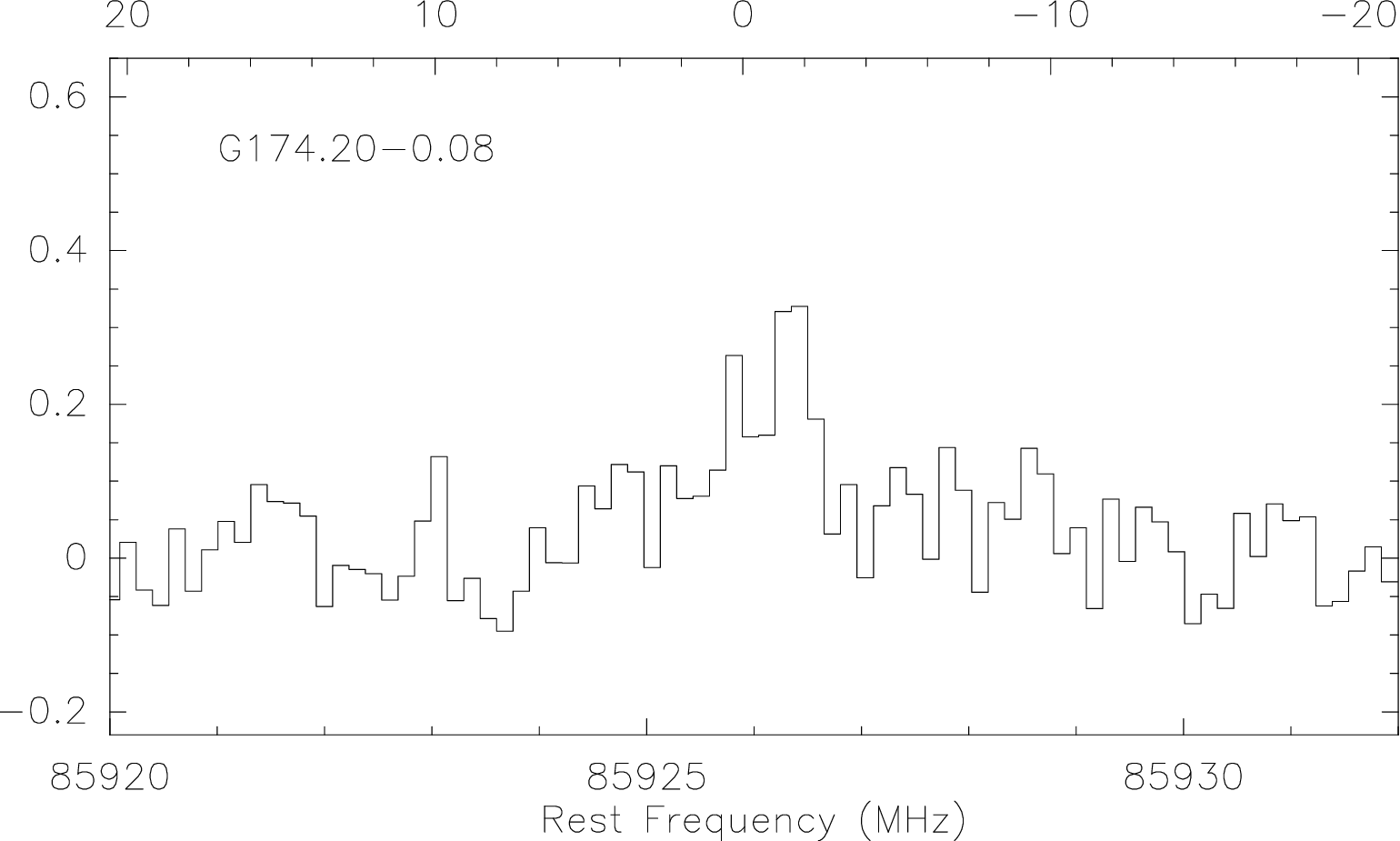}} 
\\ \vspace{1cm}
\end{minipage}

\vfill
\begin{minipage}[h]{0.3\linewidth}
\center{\includegraphics[width=1\linewidth]{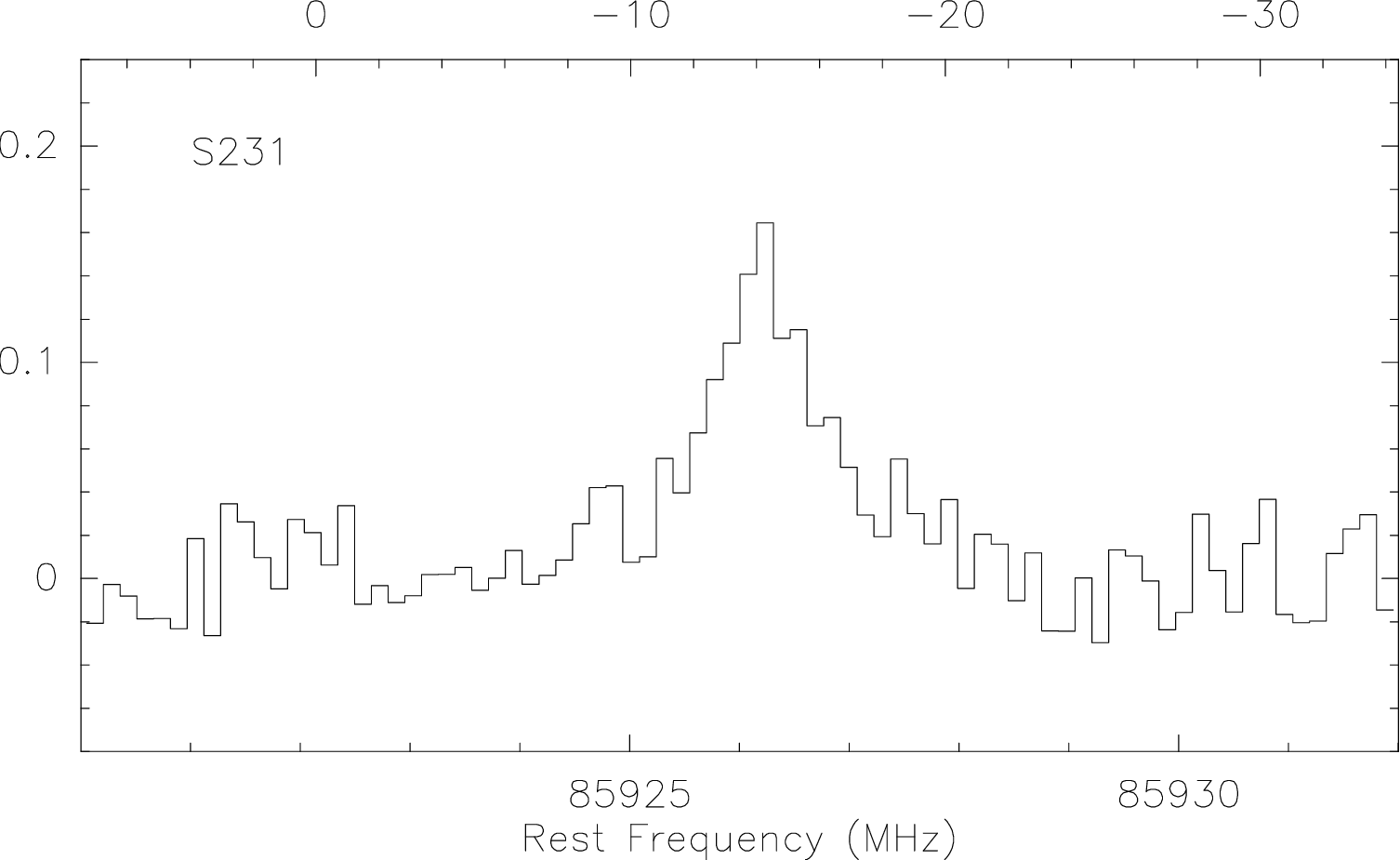}}  \\ \vspace{1cm}
\end{minipage}
\hfill
\begin{minipage}[h]{0.3\linewidth}
\center{\includegraphics[width=1\linewidth]{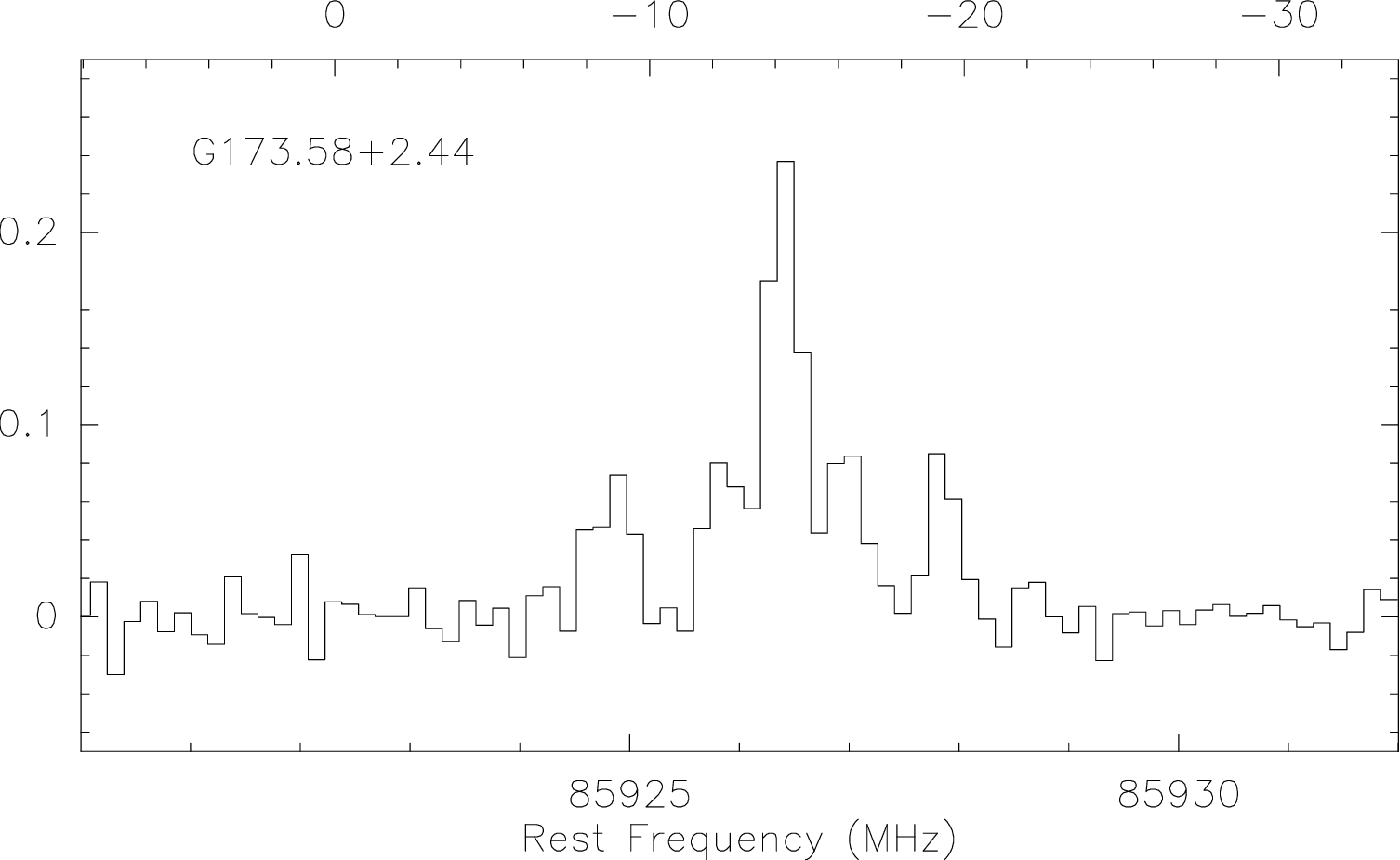}} 
\\ \vspace{1cm}
\end{minipage}
\hfill
\begin{minipage}[h]{0.3\linewidth}
\center{\includegraphics[width=1\linewidth]{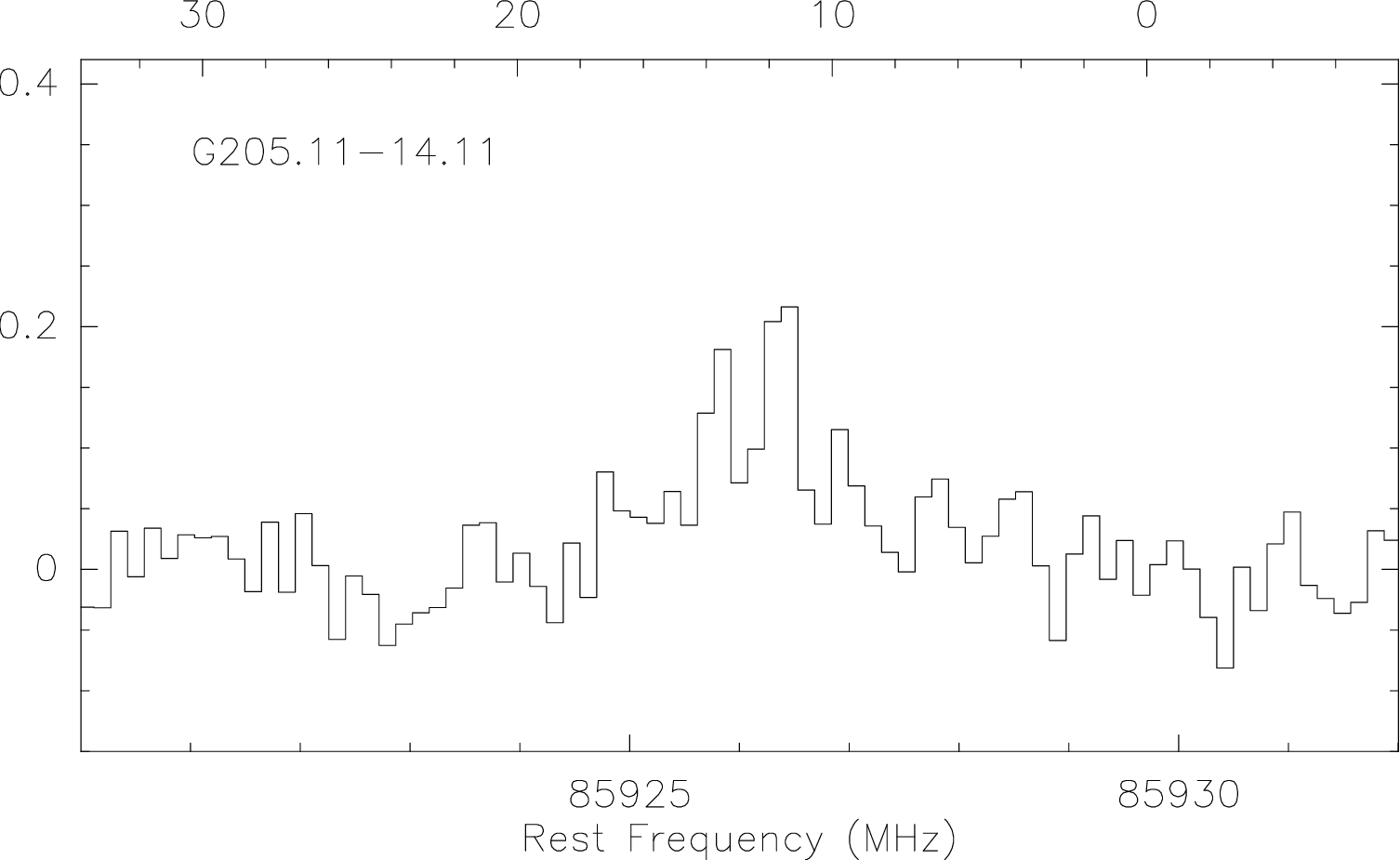}} 
\\ \vspace{1cm}
\end{minipage}

\vfill
\begin{minipage}[h]{0.3\linewidth}
\center{\includegraphics[width=1\linewidth]{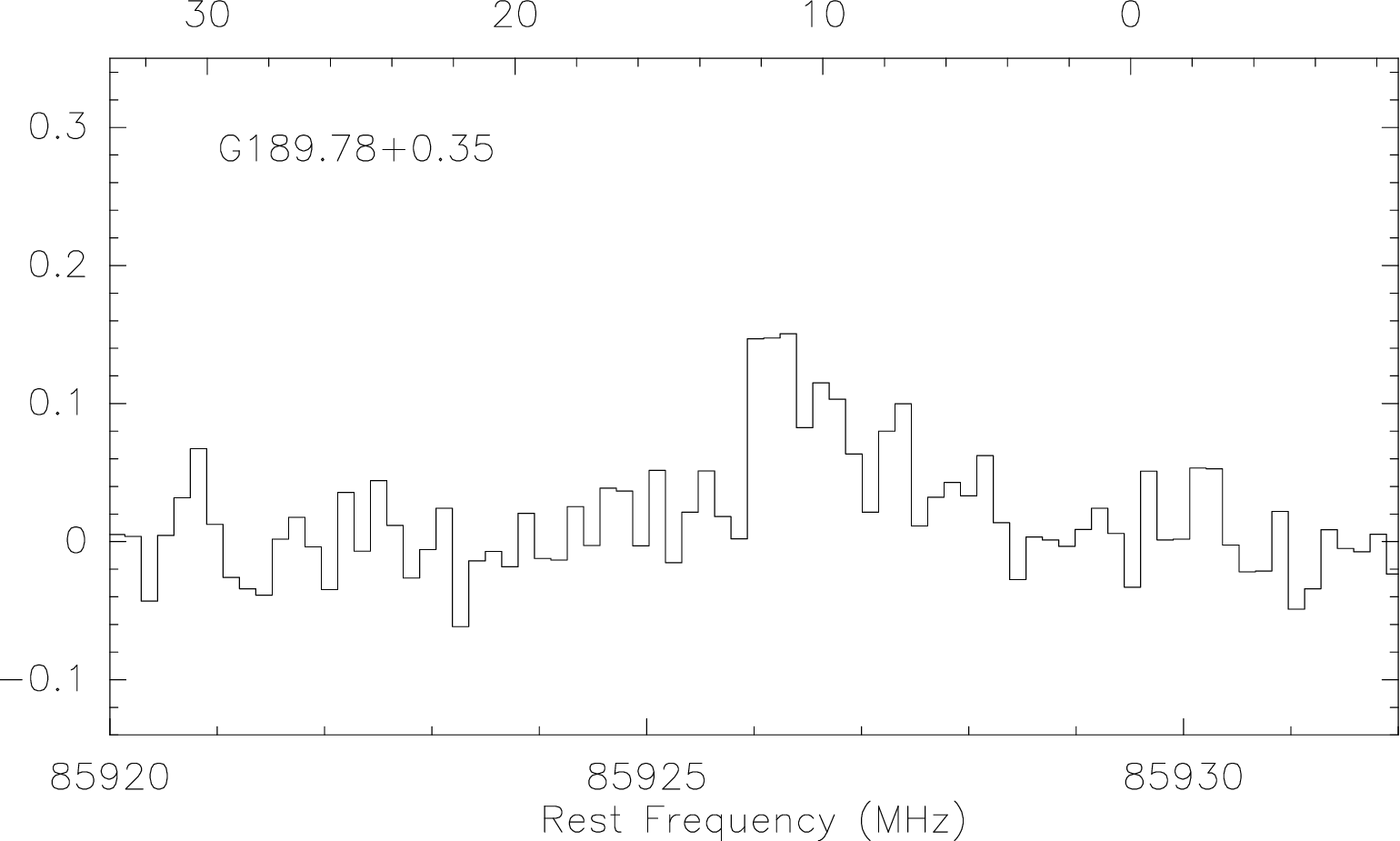}}  \\ \vspace{1cm}
\end{minipage}
\hfill
\begin{minipage}[h]{0.3\linewidth}
\center{\includegraphics[width=1\linewidth]{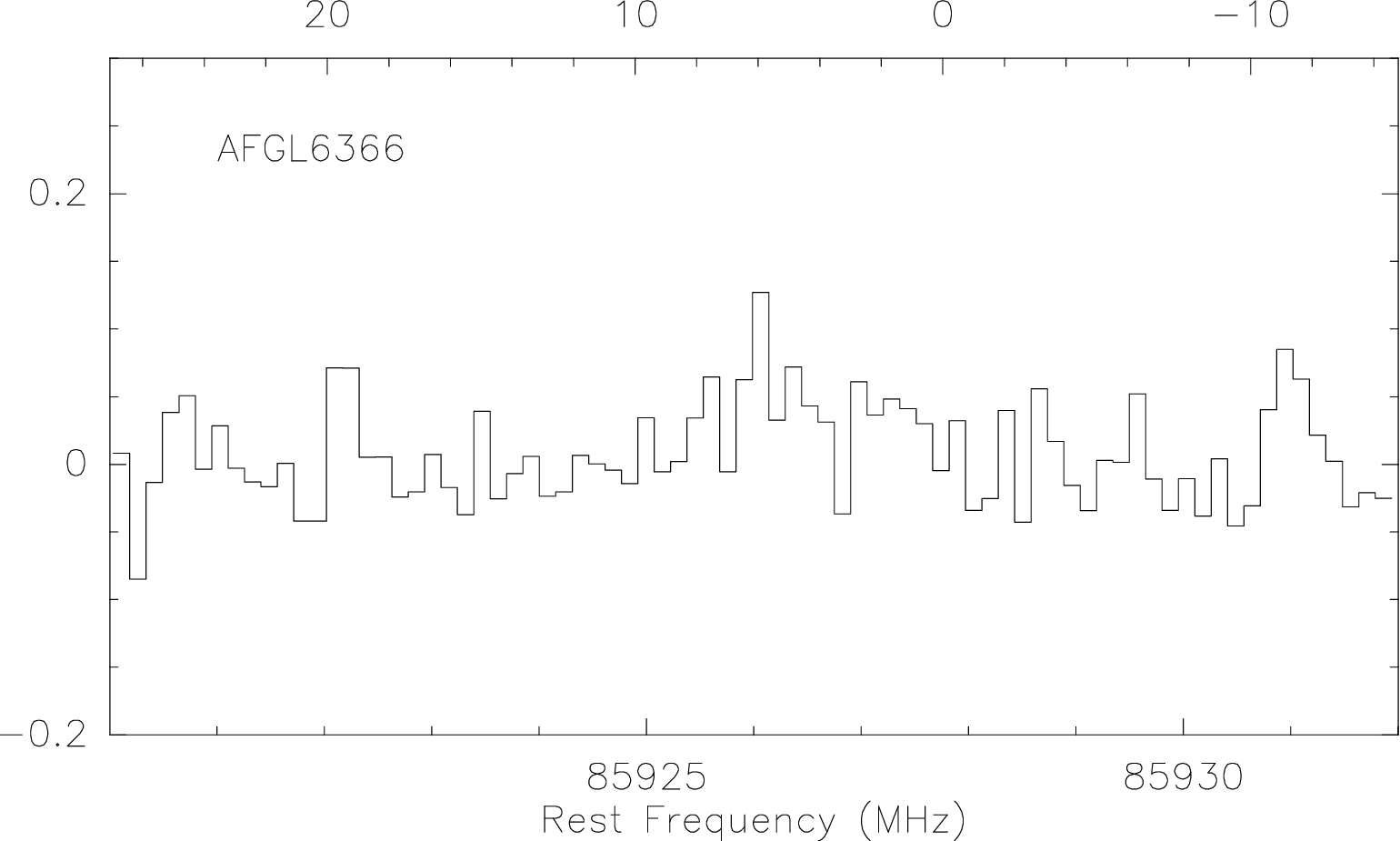}} 
\\ \vspace{1cm}
\end{minipage}
\hfill
\begin{minipage}[h]{0.3\linewidth}
\center{\includegraphics[width=1\linewidth]{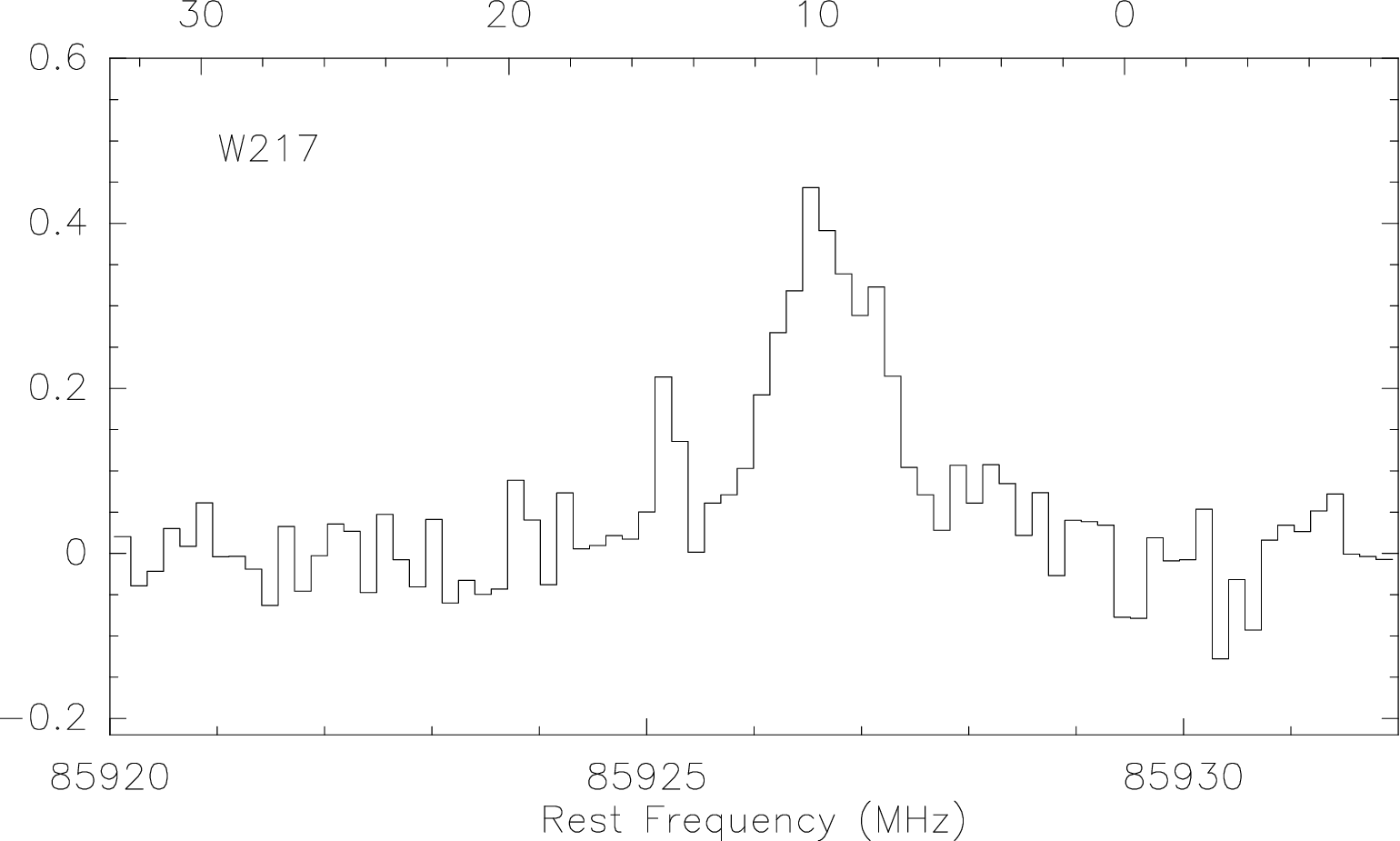}} 
\\ \vspace{1cm}
\end{minipage}

\vfill
\begin{minipage}[h]{0.3\linewidth}
\center{\includegraphics[width=1\linewidth]{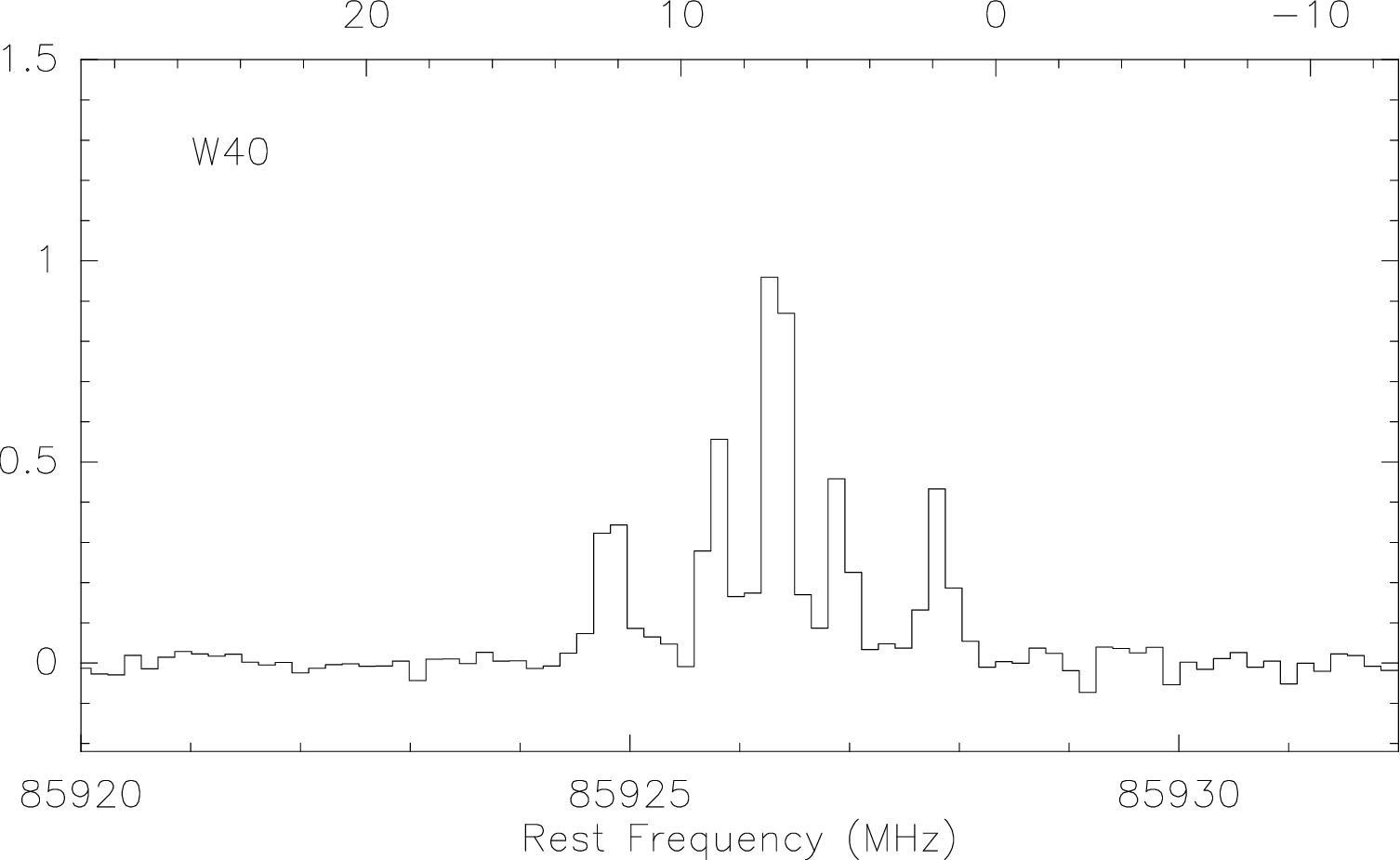}}  \\ \vspace{1cm}
\end{minipage}
\hfill
\begin{minipage}[h]{0.3\linewidth}
\center{\includegraphics[width=1\linewidth]{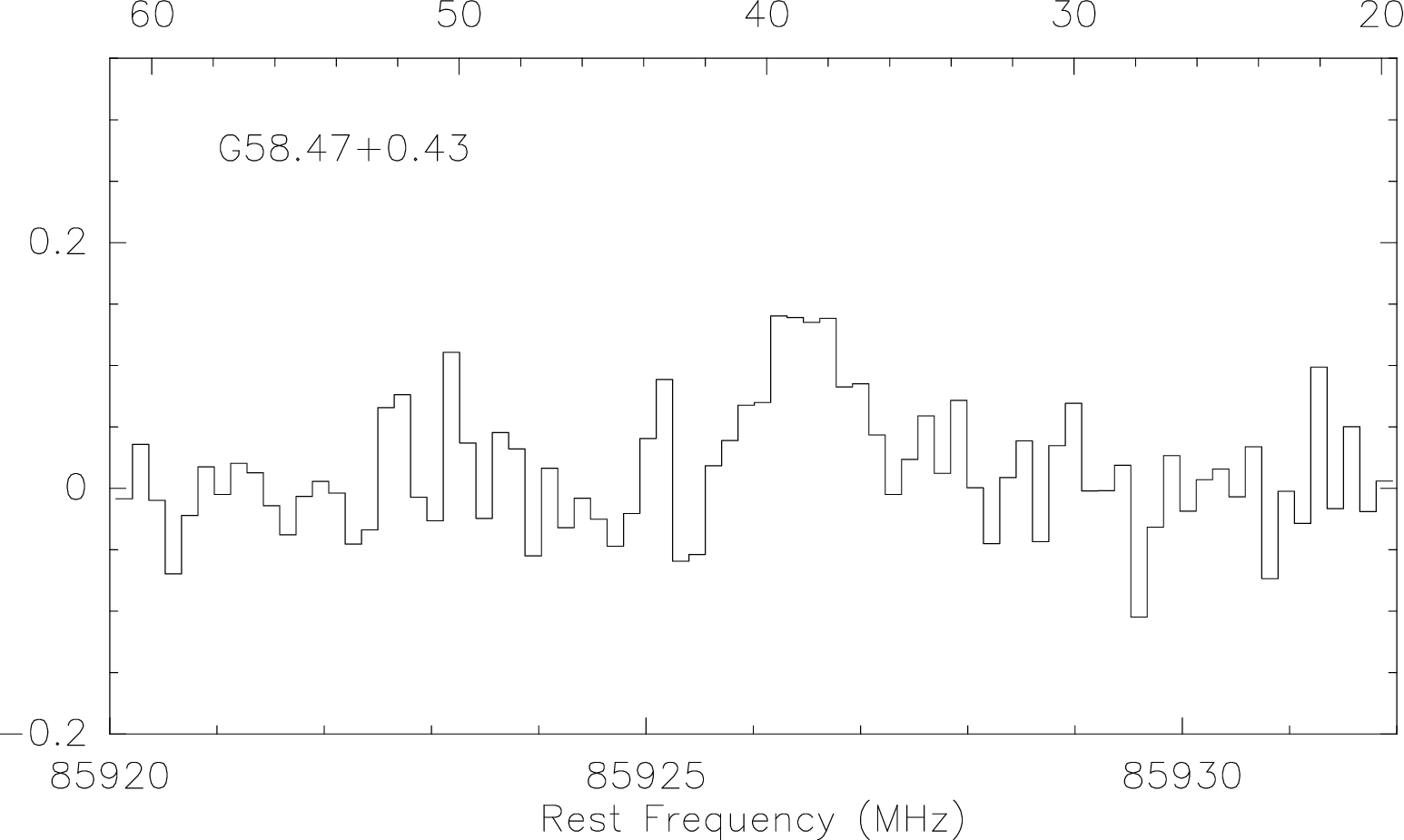}} 
\\ \vspace{1cm}
\end{minipage}
\hfill
\begin{minipage}[h]{0.3\linewidth}
\center{\includegraphics[width=1\linewidth]{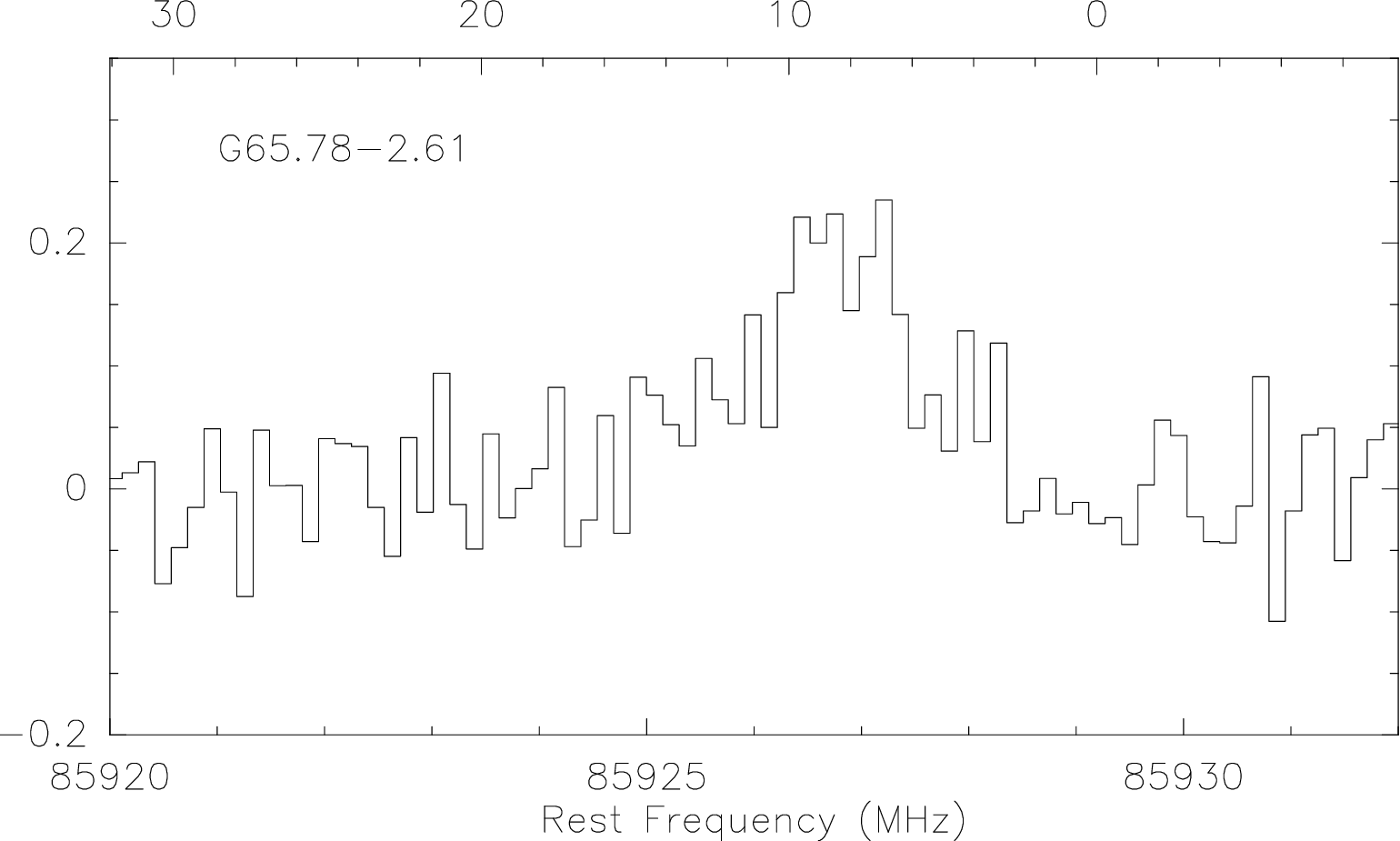}} 
\\ \vspace{1cm}
\end{minipage}

\caption{Spectra with detected NH$_2$D lines. The upper scale in the plots shows $V_\mathrm{LSR}$ in km/s.}
\label{ris:NH2D_spec_sm}
\end{figure}

\addtocounter{figure}{-1}

\begin{figure}[h]
\begin{minipage}[h]{0.3\linewidth}
\center{\includegraphics[width=1\linewidth]{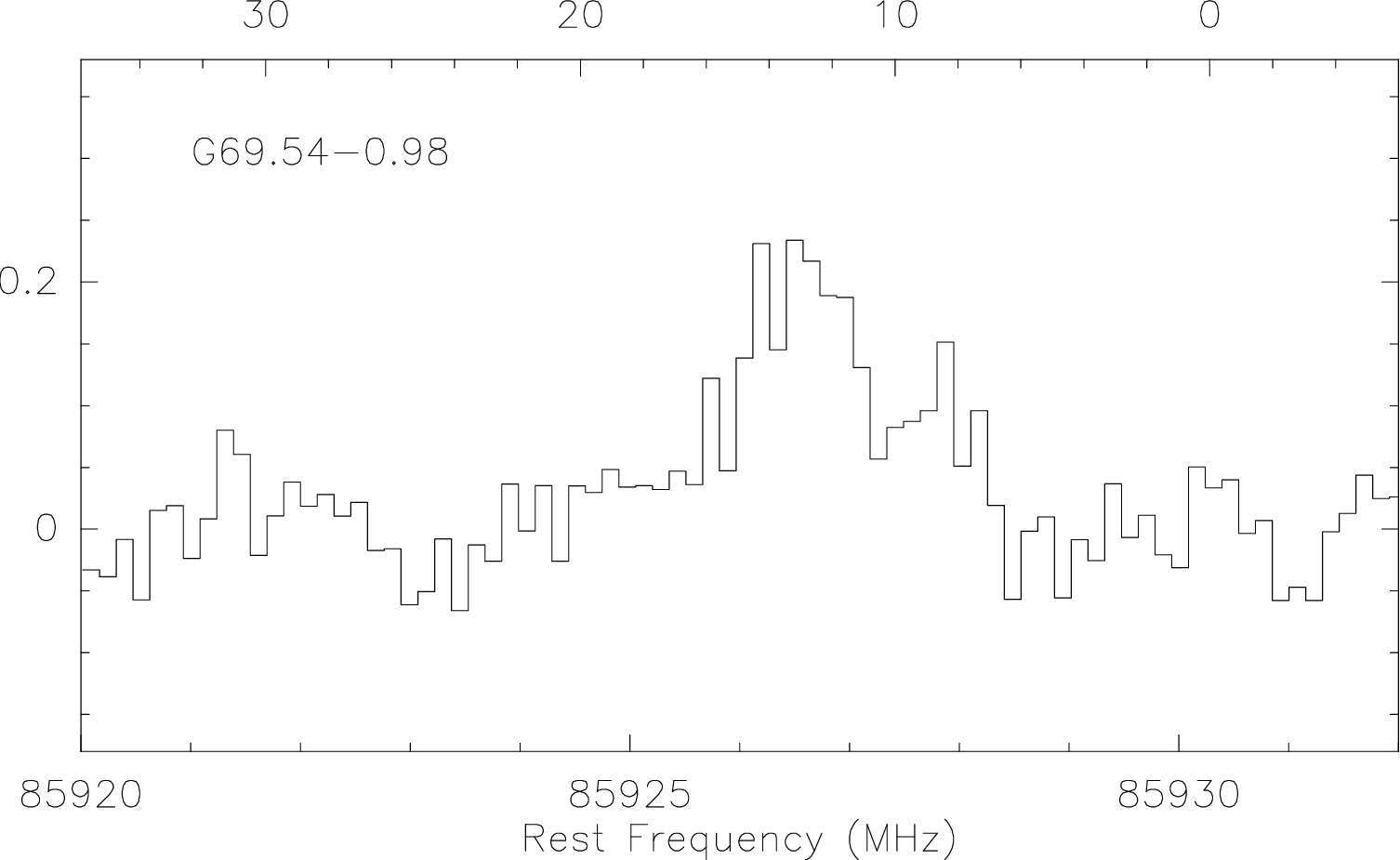}}  \\ \vspace{1cm}
\end{minipage}
\hfill
\begin{minipage}[h]{0.3\linewidth}
\center{\includegraphics[width=1\linewidth]{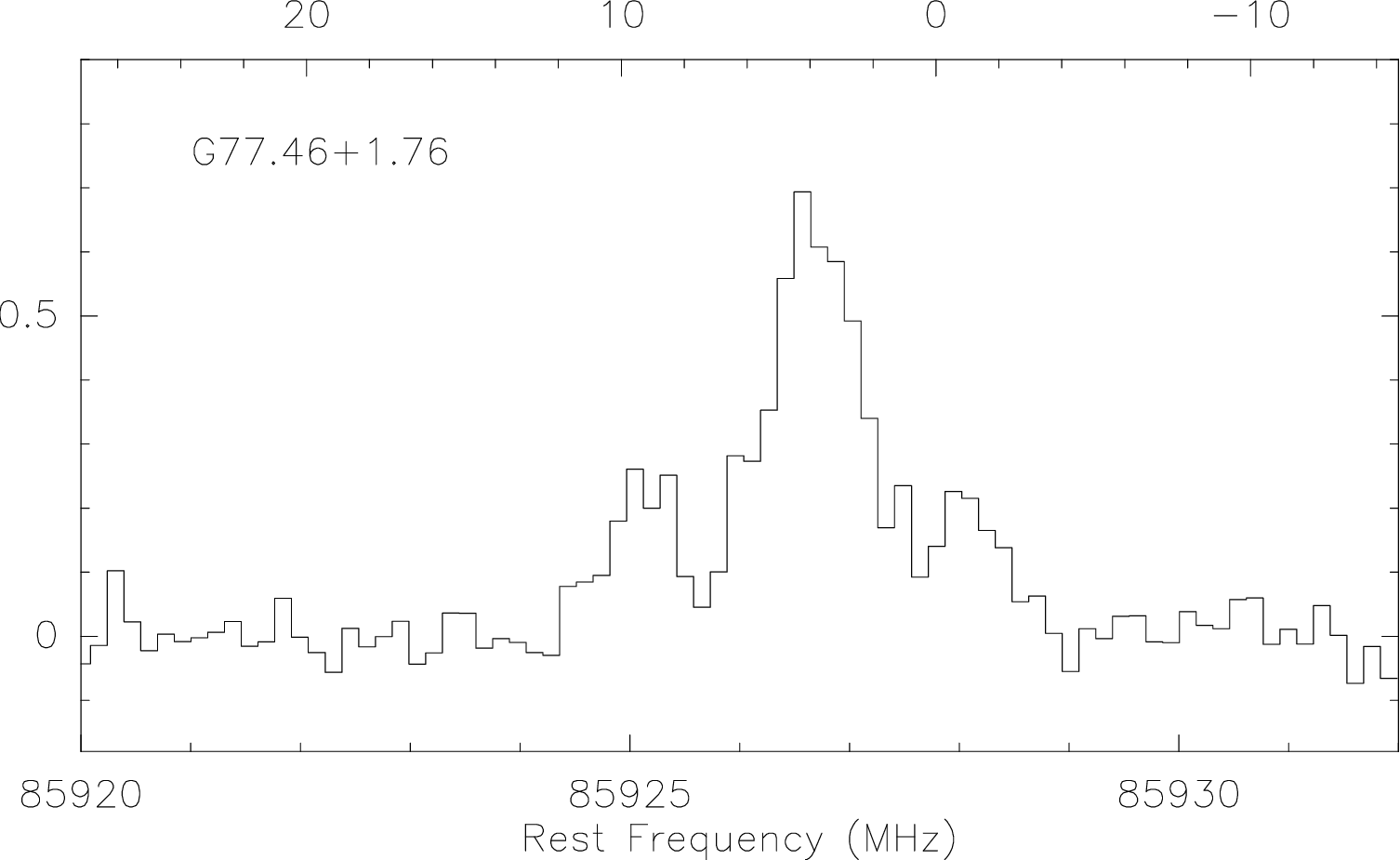}} 
\\ \vspace{1cm}
\end{minipage}
\hfill
\begin{minipage}[h]{0.3\linewidth}
\center{\includegraphics[width=1\linewidth]{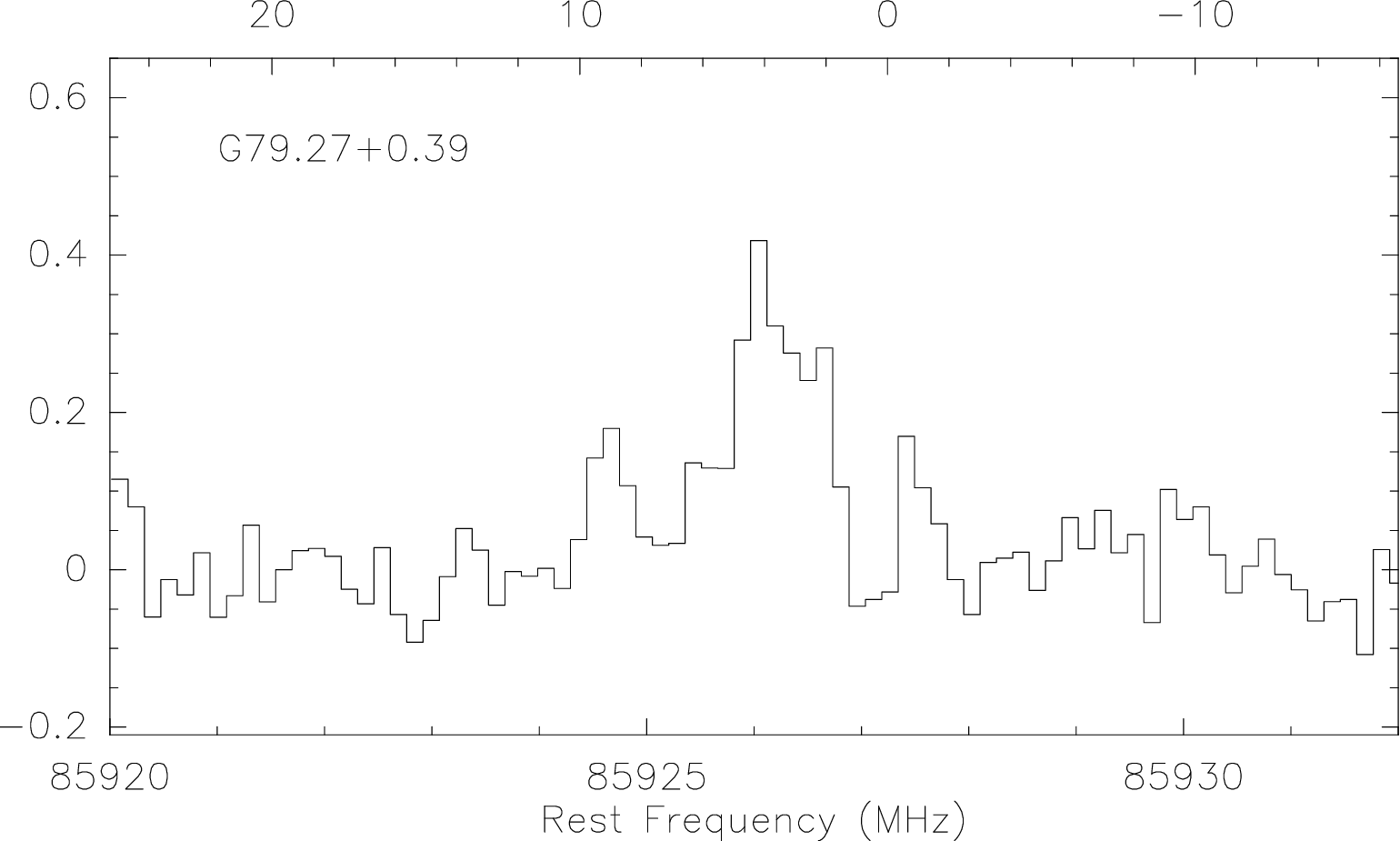}} 
\\ \vspace{1cm}
\end{minipage}

\vfill
\begin{minipage}[h]{0.3\linewidth}
\center{\includegraphics[width=1\linewidth]{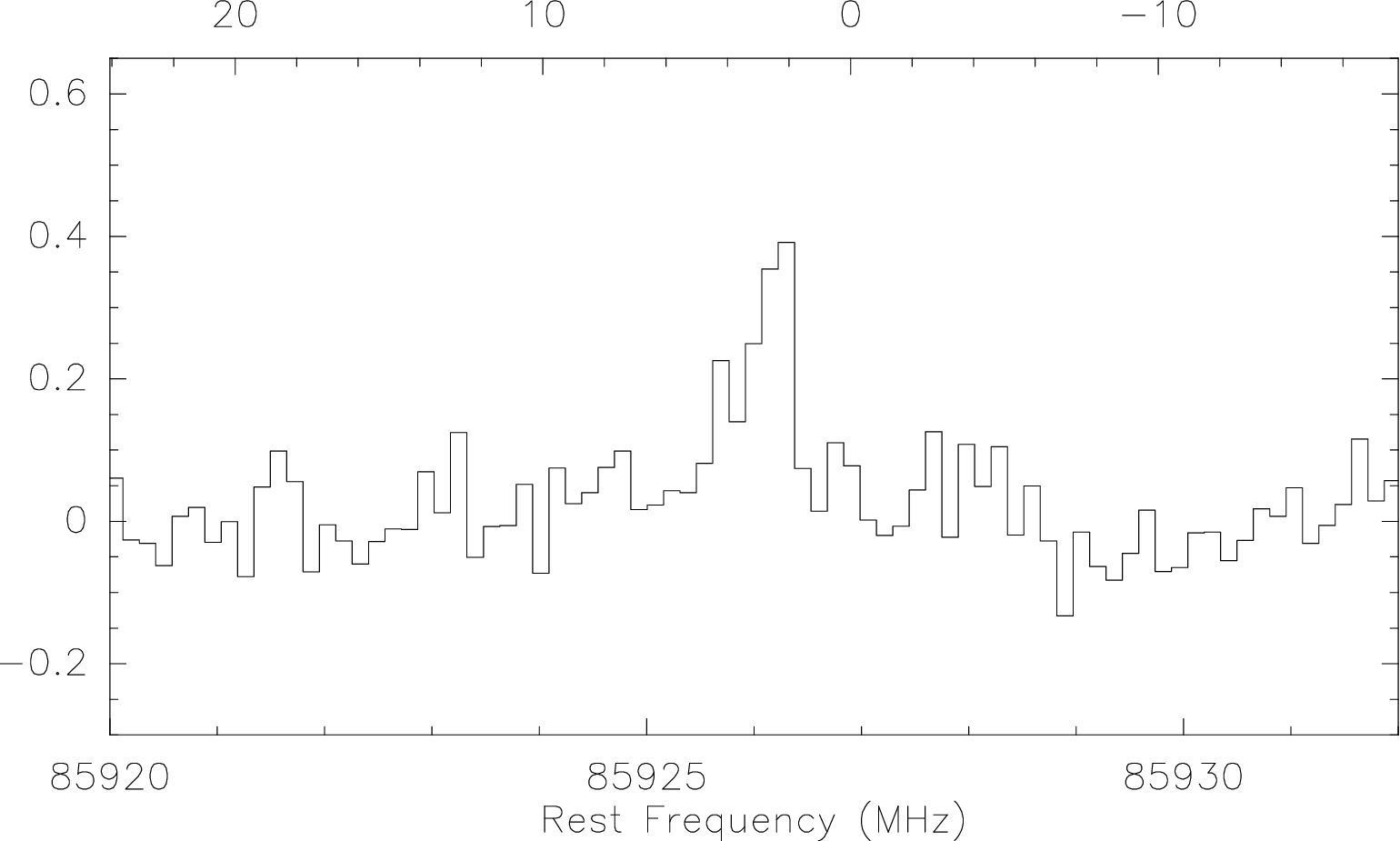}}  \\ \vspace{1cm}
\end{minipage}
\hfill
\begin{minipage}[h]{0.3\linewidth}
\center{\includegraphics[width=1\linewidth]{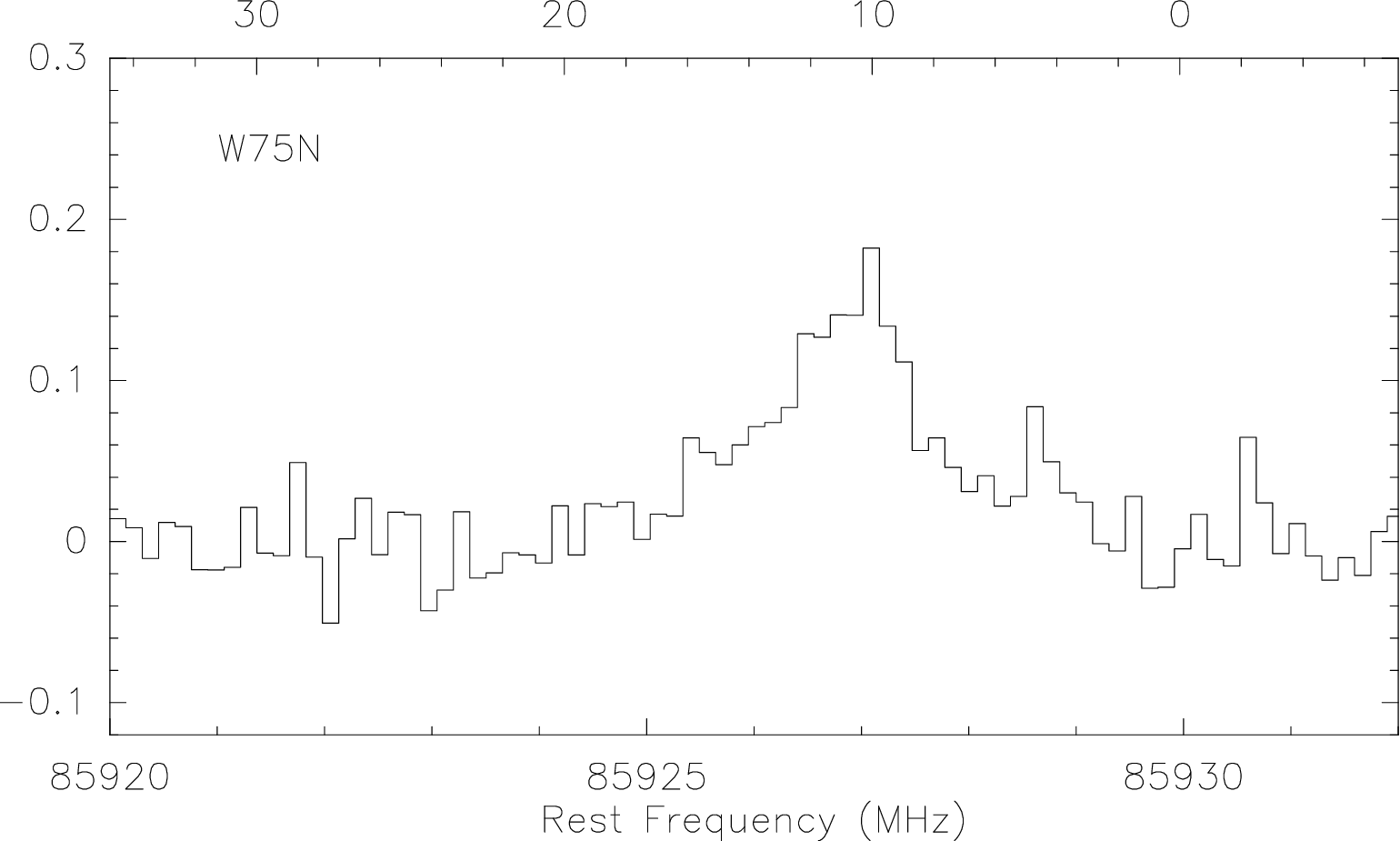}} 
\\ \vspace{1cm}
\end{minipage}
\hfill
\begin{minipage}[h]{0.3\linewidth}
\center{\includegraphics[width=1\linewidth]{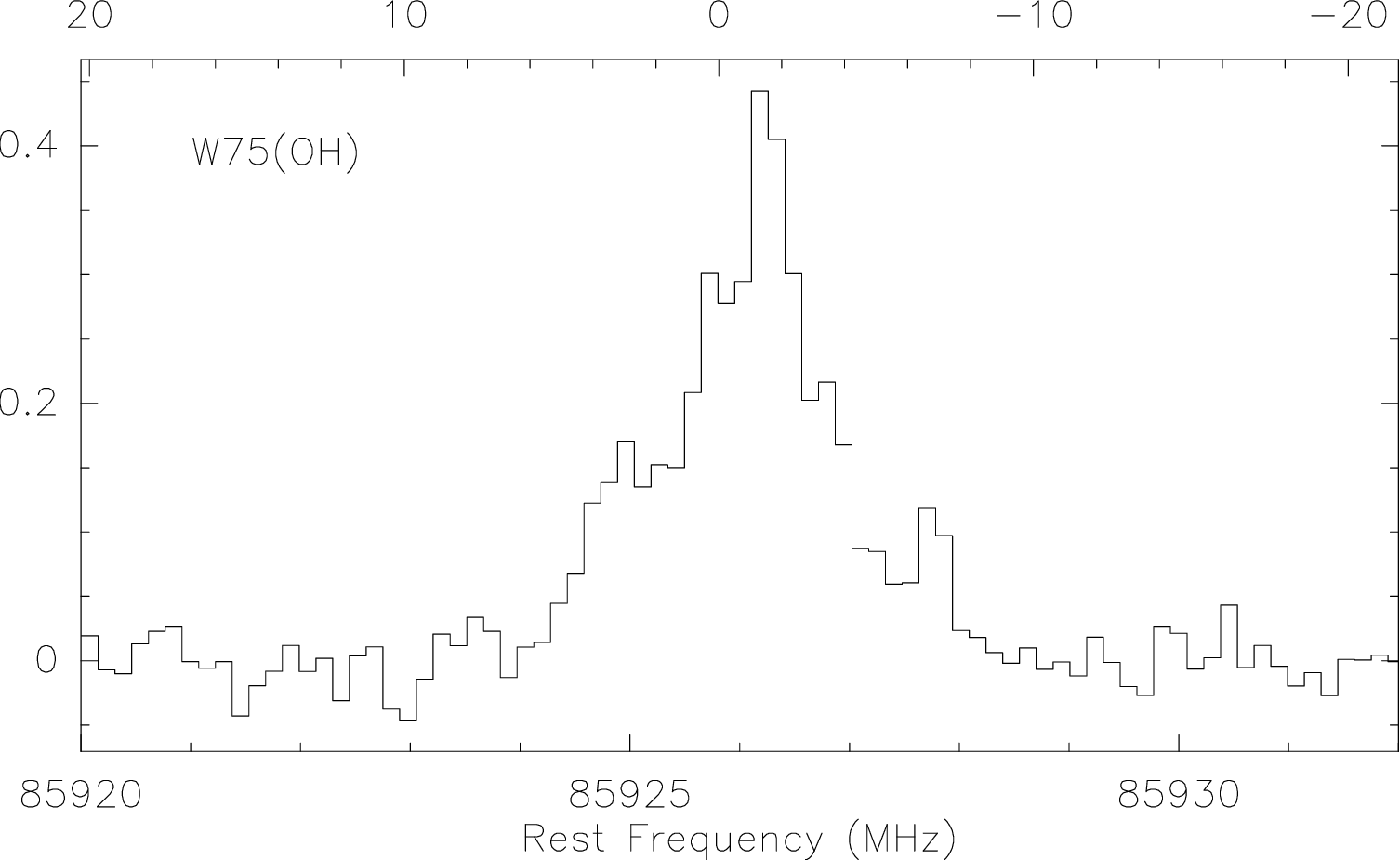}} 
\\ \vspace{1cm}
\end{minipage}

\vfill
\begin{minipage}[h]{0.3\linewidth}
\center{\includegraphics[width=1\linewidth]{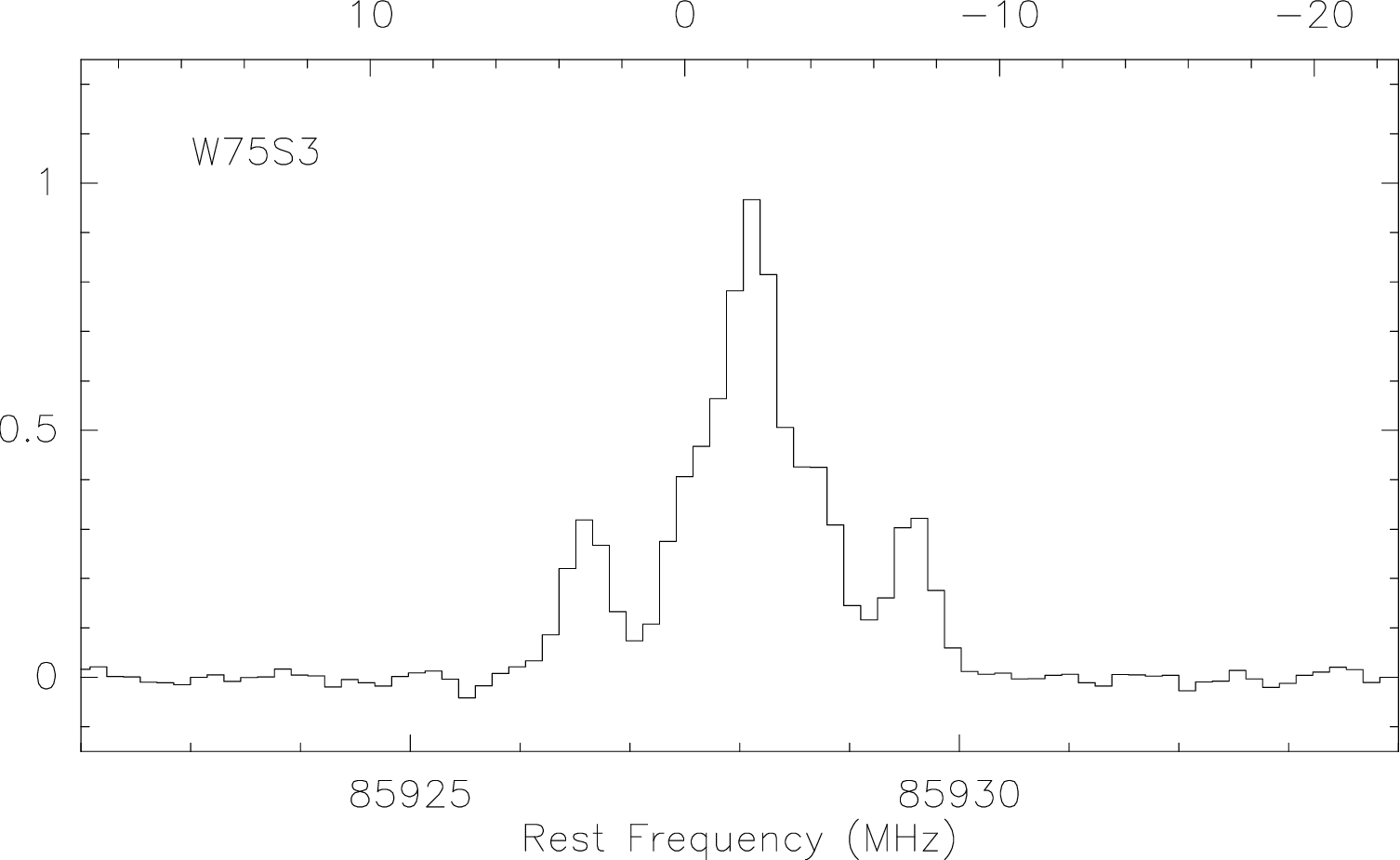}}  \\ \vspace{1cm}
\end{minipage}
\hfill
\begin{minipage}[h]{0.3\linewidth}
\center{\includegraphics[width=1\linewidth]{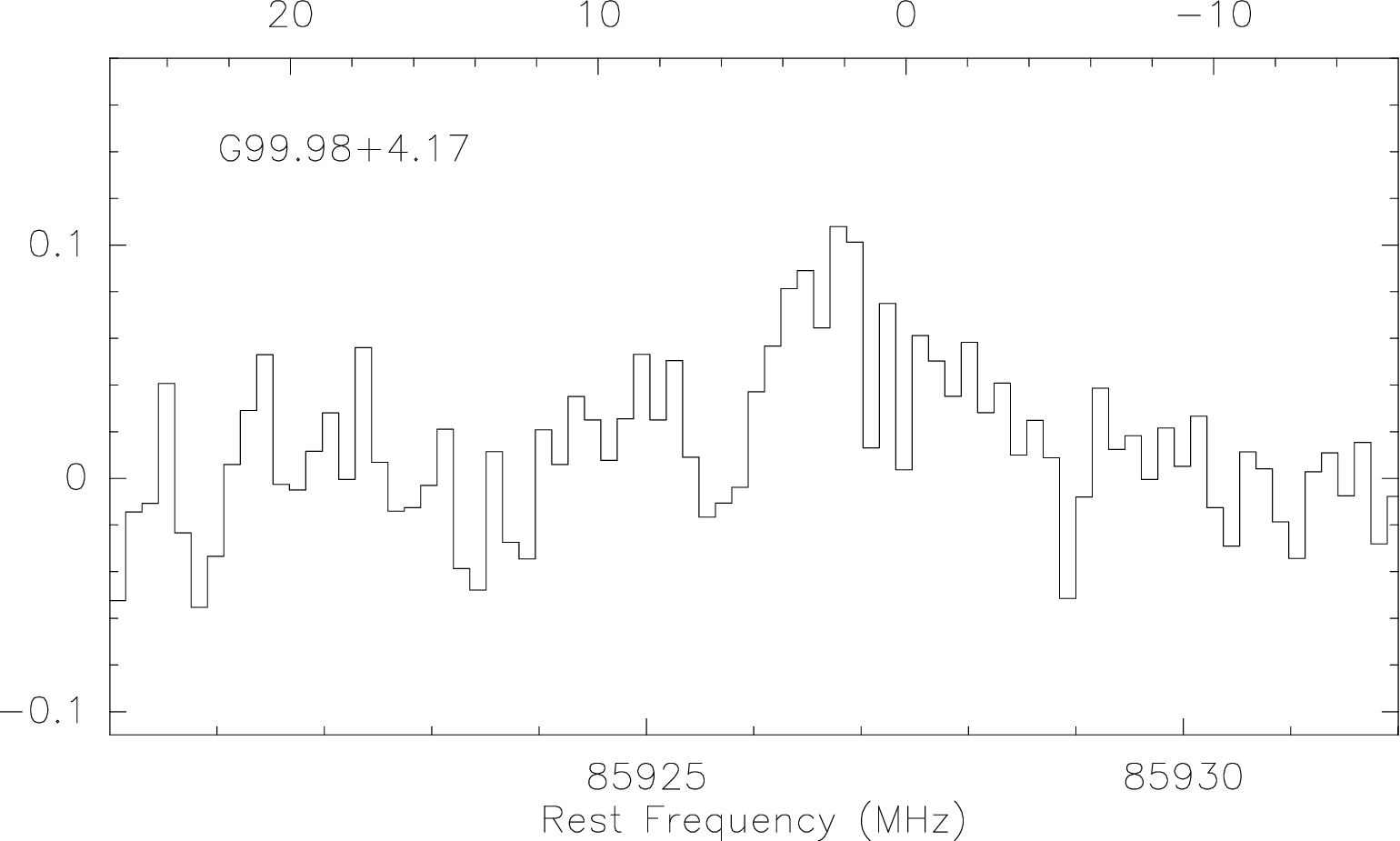}} 
\\ \vspace{1cm}
\end{minipage}
\hfill
\begin{minipage}[h]{0.3\linewidth}
\center{\includegraphics[width=1\linewidth]{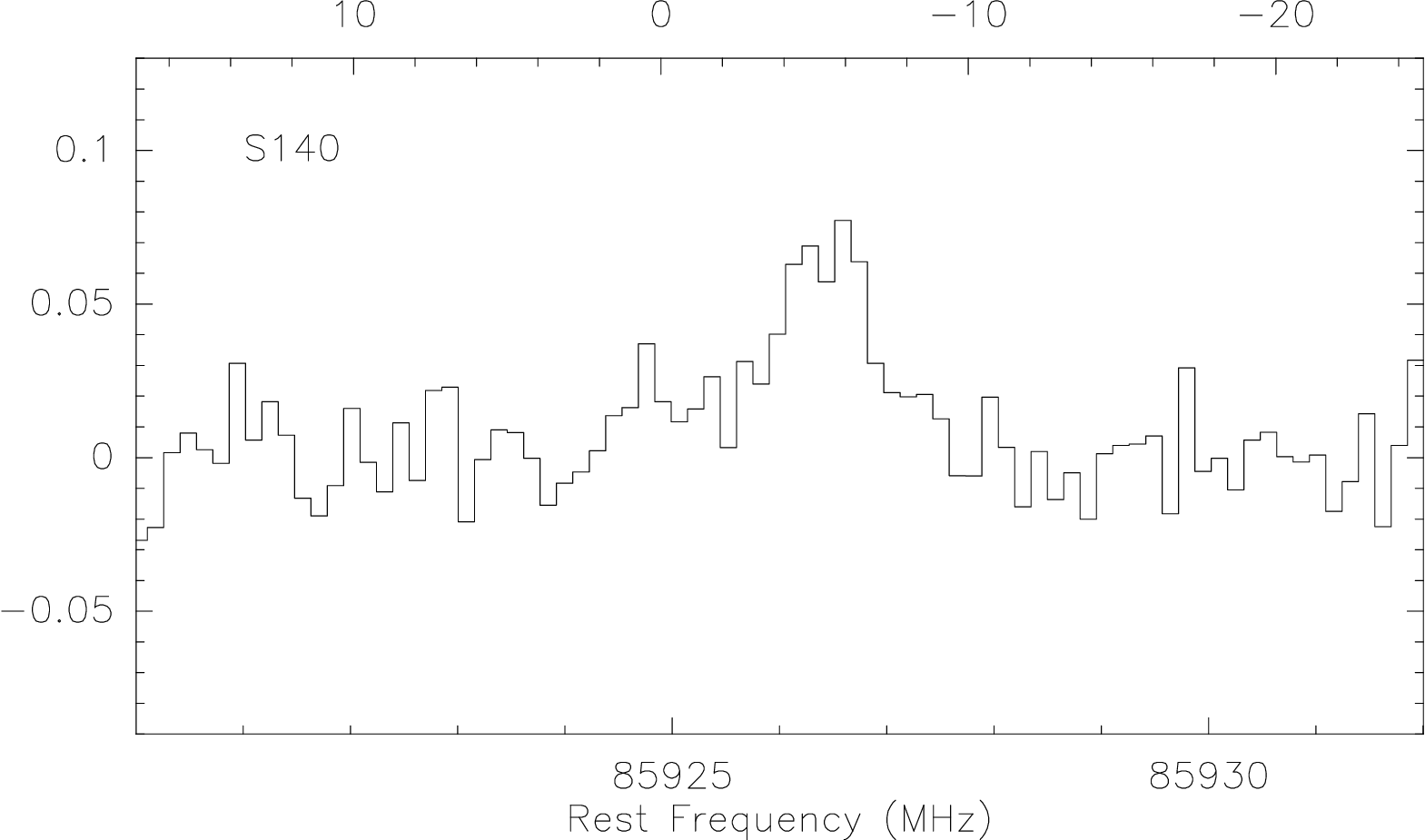}} 
\\ \vspace{1cm}
\end{minipage}

\vfill
\begin{minipage}[h]{0.3\linewidth}
\center{\includegraphics[width=1\linewidth]{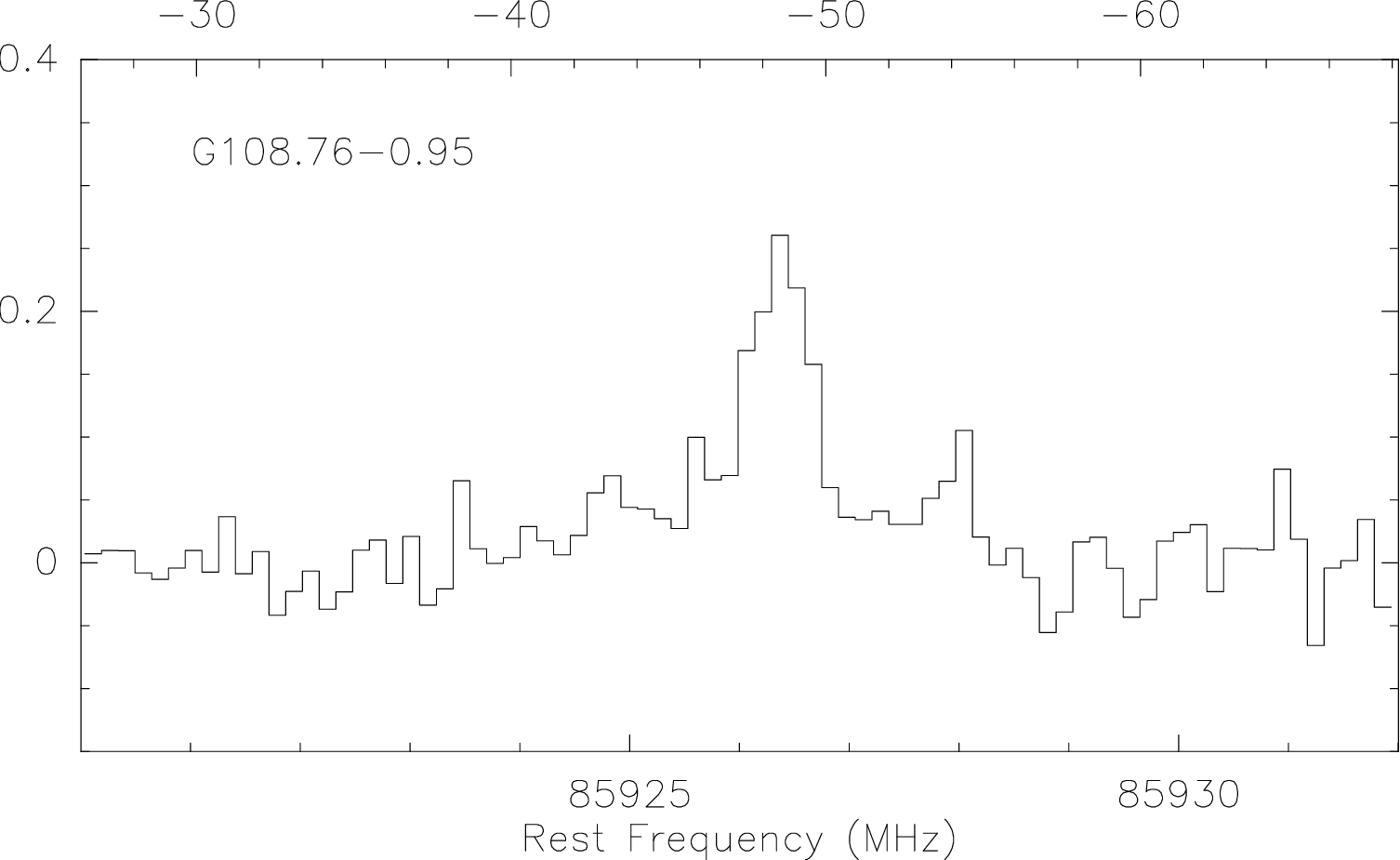}}  \\ \vspace{1cm}
\end{minipage}
\hfill
\begin{minipage}[h]{0.3\linewidth}
\center{\includegraphics[width=1\linewidth]{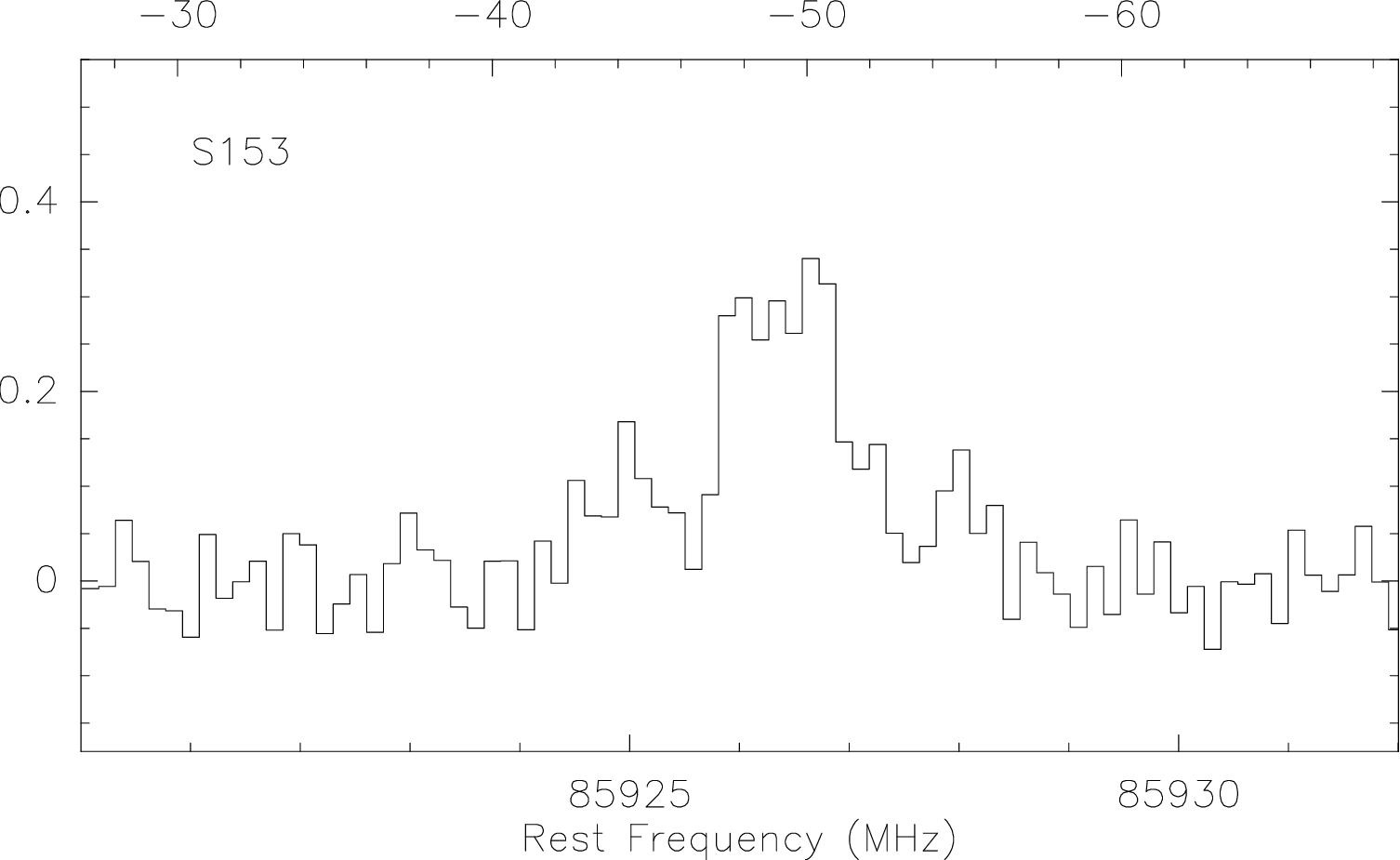}} 
\\ \vspace{1cm}
\end{minipage}
\hfill
\begin{minipage}[h]{0.3\linewidth}
\center{\includegraphics[width=1\linewidth]{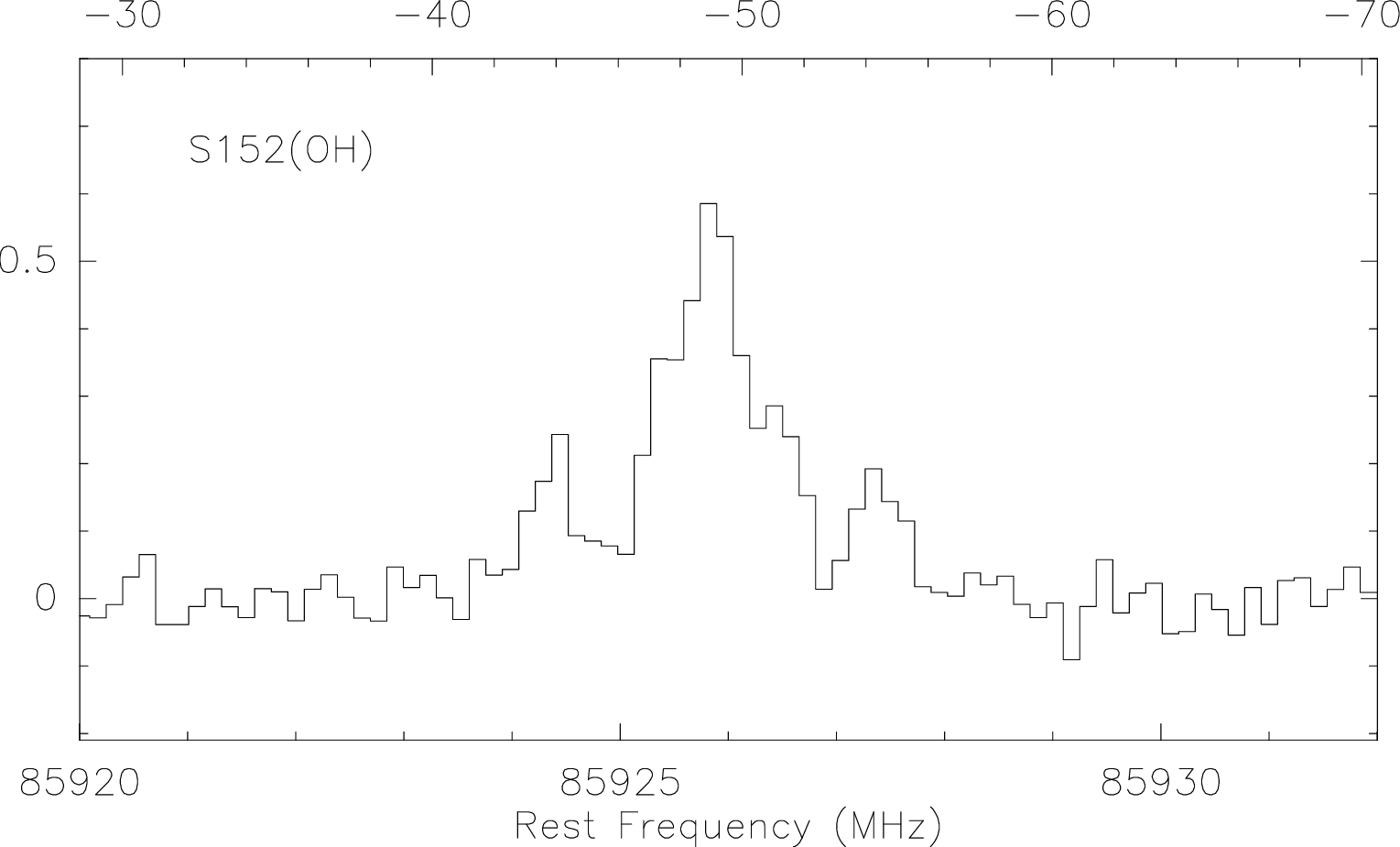}} 
\\ \vspace{1cm}
\end{minipage}

\vfill
\begin{minipage}[h]{0.3\linewidth}
\center{\includegraphics[width=1\linewidth]{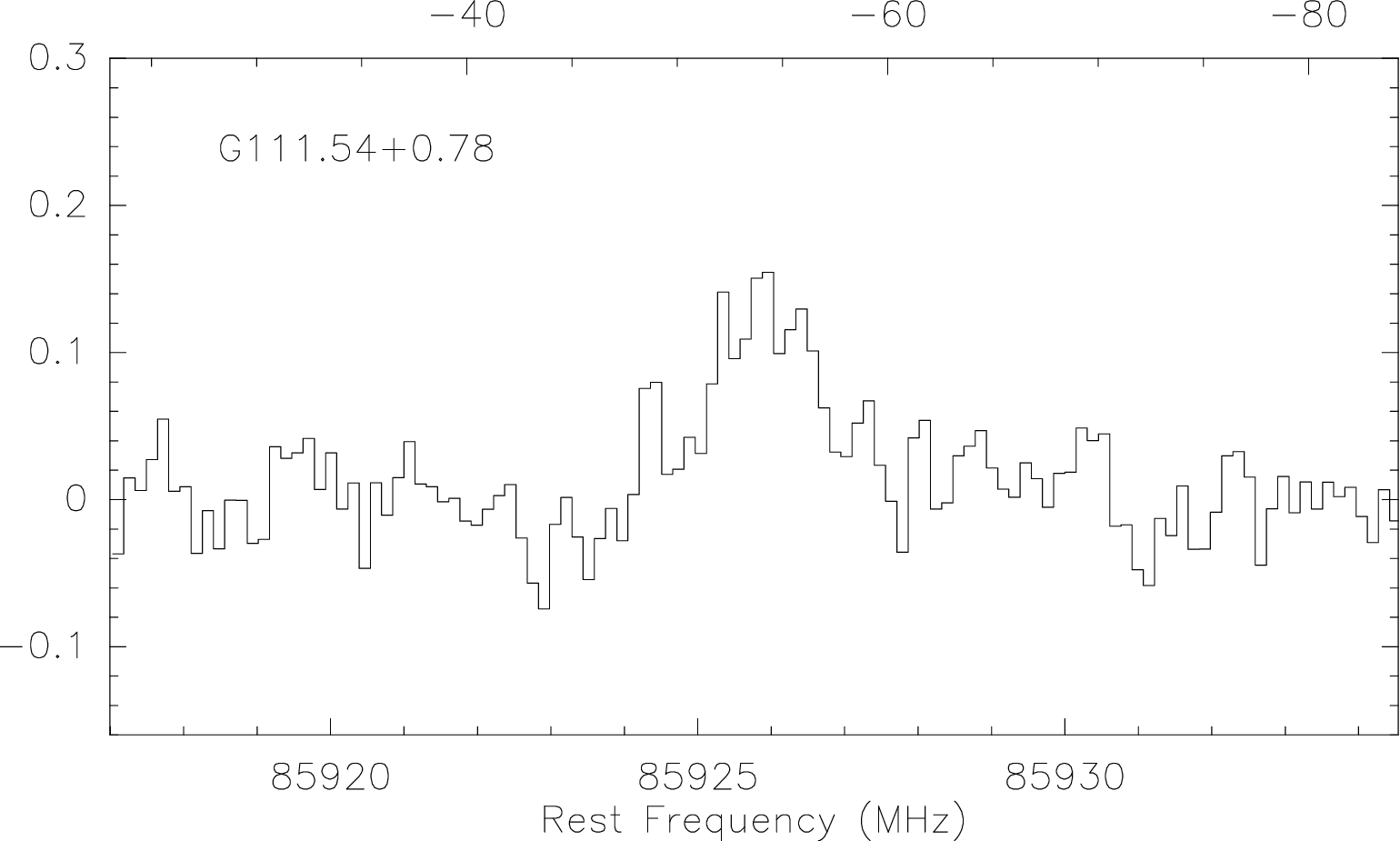}}  \\ \vspace{1cm}
\end{minipage}
\hfill
\begin{minipage}[h]{0.3\linewidth}
\center{\includegraphics[width=1\linewidth]{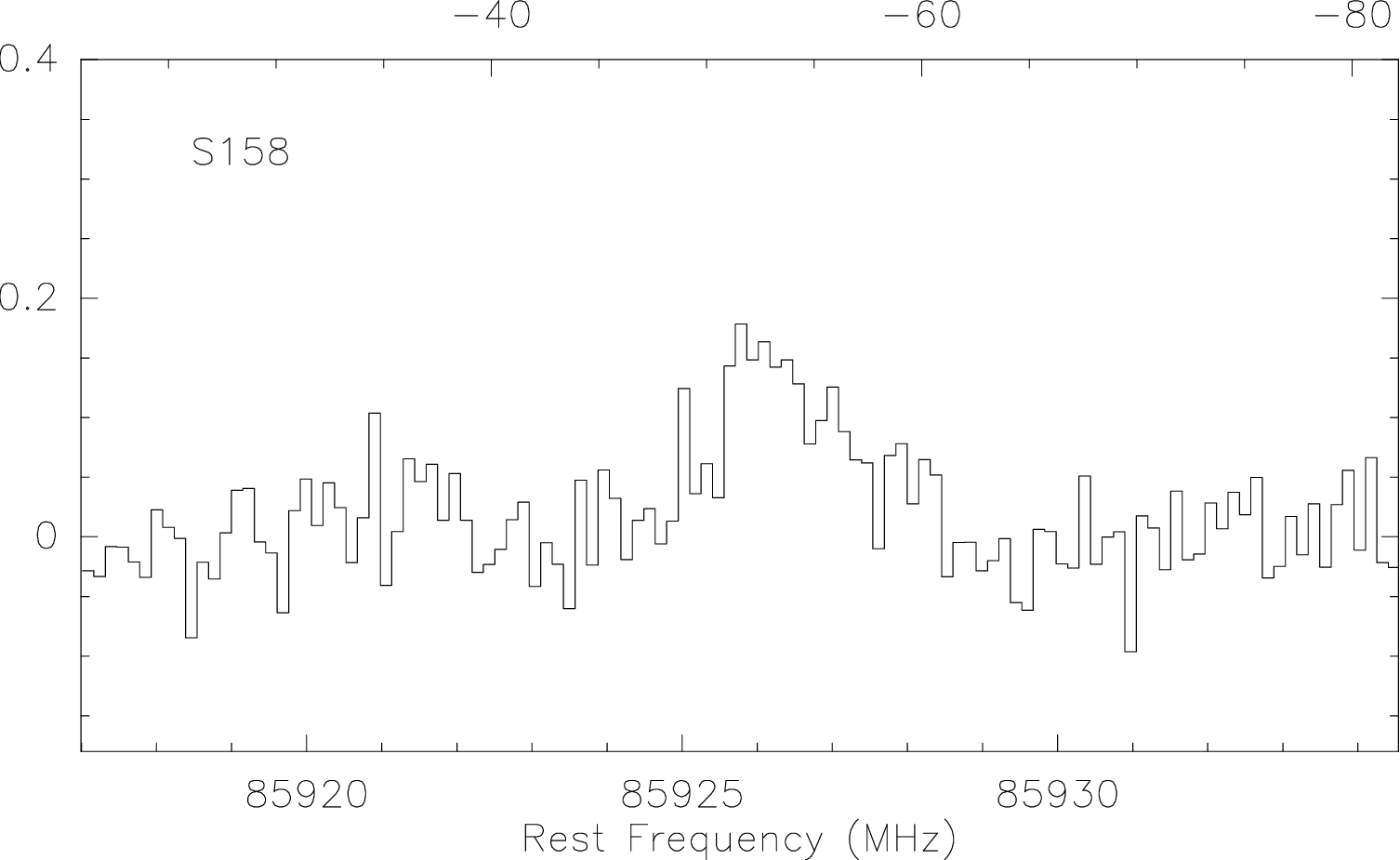}} 
\\ \vspace{1cm}
\end{minipage}

\caption{Continuation.}
\end{figure}

\newpage

\begin{figure}[h]
\begin{minipage}[h]{0.3\linewidth}
\center{\includegraphics[width=1\linewidth]{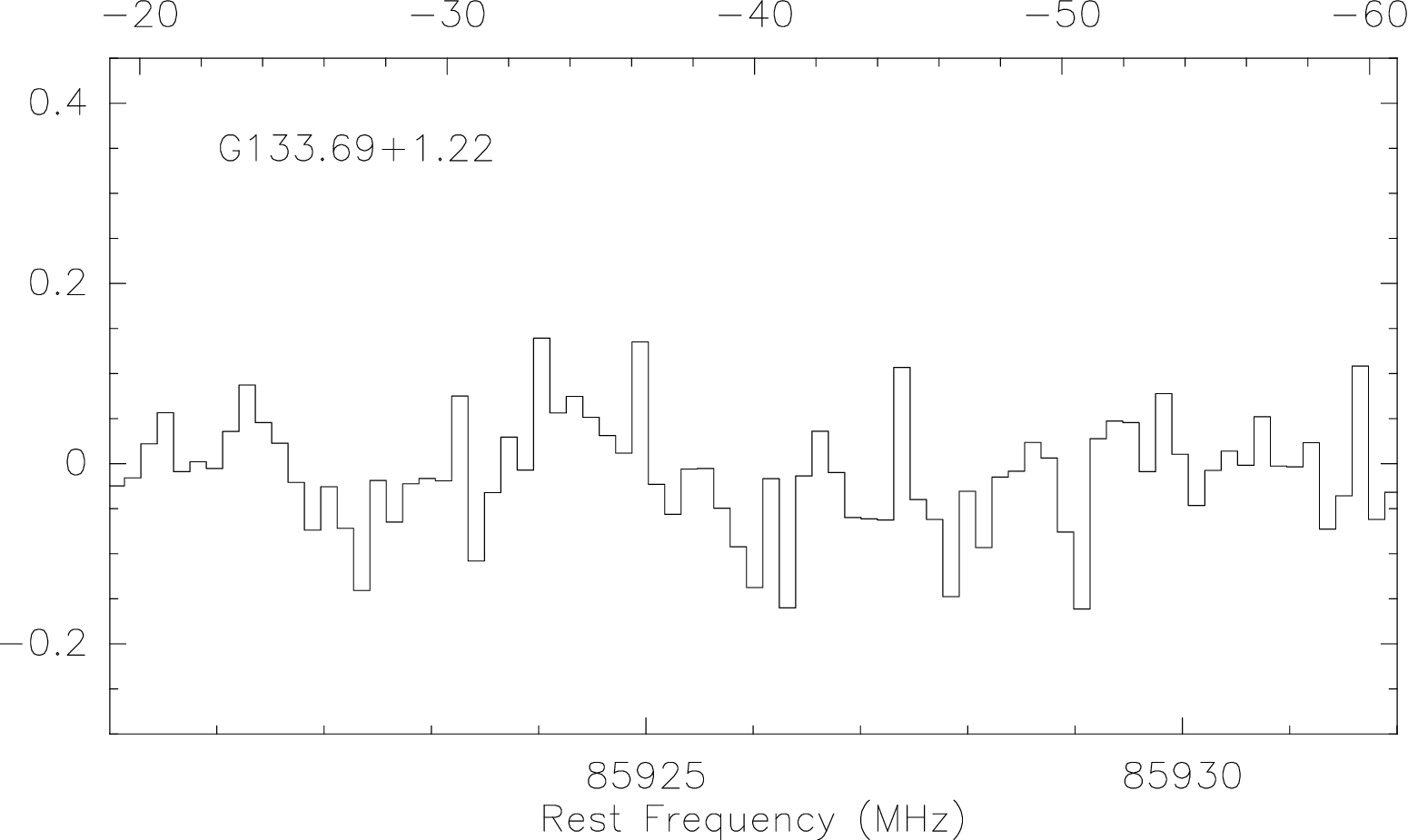}}  \\ \vspace{1cm}
\end{minipage}
\hfill
\begin{minipage}[h]{0.3\linewidth}
\center{\includegraphics[width=1\linewidth]{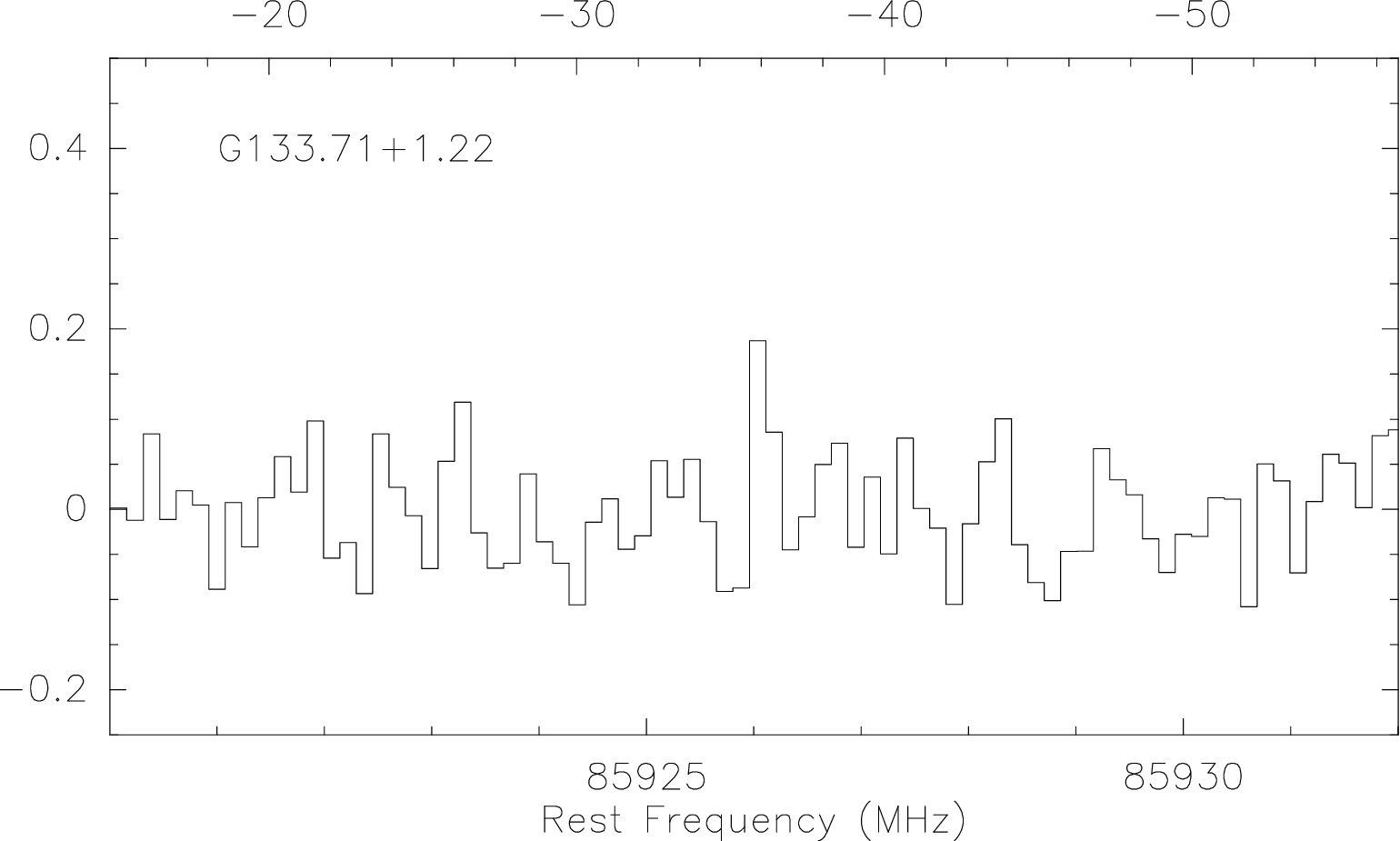}} 
\\ \vspace{1cm}
\end{minipage}
\hfill
\begin{minipage}[h]{0.3\linewidth}
\center{\includegraphics[width=1\linewidth]{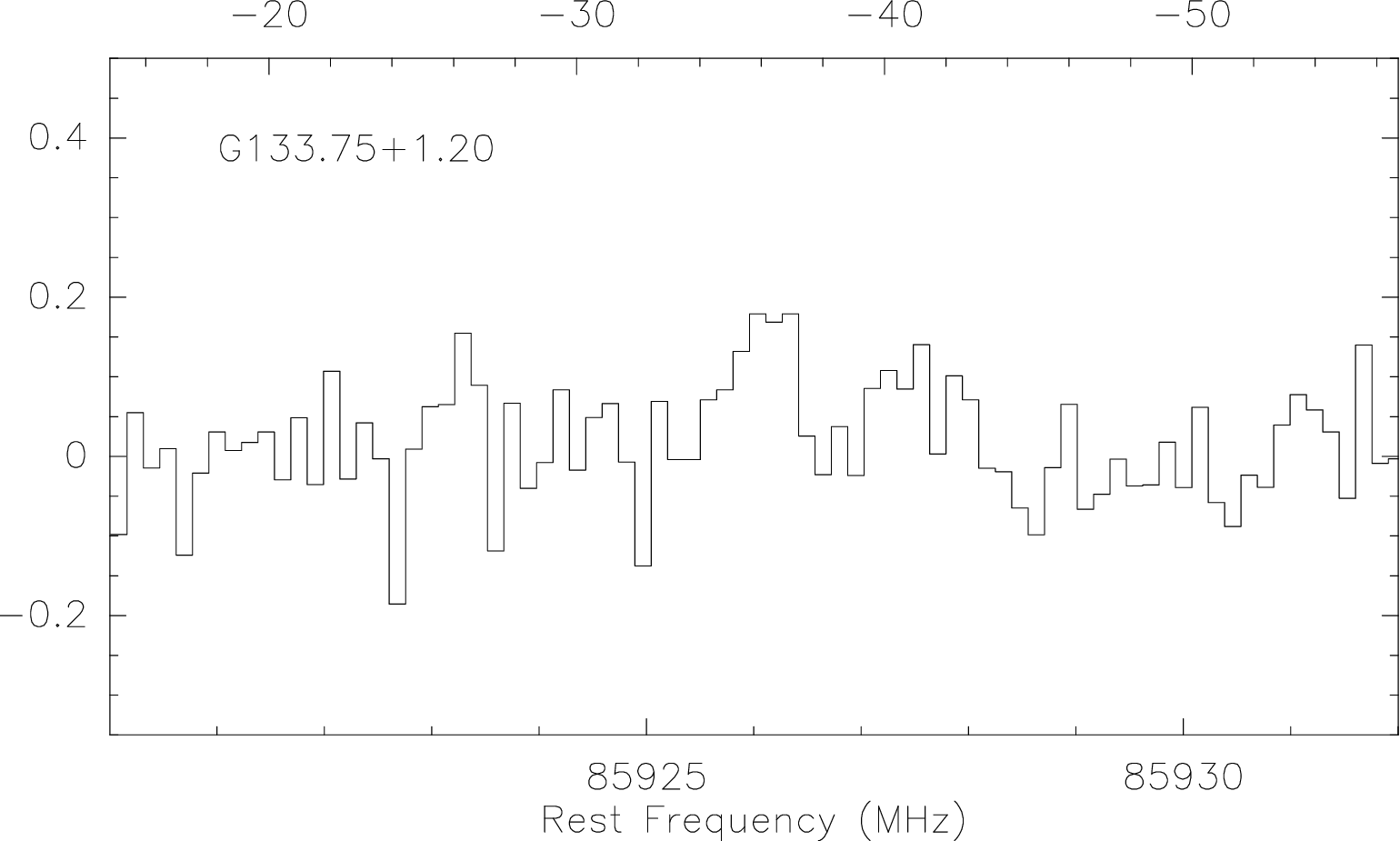}} 
\\ \vspace{1cm}
\end{minipage}

\vfill
\begin{minipage}[h]{0.3\linewidth}
\center{\includegraphics[width=1\linewidth]{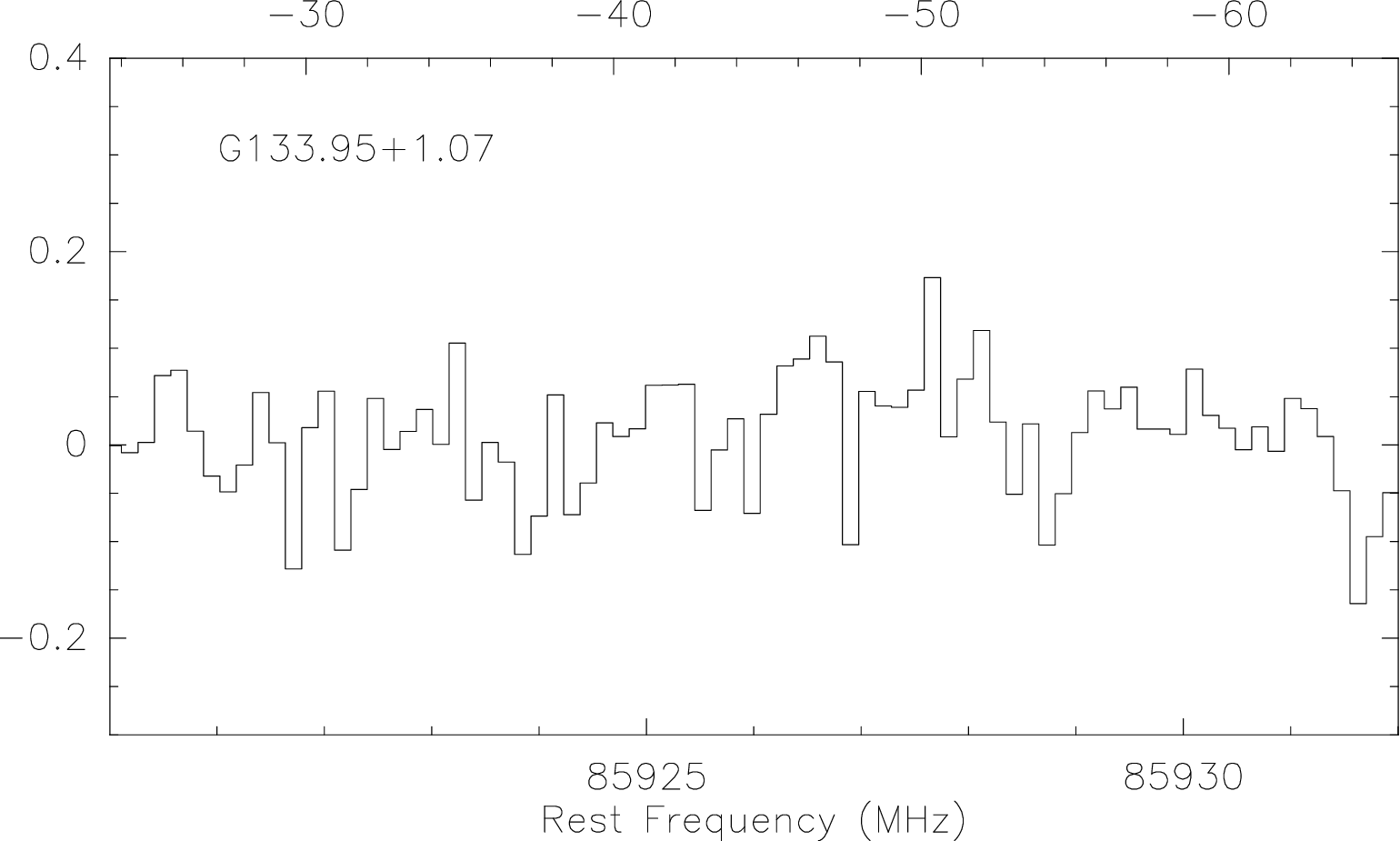}}  \\ \vspace{1cm}
\end{minipage}
\hfill
\begin{minipage}[h]{0.3\linewidth}
\center{\includegraphics[width=1\linewidth]{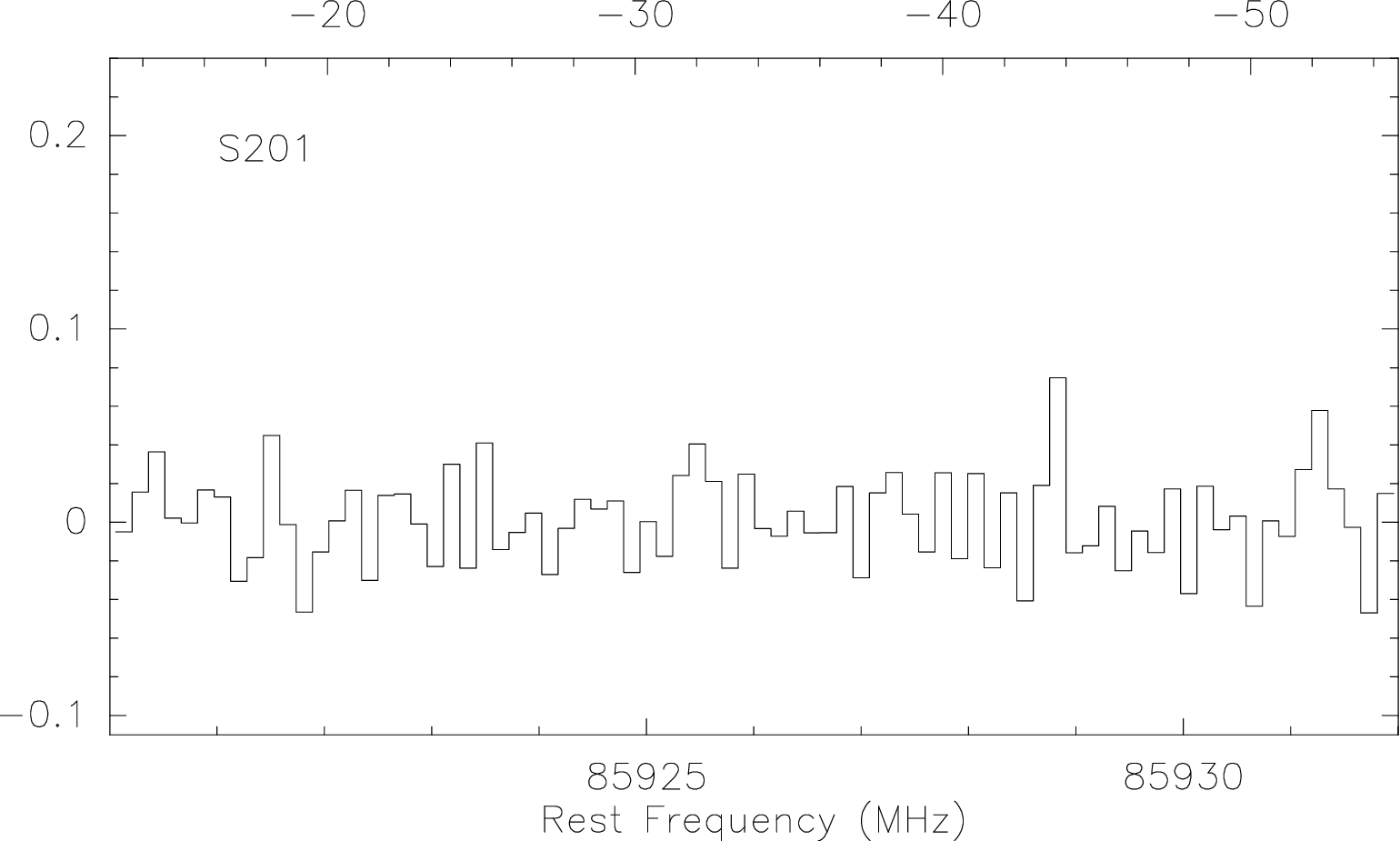}} 
\\ \vspace{1cm}
\end{minipage}
\hfill
\begin{minipage}[h]{0.3\linewidth}
\center{\includegraphics[width=1\linewidth]{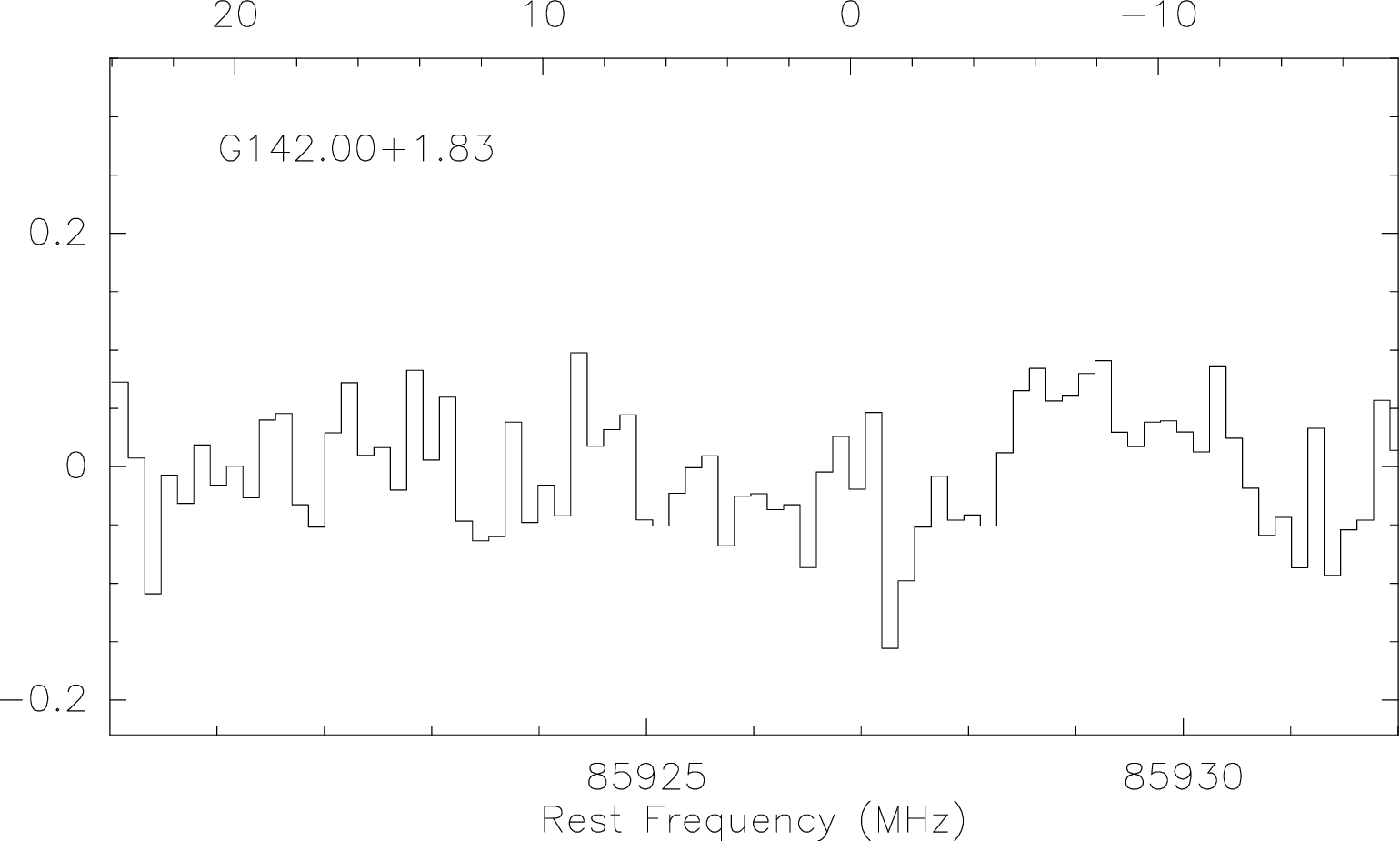}} 
\\ \vspace{1cm}
\end{minipage}

\vfill
\begin{minipage}[h]{0.3\linewidth}
\center{\includegraphics[width=1\linewidth]{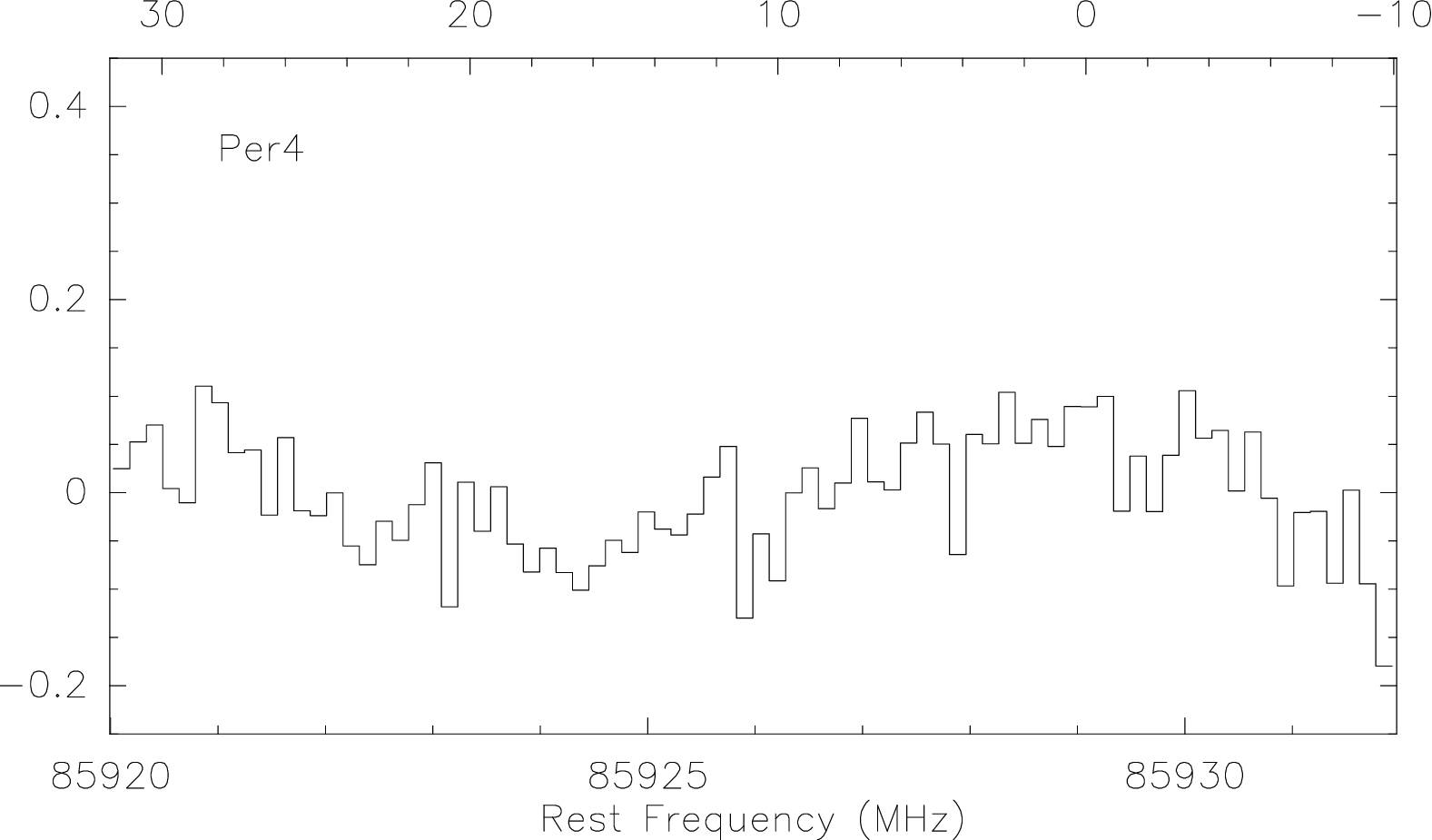}}  \\ \vspace{1cm}
\end{minipage}
\hfill
\begin{minipage}[h]{0.3\linewidth}
\center{\includegraphics[width=1\linewidth]{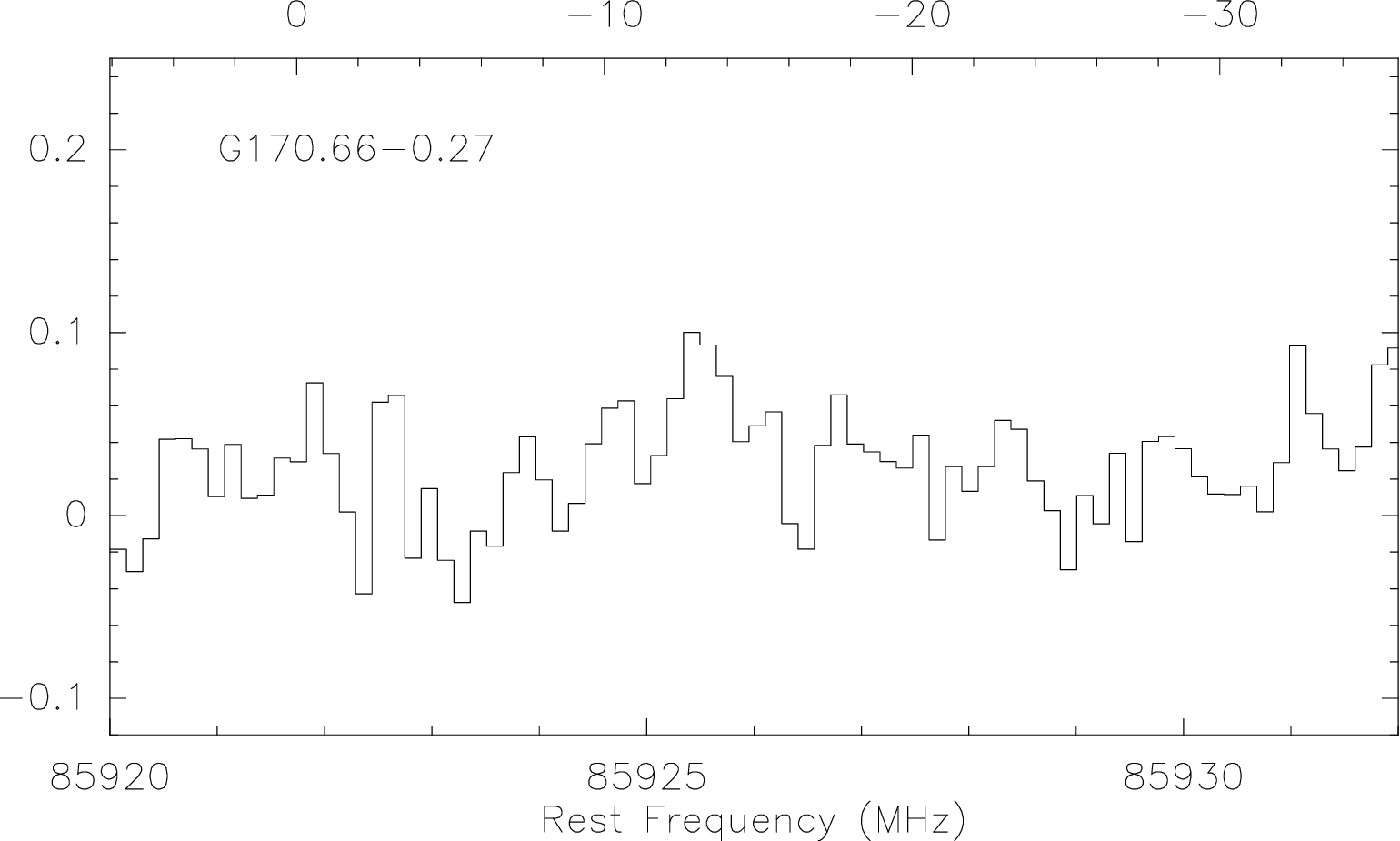}} 
\\ \vspace{1cm}
\end{minipage}
\hfill
\begin{minipage}[h]{0.3\linewidth}
\center{\includegraphics[width=1\linewidth]{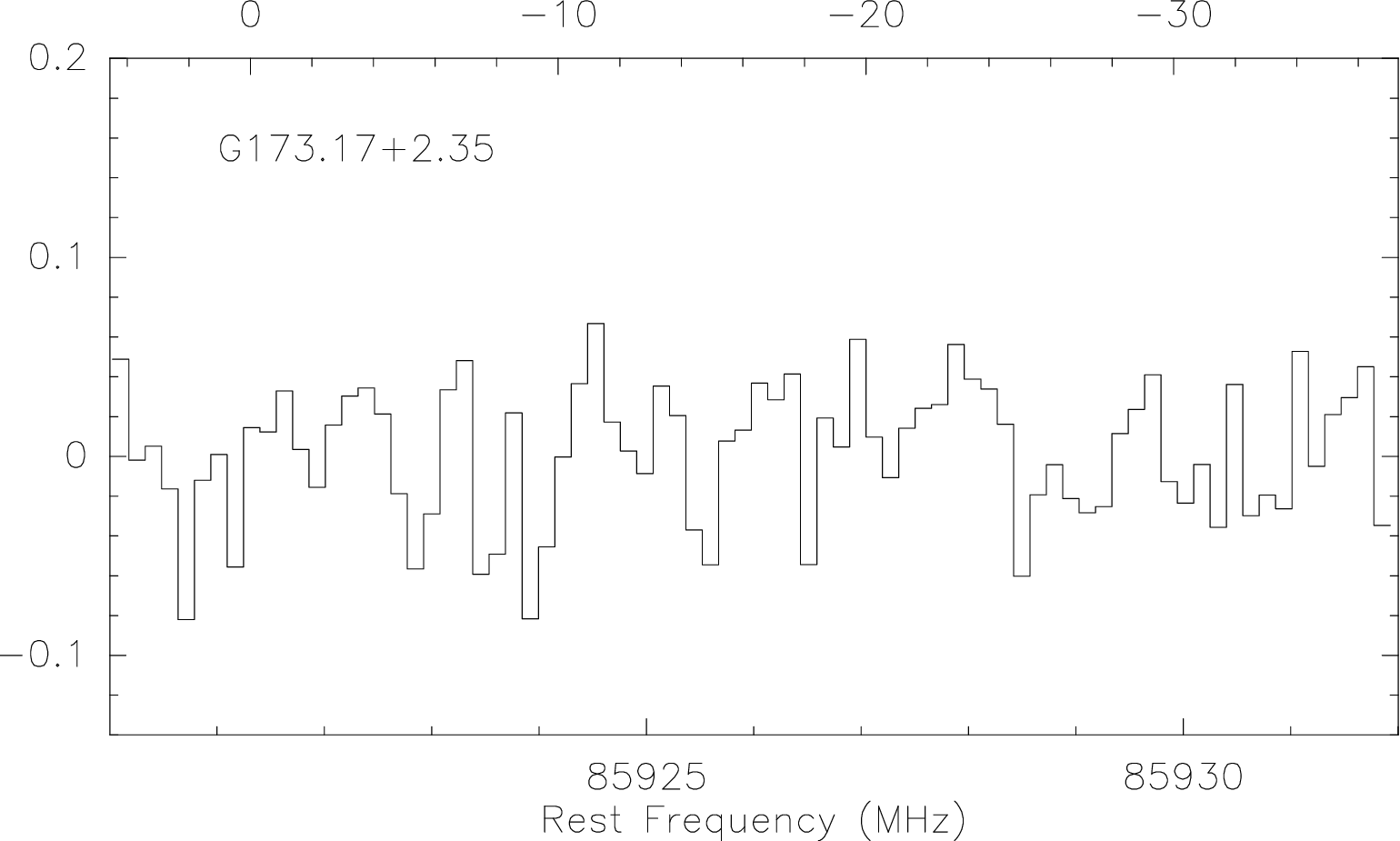}} 
\\ \vspace{1cm}
\end{minipage}

\vfill
\begin{minipage}[h]{0.3\linewidth}
\center{\includegraphics[width=1\linewidth]{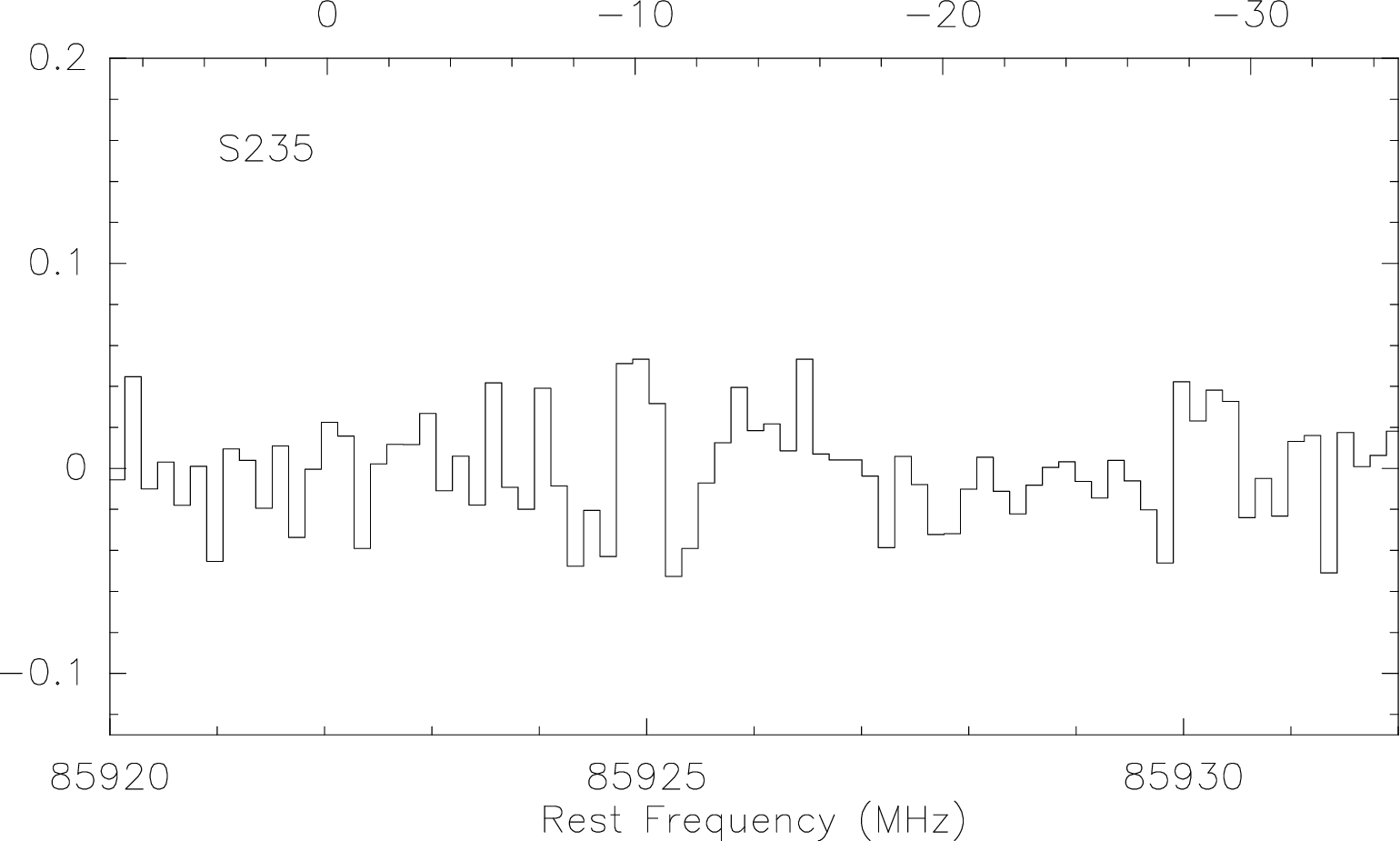}}  \\ \vspace{1cm}
\end{minipage}
\hfill
\begin{minipage}[h]{0.3\linewidth}
\center{\includegraphics[width=1\linewidth]{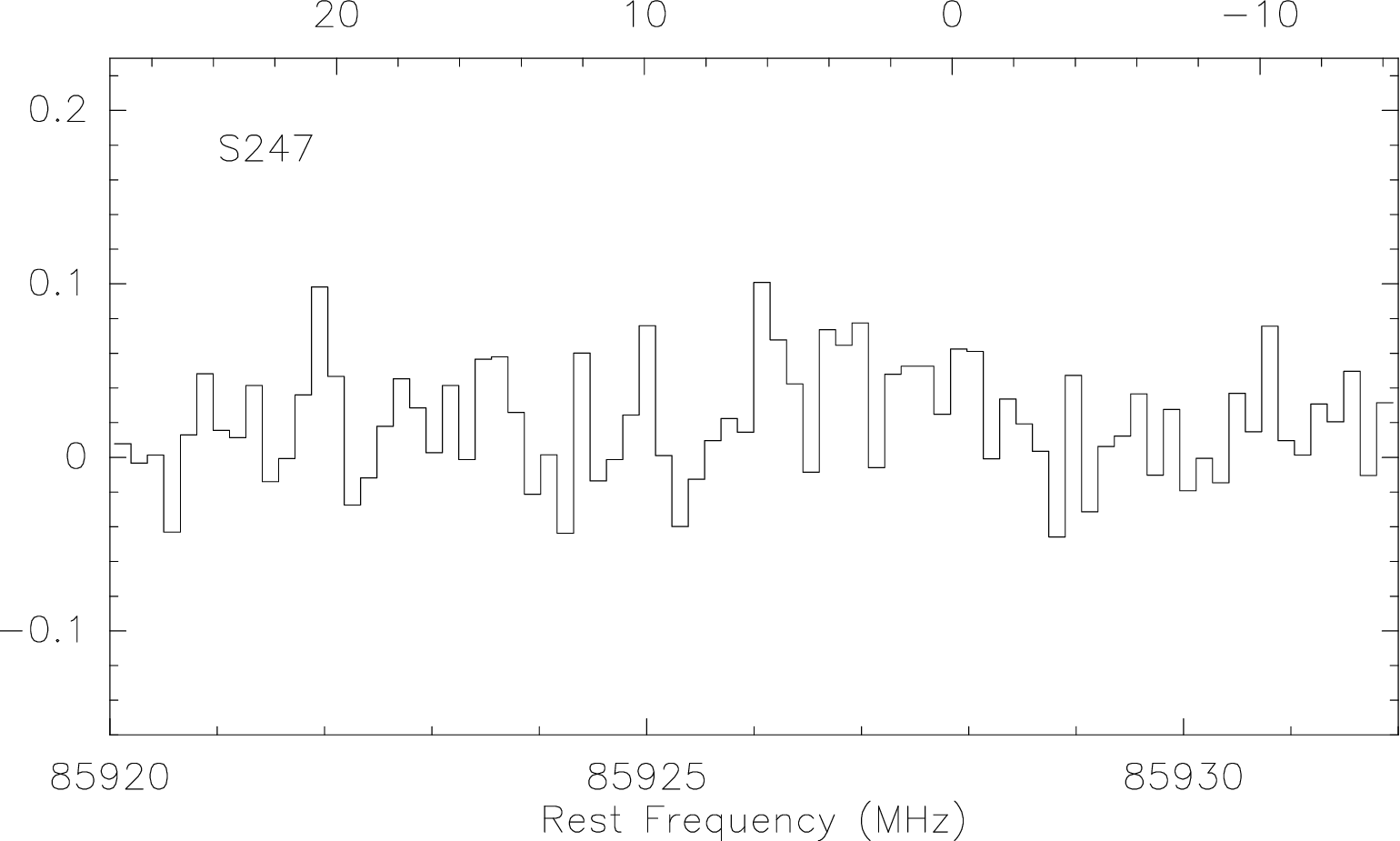}} 
\\ \vspace{1cm}
\end{minipage}
\hfill
\begin{minipage}[h]{0.3\linewidth}
\center{\includegraphics[width=1\linewidth]{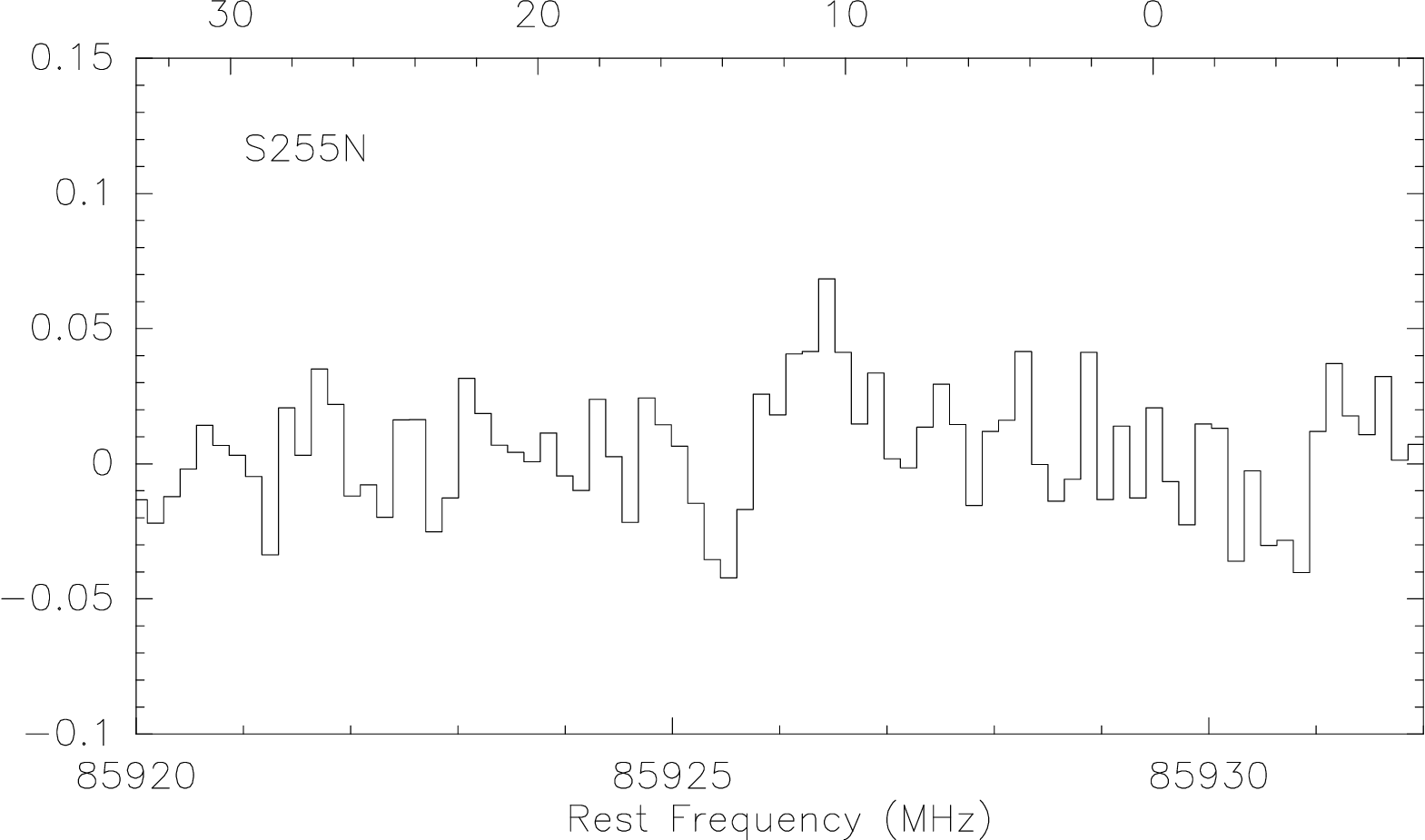}} 
\\ \vspace{1cm}
\end{minipage}

\vfill
\begin{minipage}[h]{0.3\linewidth}
\center{\includegraphics[width=1\linewidth]{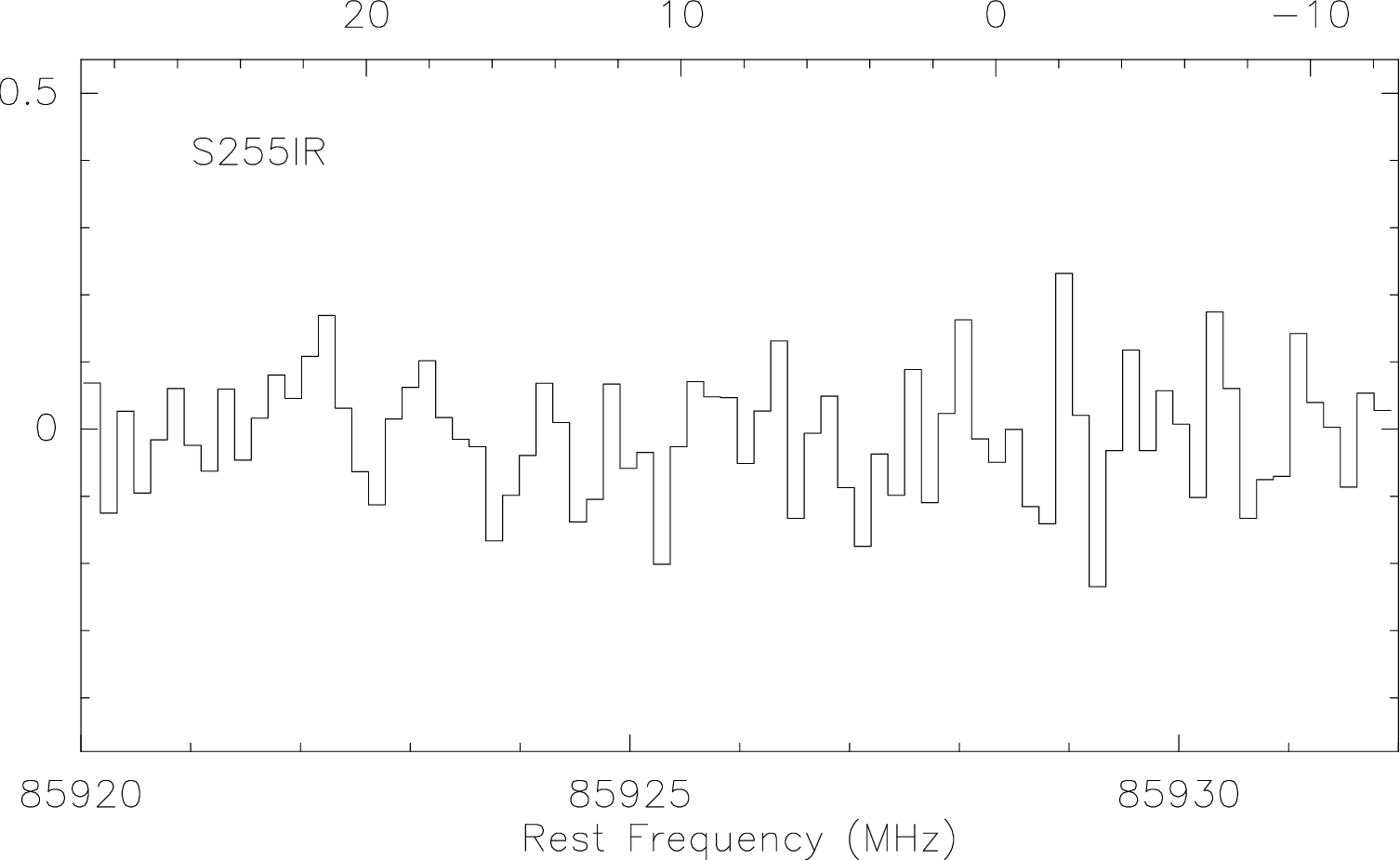}}  \\ \vspace{1cm}
\end{minipage}
\hfill
\begin{minipage}[h]{0.3\linewidth}
\center{\includegraphics[width=1\linewidth]{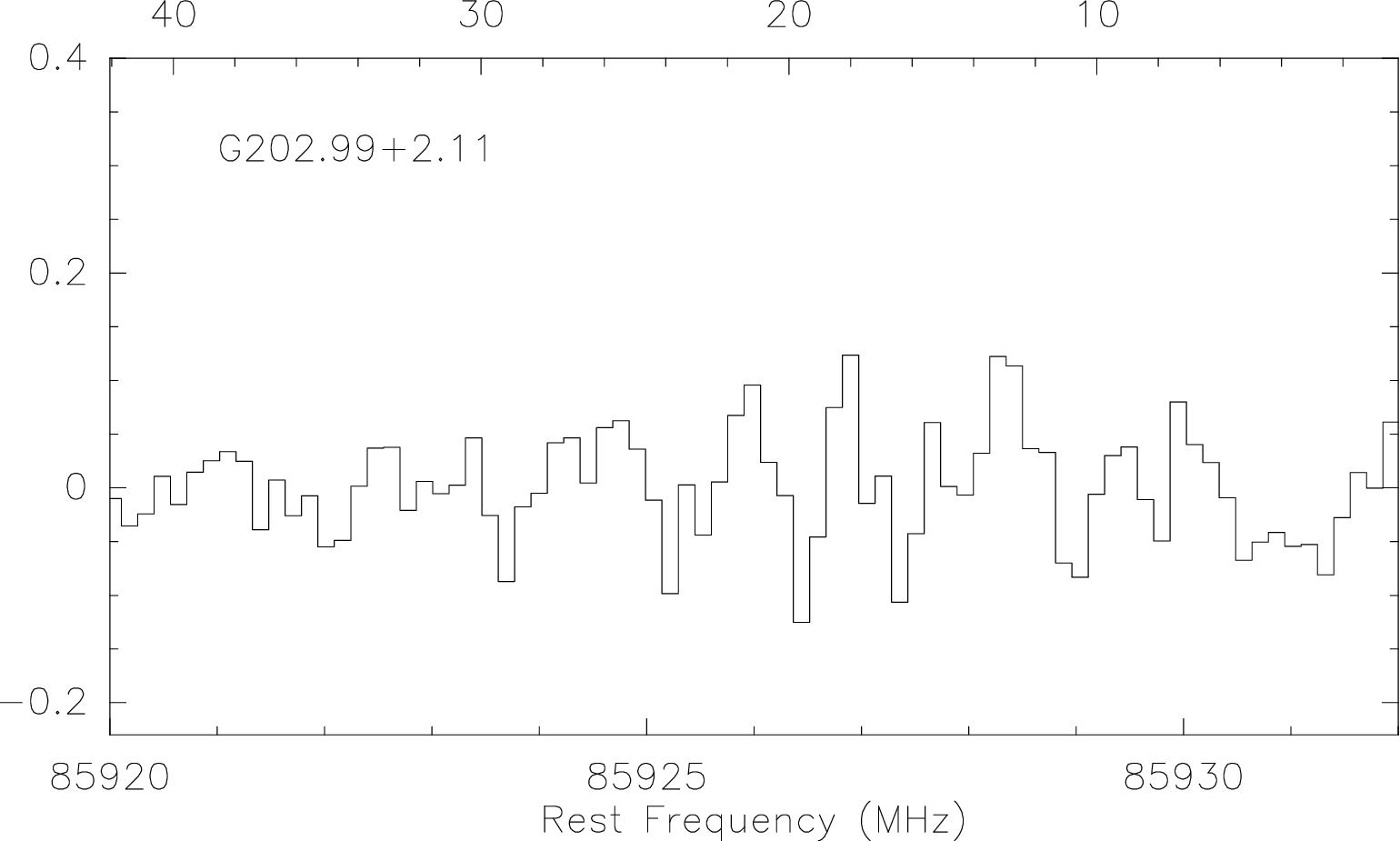}} 
\\ \vspace{1cm}
\end{minipage}
\hfill
\begin{minipage}[h]{0.3\linewidth}
\center{\includegraphics[width=1\linewidth]{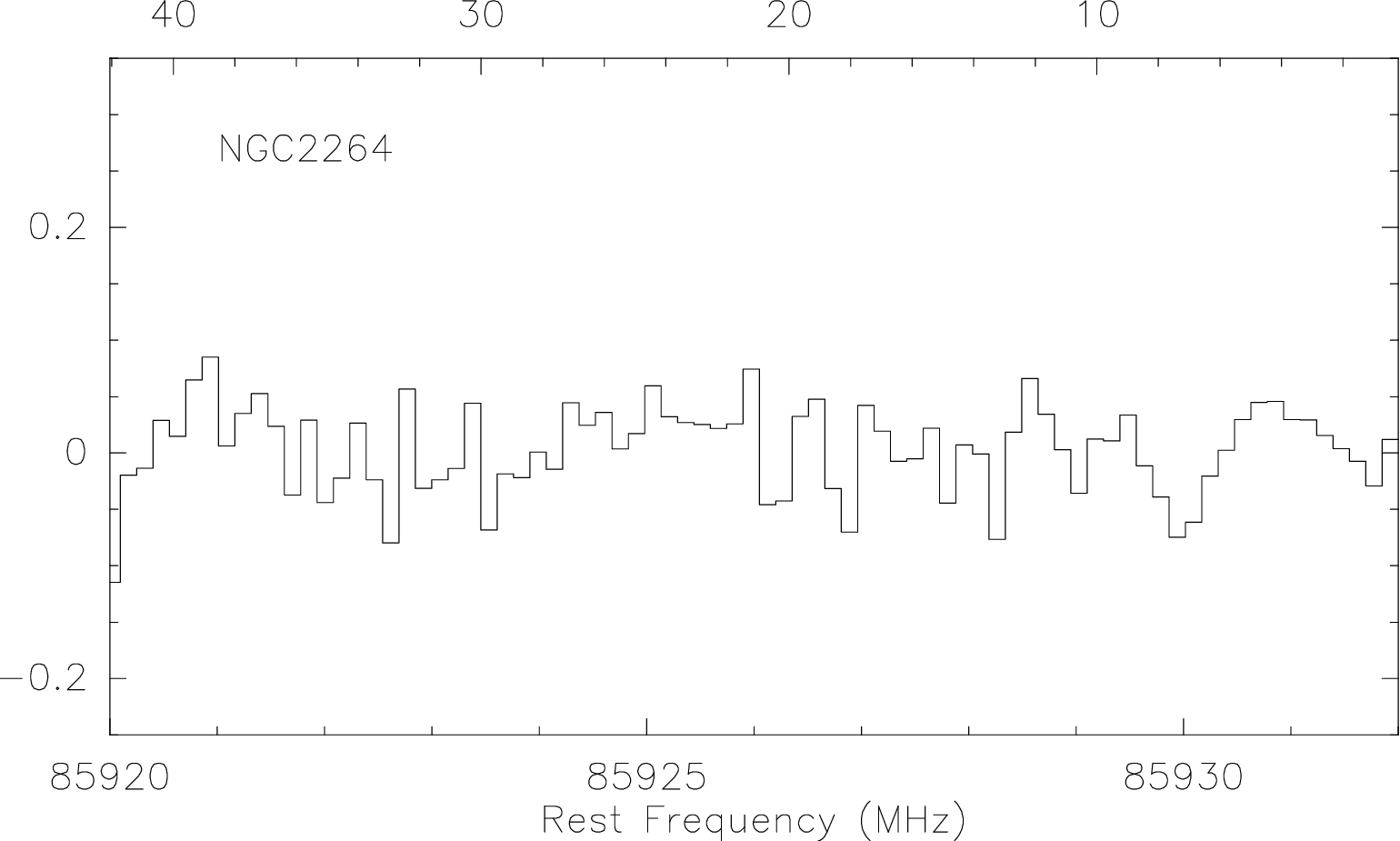}} 
\\ \vspace{1cm}
\end{minipage}

\caption{Spectra of the sources, where the NH$_2$D lines were not detected. The upper scale in the plots shows $V_\mathrm{LSR}$ in km/s.}
\label{ris:NH2D_pr_sm}
\end{figure}

\addtocounter{figure}{-1}

\begin{figure}[h]

\begin{minipage}[h]{0.3\linewidth}
\center{\includegraphics[width=1\linewidth]{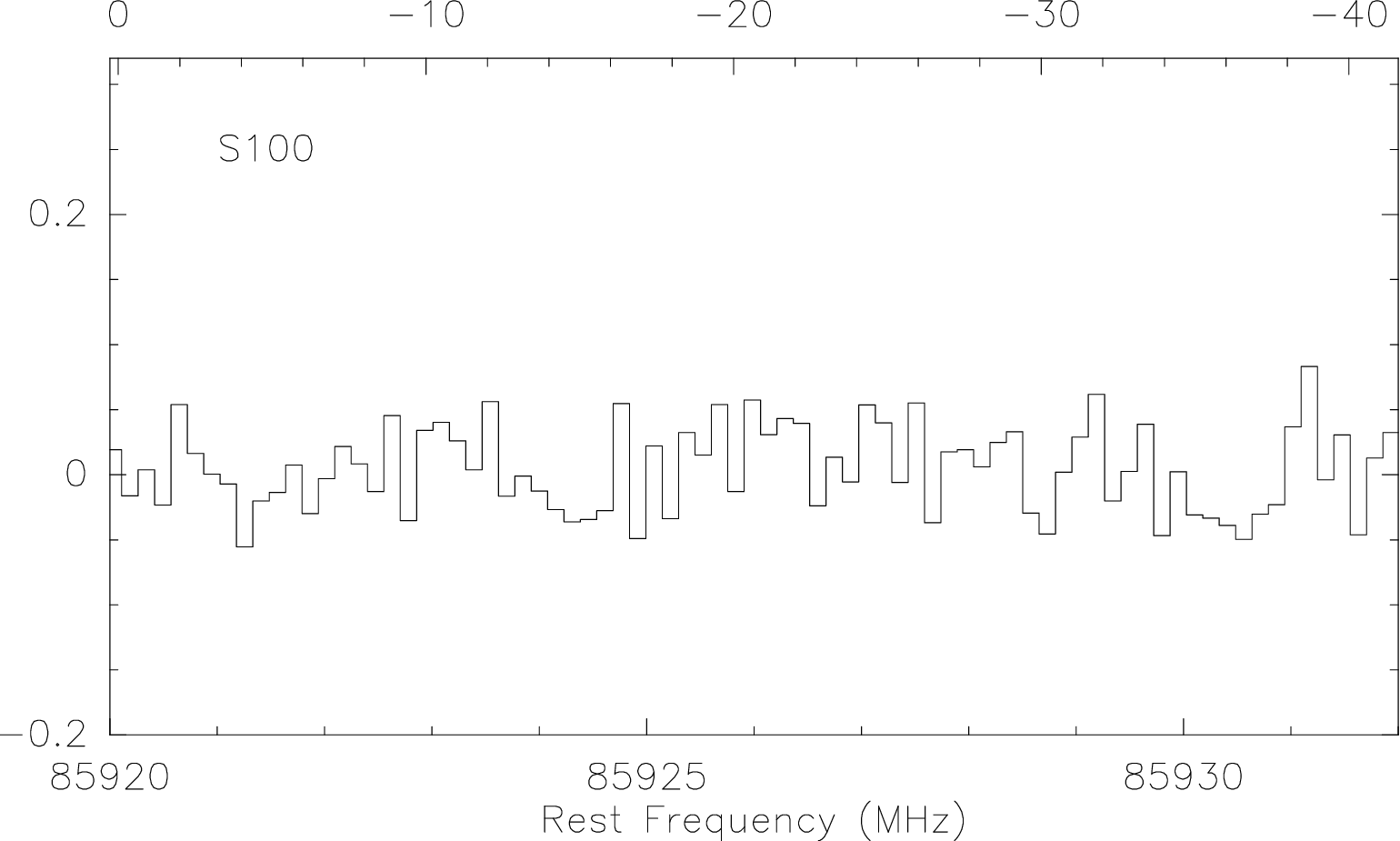}}  \\ \vspace{1cm}
\end{minipage}
\hfill
\begin{minipage}[h]{0.3\linewidth}
\center{\includegraphics[width=1\linewidth]{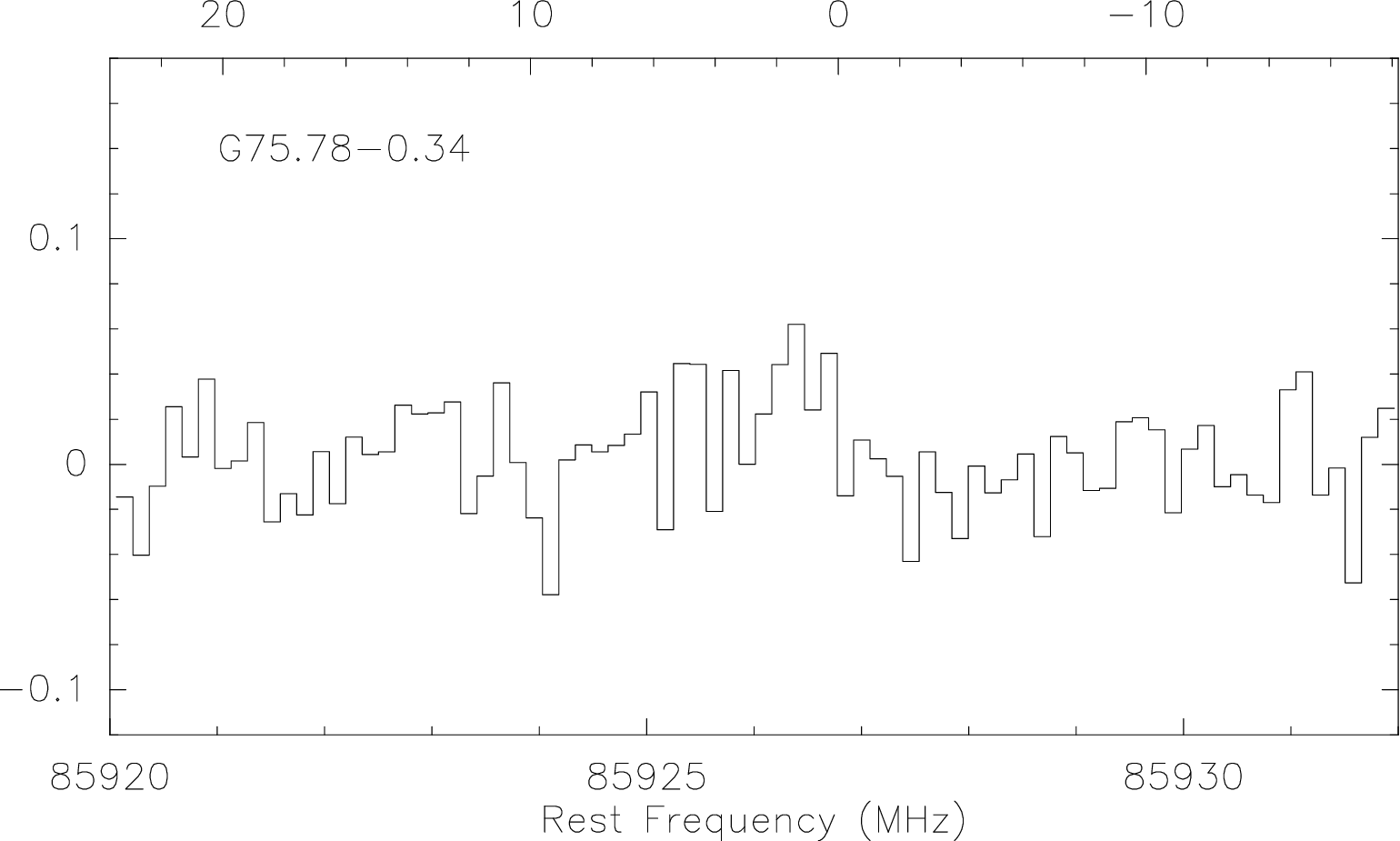}} 
\\ \vspace{1cm}
\end{minipage}
\hfill
\begin{minipage}[h]{0.3\linewidth}
\center{\includegraphics[width=1\linewidth]{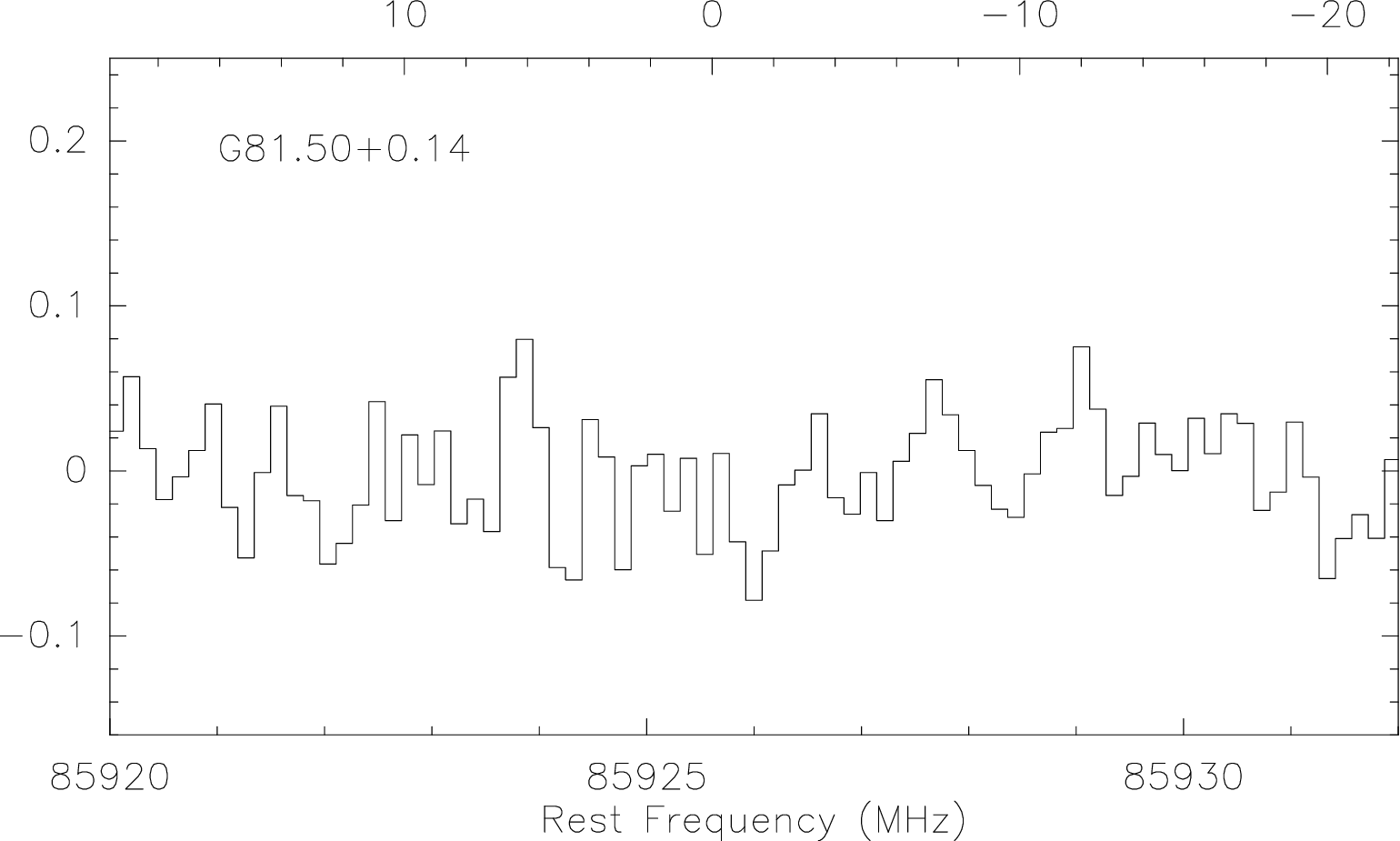}} 
\\ \vspace{1cm}
\end{minipage}

\vfill
\begin{minipage}[h]{0.3\linewidth}
\center{\includegraphics[width=1\linewidth]{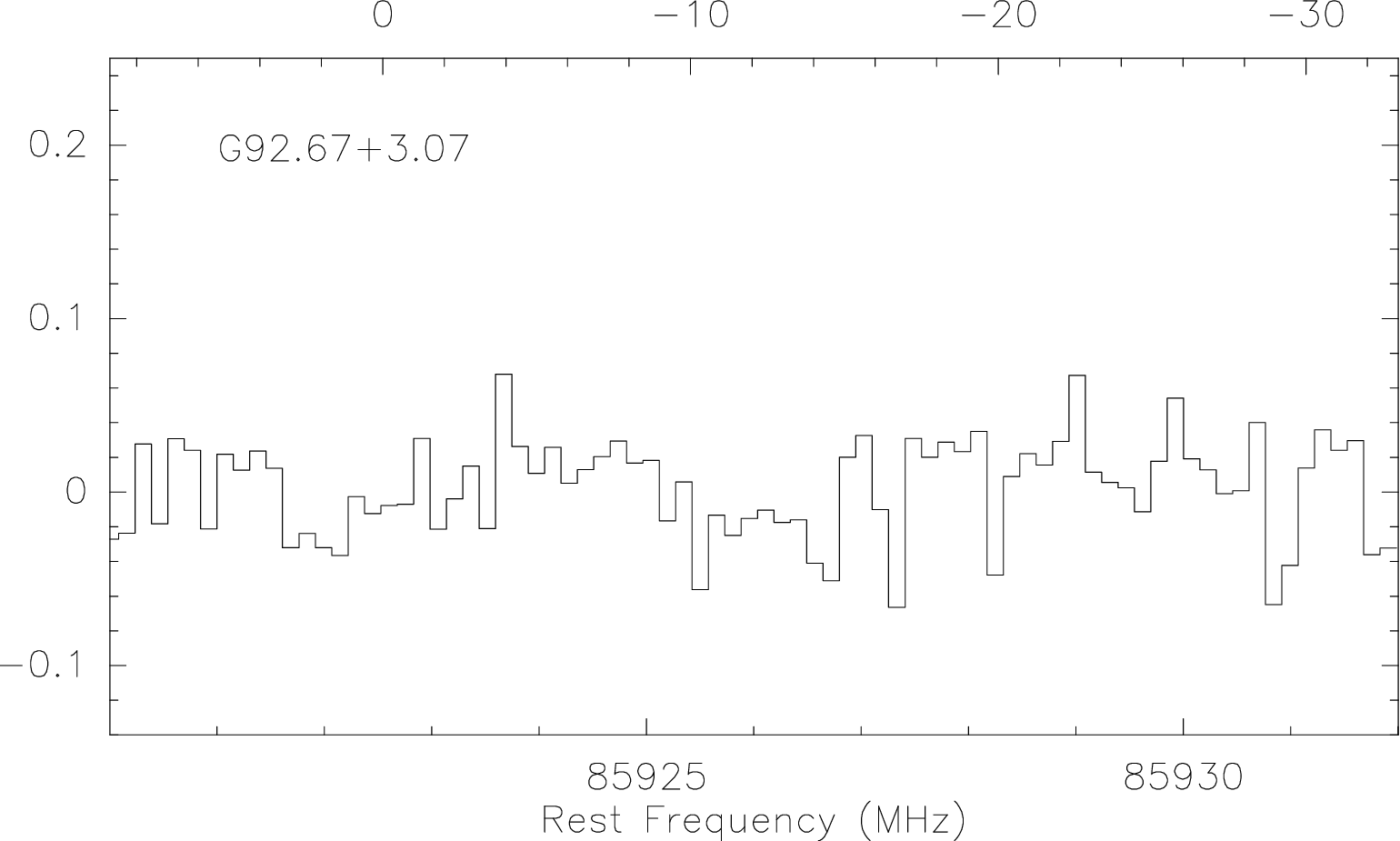}}  \\ \vspace{1cm}
\end{minipage}
\hfill
\begin{minipage}[h]{0.3\linewidth}
\center{\includegraphics[width=1\linewidth]{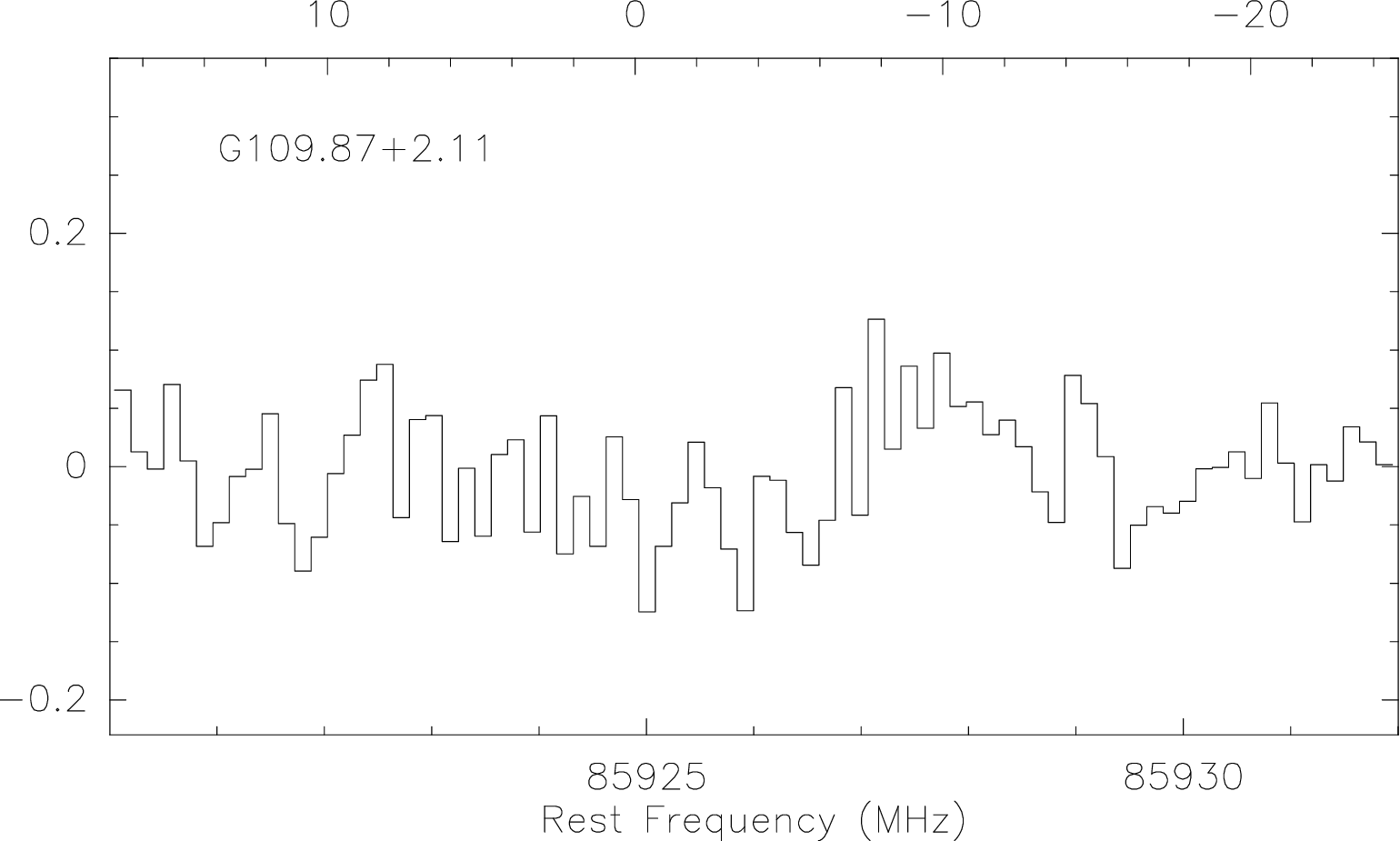}} 
\\ \vspace{1cm}
\end{minipage}
\hfill
\begin{minipage}[h]{0.3\linewidth}
\center{\includegraphics[width=1\linewidth]{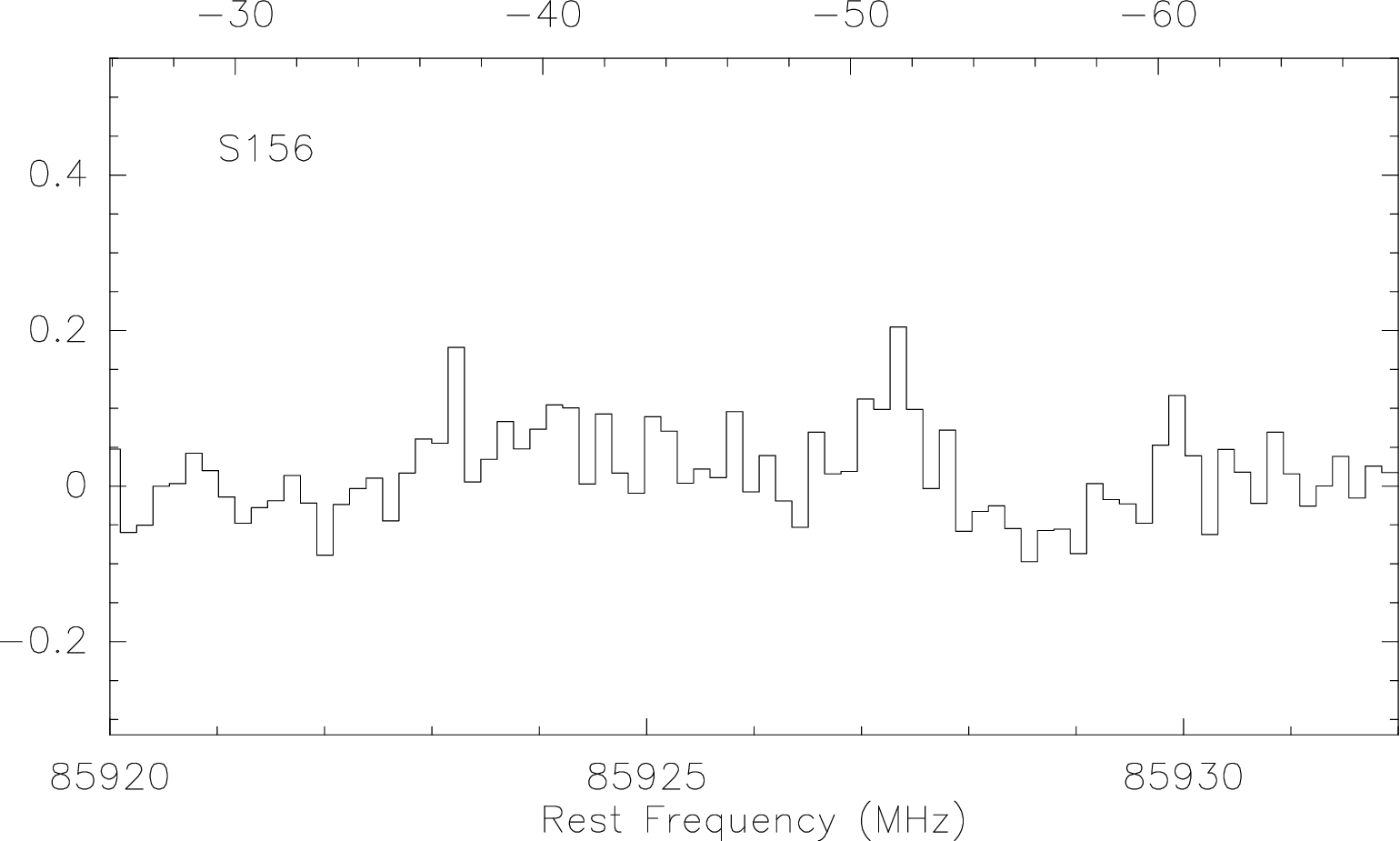}} 
\\ \vspace{1cm}
\end{minipage}

\caption{Continuation.}
\end{figure}

\begin{figure}[h]
\center{\includegraphics[width=1\linewidth]{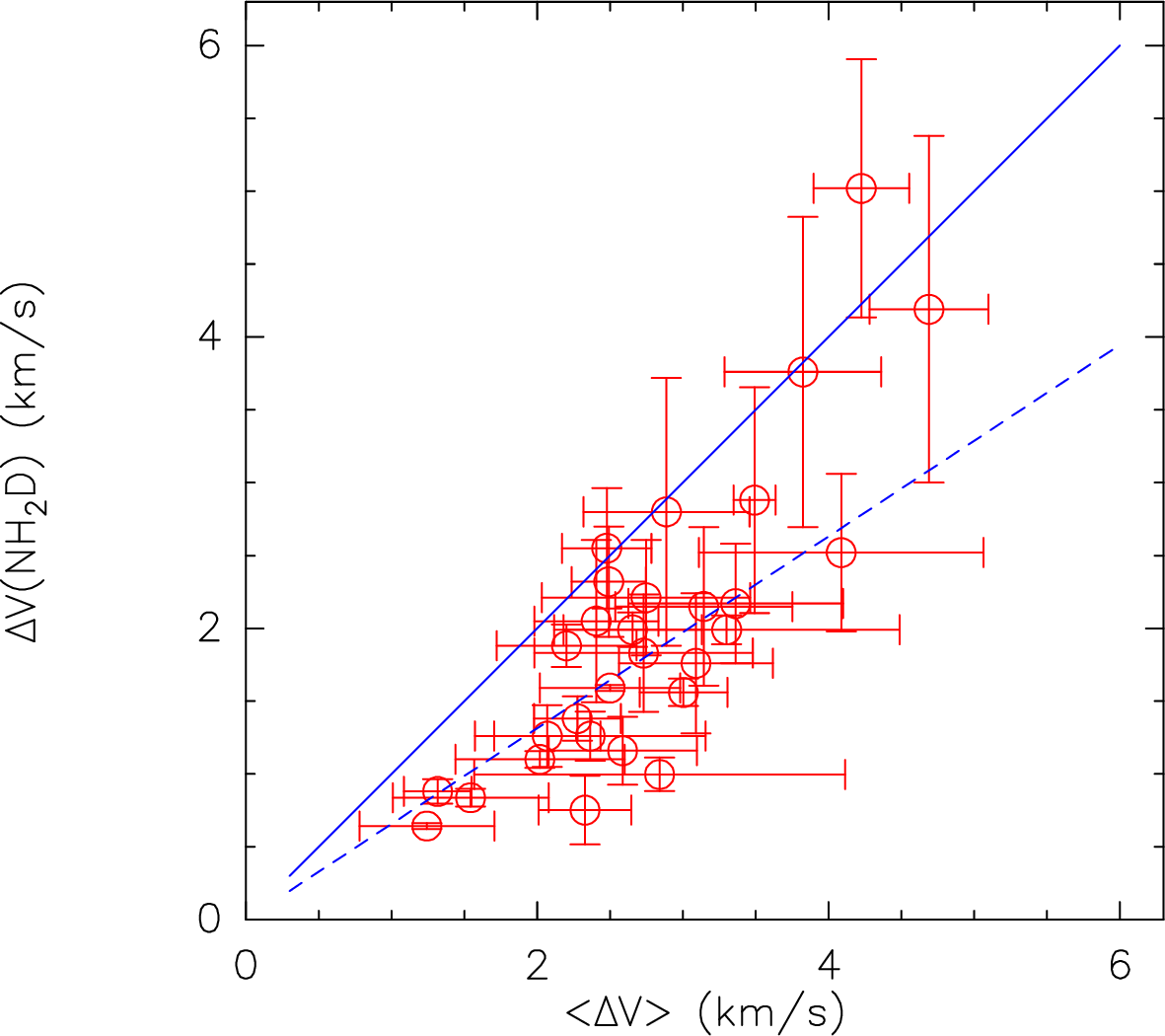}} \\
\caption{The comparison of line widths of singly deuterated ammonia NH$_2$D and the average width of narrow lines in the source $\langle \Delta V \rangle$. The solid line corresponds to $\langle \Delta V \rangle = \Delta V(\mathrm{NH_2D})$. The dashed line corresponds to $\langle \Delta V \rangle = 1.52\Delta V(\mathrm{NH_2D})$.}
\label{ris:dV-dV}
\end{figure}

\begin{figure}[h]
%\begin{minipage}[h]{0.47\linewidth}
\center{\includegraphics[width=1\linewidth]{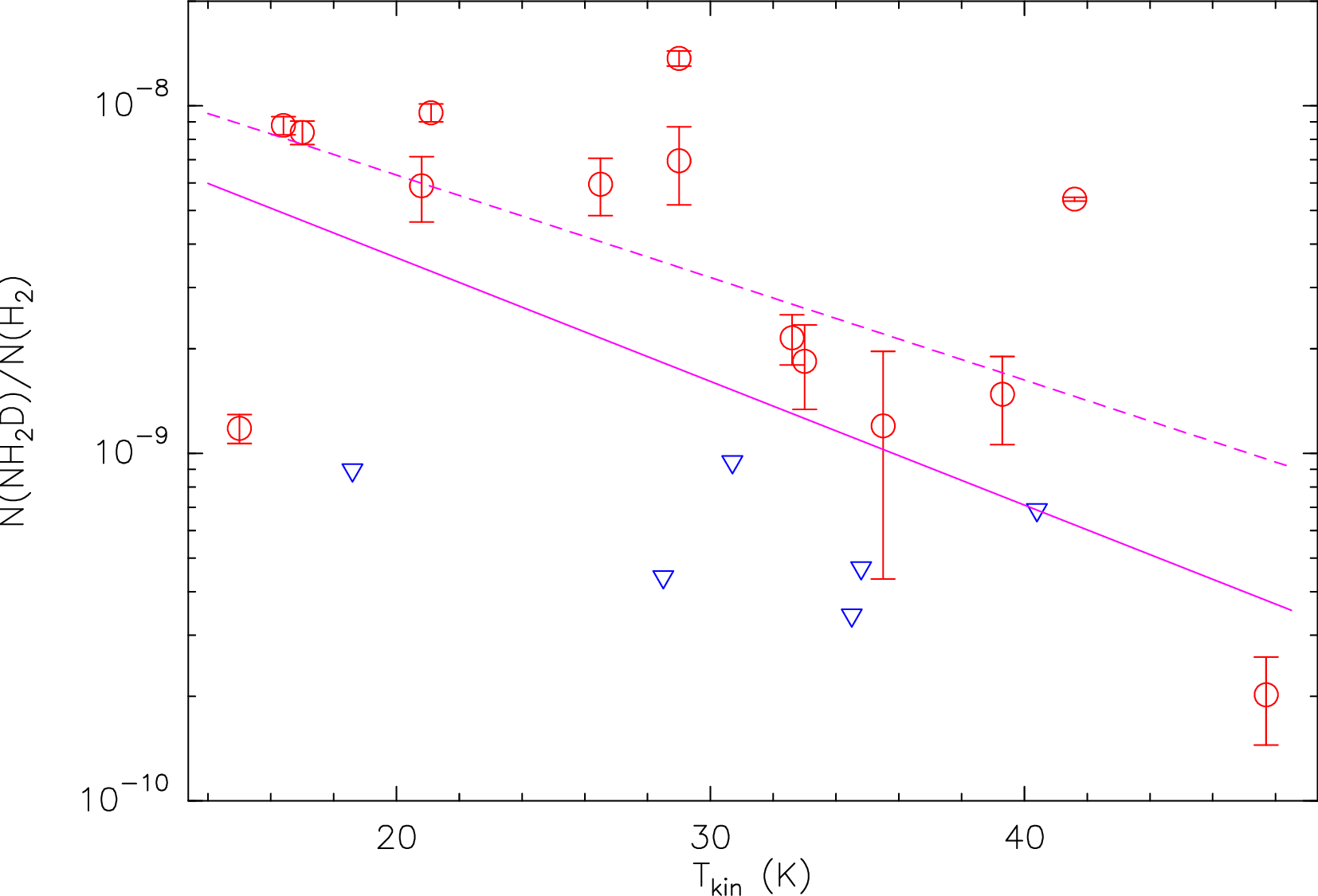}} \\
%\end{minipage} \\
\caption{The dependence of the relative abundance of singly deuterated ammonia NH$_2$D on the kinetic temperature of the source at gas density $n=10^4$ см$^{-3}$. The circles are measured values, the triangles are the upper detection limits of NH$_2$D. The solid line corresponds to the linear regression taking into account the limits. The dashed line corresponds to the linear regression without taking into account the limits.}
\label{ris:N/H-Tk}
\end{figure}

\begin{figure}[h]
%\begin{minipage}[h]{0.47\linewidth}
\center{\includegraphics[width=1\linewidth]{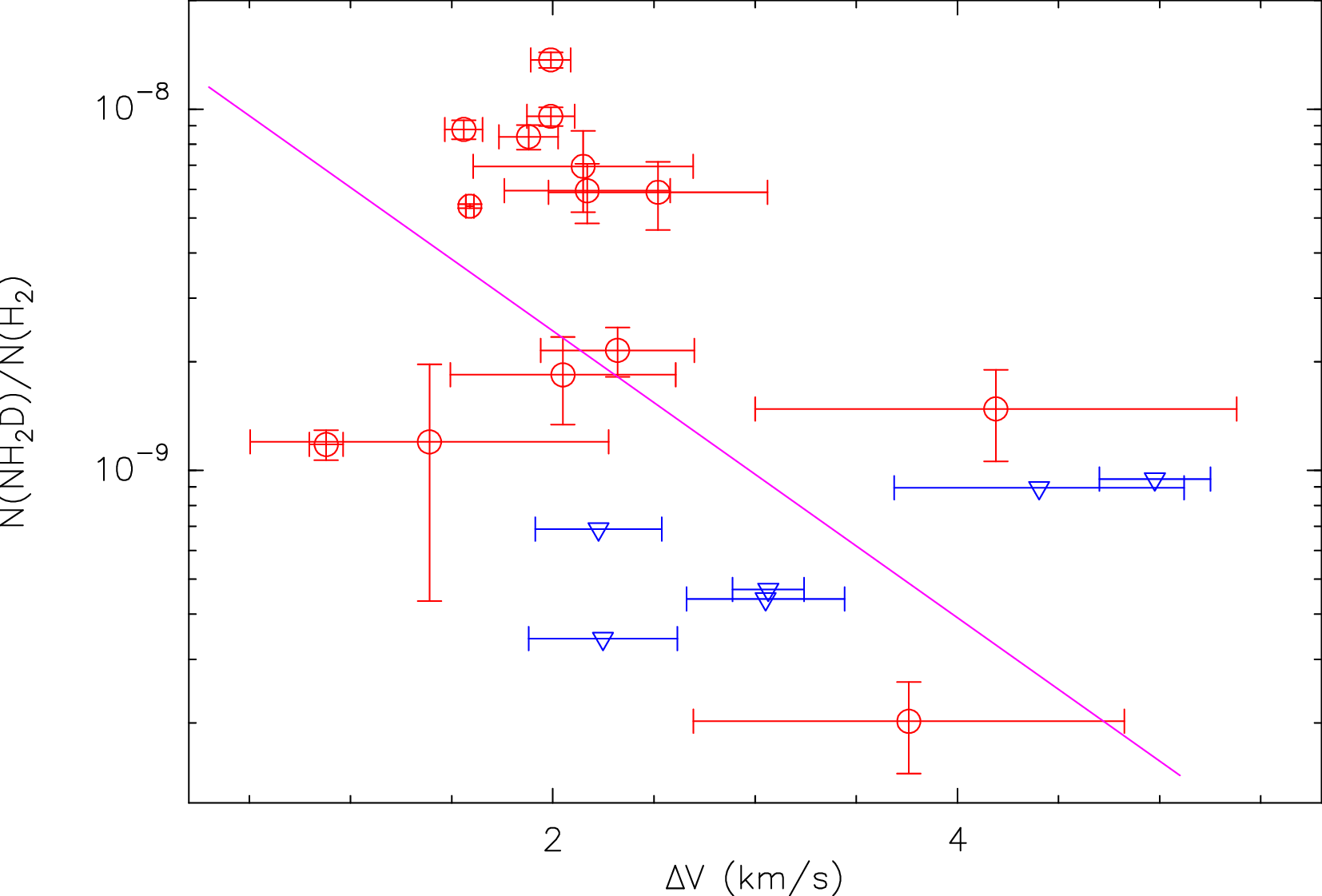}} \\
%\end{minipage} \\
\caption{The dependence of the relative abundance of singly deuterated ammonia NH$_2$D on the line width at gas density $n=10^4$ см$^{-3}$. The circles are measured values, the triangles are the upper detection limits of NH$_2$D. The solid line corresponds to the linear regression taking into account the limits.}
\label{ris:N/H-dV}
\end{figure}

\begin{figure}[h]
%\begin{minipage}[h]{0.47\linewidth}
\center{\includegraphics[width=1\linewidth]{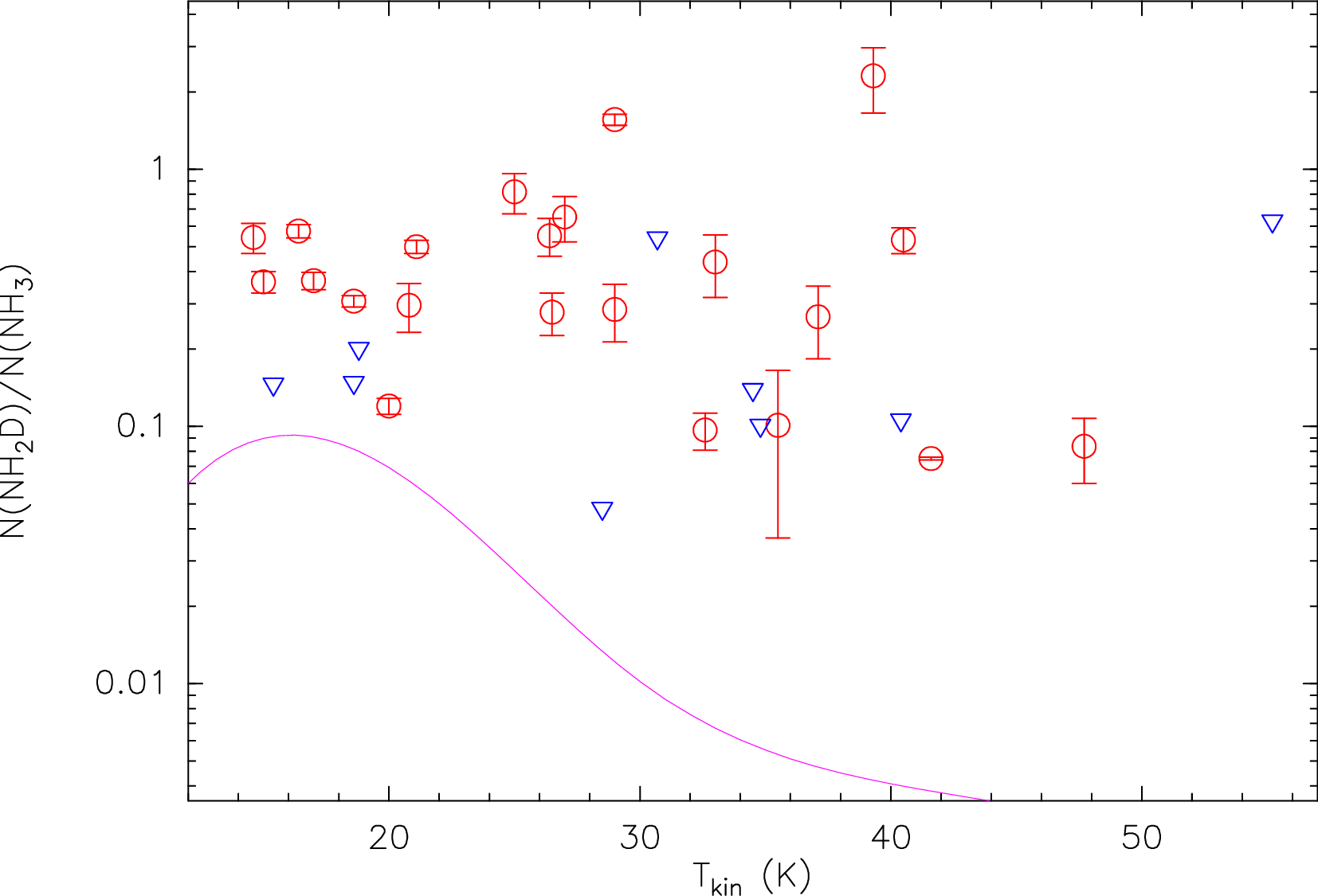}} \\
%\end{minipage} \\
\caption{The dependence of the NH$_2$D/NH$_3$ column density ratio on the kinetic temperature of the source at gas density $n=10^4$ см$^{-3}$. The circles are measured values, the triangles are the upper detection limits of NH$_2$D. The curve shows the model dependence of this ratio on the kinetic temperature of the source $T_{kin}$ according to the work \cite{Roueff07}}
\label{ris:N/N-Tk}
\end{figure}

\begin{figure}[h]
%\begin{minipage}[h]{0.47\linewidth}
\center{\includegraphics[width=1\linewidth]{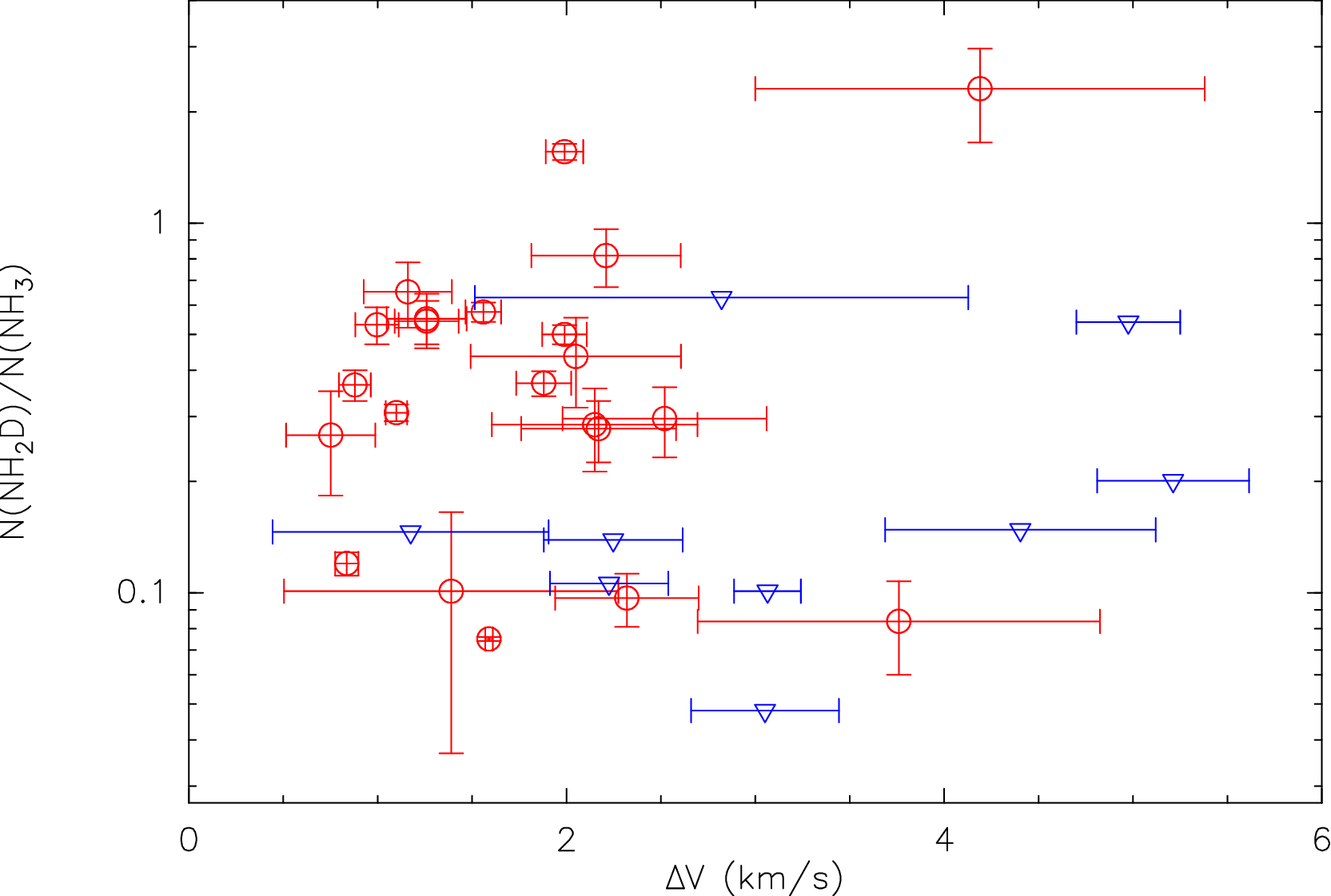}} \\
%\end{minipage} \\
\caption{The dependence of the NH$_2$D/NH$_3$ column density ratio on the line width at the gas density $n=10^4$ см$^{-3}$. The circles are measured values, the triangles are the upper detection limits of NH$_2$D.} 
\label{ris:N/N-dV}
\end{figure}

\begin{figure}[h]
\center{\includegraphics[width=1\linewidth]{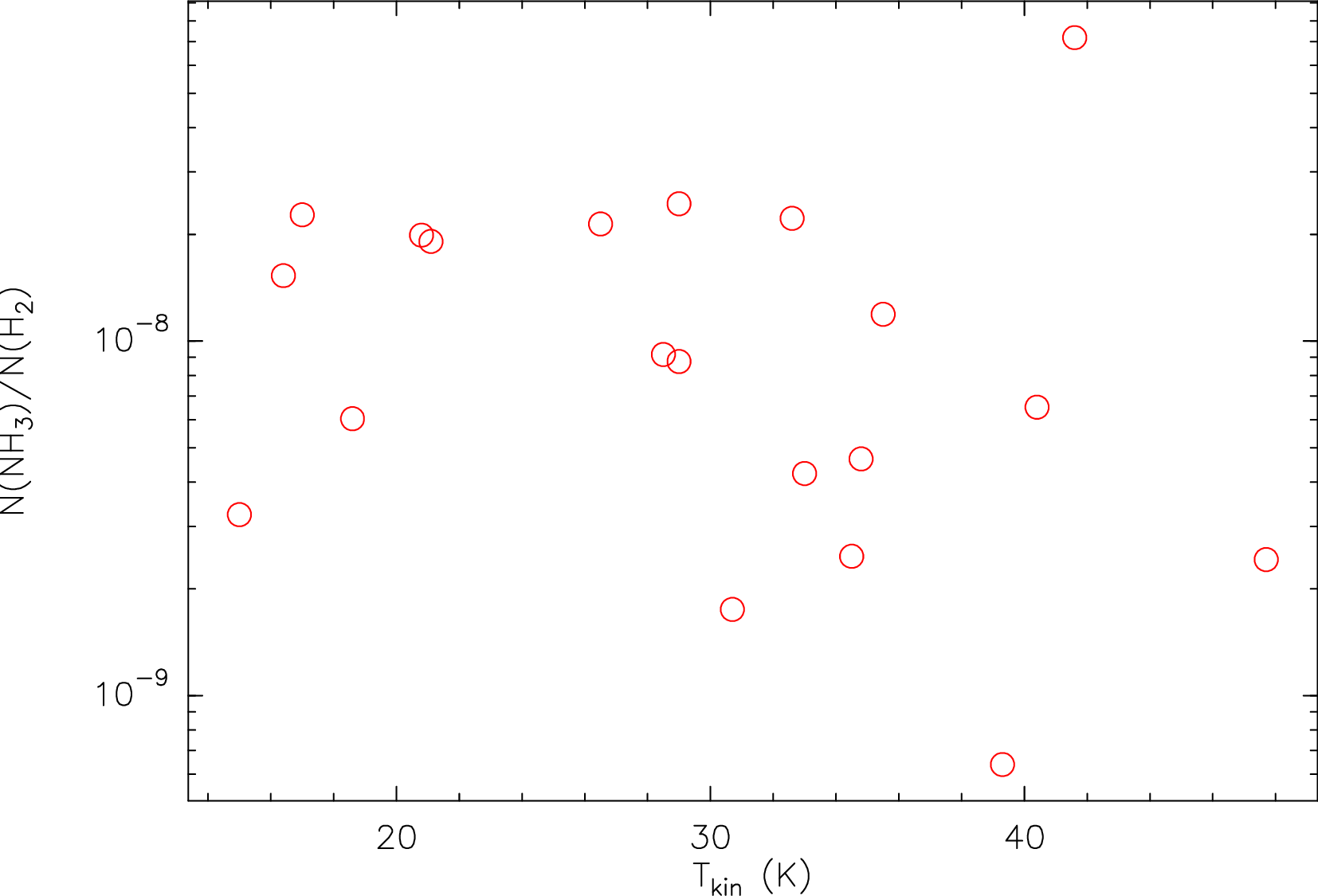}} \\
\caption{The dependence of the NH$_3$ relative abundance on the kinetic temperature of the source.}
\label{ris:NH3-Tk}
\end{figure}

\begin{figure}[h]
\center{\includegraphics[width=1\linewidth]{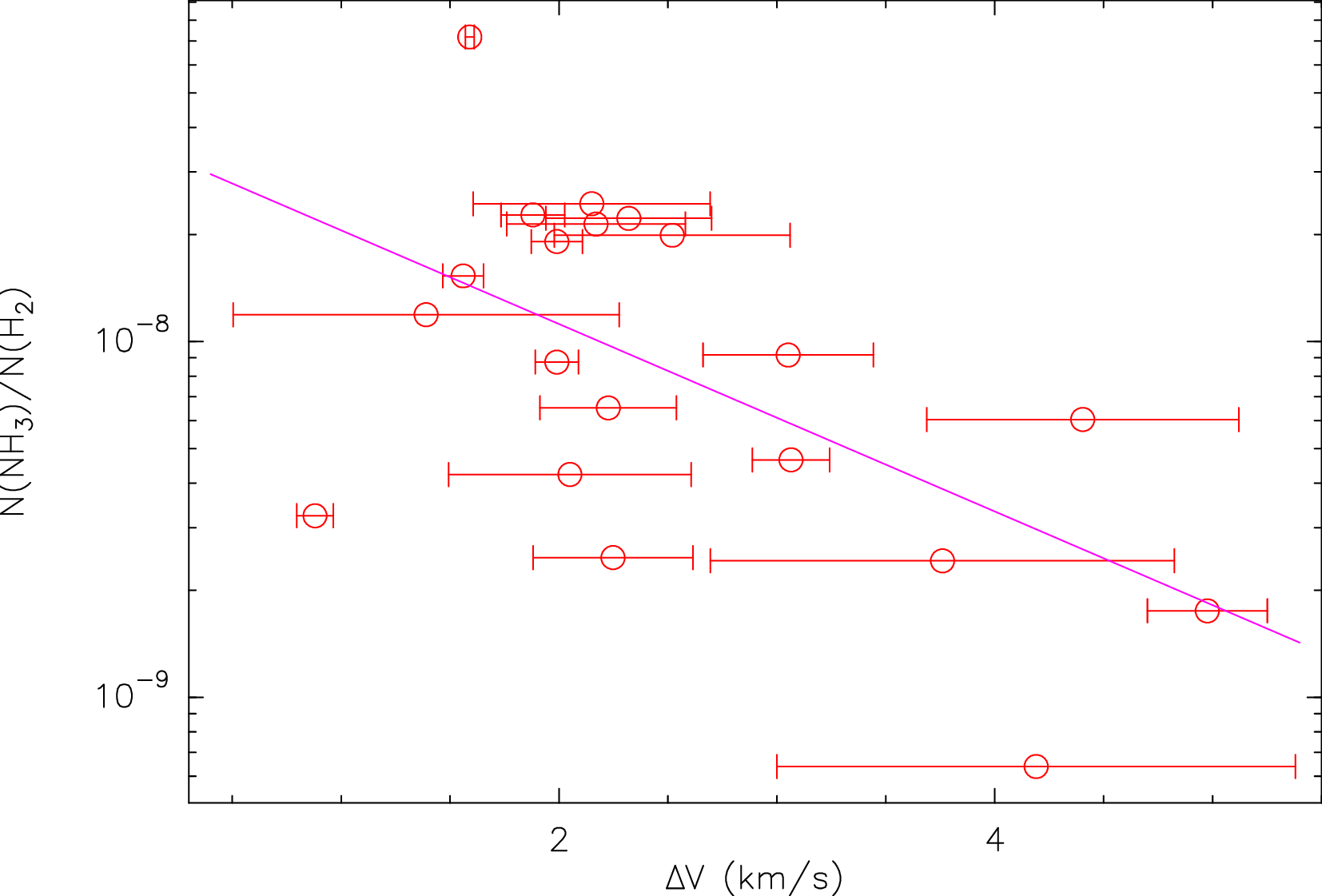}} \\
\caption{The dependence of the NH$_3$ relative abundance on the line width. The solid line is the line of linear regression.}
\label{ris:NH3-dV}
\end{figure}

% Для статей на русском языке далее следуют
% на английском языке название статьи, список авторов и краткая аннотация.
%

\end{document}